\documentclass[11pt]{article}

\usepackage{amsmath,amsthm,amssymb}
\renewcommand{\baselinestretch}{1.1}
 
\usepackage{graphicx,subfigure}
\newcommand{\nc}{\newcommand}
\nc{\ben}{\begin{eqnarray*}}
\nc{\een}{\end{eqnarray*}}
\nc{\bec}{\begin{equation}\begin{array}{lll}}
\nc{\eec}{\end{array}\end{equation}}
\def\bea{\begin{eqnarray}}
\def\eea{\end{eqnarray}}
 \def\bpm{\begin{pmatrix}}
 \def\epm{\end{pmatrix}}
 \def\bdet{\left|\begin{array}}
 \def\edet{\end{array}\right|}

\nc{\lb}[1]{\label{#1}}
\nc{\myinsert}[1]{\vspace*{2mm}\noindent{\small{\bf Insert: {#1}}}}
\nc\ds{\displaystyle}
\nc{\pd}{\partial}
\nc{\mref}[1]{(\ref{#1})}
\nc{\bra}[1]{\langle #1 |}
\nc{\ket}[1]{| #1 \rangle}
\nc{\ot}{\otimes}
\nc{\ed}{\end{document}}
\nc{\half}{\ensuremath{\frac{1}{2}}}
 \nc{\pvac}{|\textrm{vac};pr\rangle}
 \nc{\dpvac}{\langle\textrm{vac};pr}
 \nc{\vac}{|\textrm{vac}\rangle}
 \nc{\dvac}{\langle\textrm{vac}}
\nc{\id}{\mathbb{I}}
\nc{\ra}{\rightarrow}  
\nc{\lra}{\longrightarrow}
\nc{\uqp}{U^{\prime}_q (\widehat{sl}_2)}
\nc{\ub}{U^{\prime}_q (b_+)}
\nc{\vsl}{V(\sigma(\lambda))}
\nc{\vl}{V(\lambda)}  
\nc{\bu}{\bullet}
\nc{\an}{{\ell}}
\nc{\slth}{\widehat{\mathfrak{sl}}_2\hskip 1pt}
\newcommand{\uq}{U_q\bigl(\slth\bigr)}
 \nc{\ws}{\;\;}
\nc{\ad}{{\rm Ad}}
\nc{\hb}{\hbox}
\nc{\nn}{\nonumber} 
\nc{\curlra}{\buildrel{\sim}\over\longrightarrow}
\nc{\epp}{\varepsilon^{\prime}} 
\nc{\ol}{\overline}
\nc{\pl}{\prod\limits} 
\nc{\sli}{\sum\limits} 
\nc{\nin}{\noindent}
\newcommand{\Aqp}{{\cal A}_{q,p}}

\nc{\ga}{\alpha}
\nc{\gb}{\beta}
\nc{\gd}{\delta}
\nc{\gep}{\varepsilon}
\nc{\gz}{\zeta}
\nc{\gt}{\theta}
\nc{\gk}{\kappa}
\nc{\gl}{\lambda}
\nc{\gp}{\phi}
\nc{\gs}{\sigma}
\nc{\go}{\omega}
\nc{\gn}{\nu}
\nc{\gr}{\rho}

\nc{\gou}{\underline{\go}}
\nc{\un}{\underline{n}}
\nc{\um}{\underline{m}}
\nc{\uw}{\underline{w}}

\nc{\s}{\sigma}
\nc{\ep}{\varepsilon}
\nc{\z}{\zeta}
\nc{\g}{\gamma}
\nc{\zi}{\zeta^{-1}}

\nc{\gG}{\Gamma}
\nc{\gD}{\Delta}
\nc{\gT}{\Theta}
\nc{\gL}{\Lambda}
\nc{\gO}{\Omega}
\nc{\gP}{\Phi}

\nc{\cJ}{{\mathcal{J}}}
\nc{\cL}{\mathcal{L}}
\nc{\cF}{\mathcal{F}}
\nc{\cP}{\mathcal{P}}
\nc{\cS}{\mathcal{S}}
\nc{\cN}{\mathcal{N}}
\nc{\cH}{\mathcal{H}}
\nc{\cO}{\mathcal{O}}
\nc{\cT}{\mathcal{T}}
\nc{\cQ}{\mathcal{Q}}
\nc{\cW}{\mathcal{W}}
\nc{\cR}{\mathcal{R}}

\newcommand{\ZZ}{\mathbb{Z}}

\newcommand{\RR}{\mathbb{R}}

\nc{\fg}{\mathfrak{g}}

\nc{\barx}{\bar{x}}
\nc{\bi}{\bar{i}}
\nc{\bj}{\bar{j}}
\nc{\bgr}{\bar{\rho}}
\nc{\bA}{\bar{\alpha}}
\nc{\bB}{\bar{\beta}}
\nc{\bC}{\bar{\gamma}}
\nc{\by}{\bar{y}}
\nc{\brv}{\overline{V}}
\nc{\brp}{\overline{P}}

\nc{\tf}{\tilde{f}}
\nc{\te}{\tilde{e}}
\nc{\ts}{\tilde{s}}
\nc{\tgP}{\widetilde{\Phi}}
\nc{\tgPs}{\tilde{\Psi}}
\nc{\tgn}{\tilde{\nu}}
\nc{\tgl}{\tilde{\lambda}}
\nc{\tge}{\tilde{\eta}}
\nc{\txi}{\tilde{\xi}}
\nc{\tep}{\tilde{\epsilon}}

\newcommand{\mmatrix}[1]{\begin{matrix} #1 \end{matrix}}
\newcommand{\mat}[1]{\left(\mmatrix{#1}\right)}

\newcommand{\ena}{\end{eqnarray}}

\newcommand{\be}{\begin{eqnarray*}}
\newcommand{\en}{\end{eqnarray*}}

\newcommand{\la}{\lambda}
\newcommand{\snh}{{\rm snh}}
\newcommand{\sn}{{\rm sn}}

\newcommand{\cn}{{\rm cn}}

\newcommand{\dn}{{\rm dn}}
\newcommand{\am}{{\rm am}}

\newcommand{\End}{{\rm End}}
\newcommand{\Ad}{{\rm Ad}}

\newcommand{\C}{{\Bbb C}} %??
\def\H{{\cal H}}
\def\F{{\cal F}}

\newcommand{\vep}{{\varepsilon}}

\newcommand{\tr}{{\rm tr}}
\nc{\G}{\mathcal{G}}

\newcommand{\tPsi}{\widetilde{\Psi}}
\newcommand{\tPhi}{\widetilde{\Phi}}

\begin{document}
\bibliographystyle{unsrt}
%%%%%%%%%%%%%%%%%%%%%%%%%%%%%%%
\title{Exact Form-Factor Results for the Longitudinal Structure Factor of the Massless XXZ Model in Zero Field}
\author{Jean-S{\'e}bastien Caux$^1$,  Hitoshi Konno$^2$, Mark Sorrell$^3$, and Robert Weston$^4$\\[5mm]
{\it \small$^1$Institute for Theoretical Physics, Universiteit van Amsterdam,
Science Park 904,} \\ 
{\it \small Postbus 94485, 1090 GL Amsterdam, The Netherlands}\\
{\it\small $^2$Department of Mathematics, Hiroshima University, Higashi-Hiroshima 739-8521, Japan}\\
{\it\small $^3$Department of Mathematics \& Statistics,
The University of Melbourne,
Parkville, VIC, 3010,
Australia }\\
{\it \small$^4$Department of Mathematics, Heriot-Watt University, Edinburgh EH14 4AS, UK}
}
\date{\today}
\maketitle
\begin{abstract}
We consider the XXZ quantum spin chain in its massless, disordered regime at zero field. We derive an exact expression for the two-spinon form-factor of $S^z=\half\sigma^z$ by taking a limit of the massive XYZ form-factors found by Lashkevich and by Lukyanov and Terras. This result is used to find the two-spinon contribution to the spectral decomposition of the longitudinal structure factor 
$S^{zz}(k,w)$. We find that this contribution provides an  accurate approximation to the full structure factor over a wide range of the anisotropy parameter. The  asymptotic behaviour of $S^{zz}(k,w)$ is computed as the upper and lower $w$ thresholds of the two-spinon $(w,k)$ band are approached, and an analysis of the region of validity of this threshold behaviour is performed. Our results reproduce and refine existing threshold behaviour predictions and extend these results to an accurate description throughout the two-spinon continuum.
 
\end{abstract}
\begin{center}{\it Dedicated to Professor Michio Jimbo on his sixtieth birthday}
\end{center}

\section{Introduction}

Interacting quantum systems have unique properties when space is one-dimensional \cite{GiamarchiBOOK}. On the one hand, the simple fact that 
particles cannot avoid each other means that the nature of quantum dynamics is
 complicated in one dimension. 
The inevitability of particle interactions means that all excitations are collective ones. 
In particular,  
the quasi-free excitations of Fermi liquids in higher dimensions are replaced by the non-perturbative excitations described at low energies by the theory of Tomonaga-Luttinger liquids in one dimension. 
On the other hand, the tools for dealing with non-perturbative systems are far more sophisticated for 
one dimension than for higher dimensions. For the class of systems that are quantum integrable, the mathematical toolbox is particularly full.

 The most studied interacting, one-dimensional quantum integrable system is the Heisenberg, or XXZ, quantum spin chain
\cite{1928_Heisenberg_ZP_49,1958_Orbach_PR_112}
\bea H_{XXZ}=
-\frac{J}{4}  \sli_{i=1}^{N} (\gs^x_i \gs^x_{i+1} + \gs^y_i \gs^y_{i+1}- \Delta  \gs^z_i \gs^z_{i+1})
\lb{HXXZ}\eea
with $\Delta=\cos(\pi/(\xi+1))$. 
This model has two nice properties: many exact non-perturbative results exist for both finite and infinite $N$ (see for example \cite{Takahashi99,McCoy10} and the many references they contain); and it is
experimentally realised. When $J>0$, the XXZ chain has a massive antiferromagnetic phase for $\Delta>1$ and 
is realised for example by $\mbox{CsCoCl}_3$ \cite{1995_Goff_PRB_52}.
When $|\Delta|\leq 1$ the model has a massless disordered phase and has been realised 
experimentally by frustrated spin ladder systems \cite{1998_Totsuka_PRB_57,2001_Watson_PRL_86,2009_Thielemann_PRL_102} 
and, very recently, has become in principle accessible using optical lattices \cite{2003_Kuklov_PRL_90}-\nocite{2003_Duan_PRL_91,2003_Garcia-Ripoll_NJP_5,2007_Lewenstein_AP_56}\cite{2011_Bloch_TBP}.

In paper \cite{2011_Caux_PRL_106}, we have considered the  $N\ra \infty$ limit of the Hamiltonian \mref{HXXZ} in the massless phase with  $0\leq \Delta\leq 1$ (the change in gauge from $H_{XXZ}$ to the Hamiltonian $H$ considered in \cite{2011_Caux_PRL_106} is given in Section 4 of the current paper). In this phase, the system is a Tomonaga-Luttinger liquid \cite{1981_Haldane_JPC_14,1981_Haldane_PLA_81} whose fundamental particles are `spinons': spin-1/2 excitations that can be viewed as domain walls dressed with quantum fluctuations \cite{1981_Faddeev_PLA_85}. When $\Delta=0$ these excitations are non-interacting and are described by free fermions. Away from $\Delta=0$ the spinons are shaped by the interactions in the bulk, and these interactions can be probed by determining how spinons contribute to correlation functions. The correlation function
 we have considered in detail in \cite{2011_Caux_PRL_106} is the longitudinal structure factor (LSF)
\bea S^{zz}(k,\omega)= \sli_{j\in \ZZ} e^{-ikj} \int_{-\infty}^{\infty} dt e^{i\omega t} \dvac| S_j^z(t) S_0^z(0)\vac\lb{lsf}\eea
where $S^z=\half \gs^z$. The LSF can be measured directly in neutron scattering experiments (see \cite{NEUTRONSCATTERINGBOOK} and references therein).
 $S^{zz}(k,\omega)$ can be  computed  by resolving the identity in terms of a complete set of spinon states
$\id =  \sli_{\alpha} \ket{\alpha}\bra{\alpha}$
and inserting into \mref{lsf} to give the spectral decomposition
\ben S^{zz}(k,w)=\sli_{\alpha} (2\pi)^2 \delta(k-K(\alpha)) \delta(w-W(\alpha) \, |\dvac| S_0^{z}|\alpha\rangle|^2,\een
where $K(\alpha)$ and $W(\alpha)$ are the momentum and energy of the state $\ket{\alpha}$. In our paper \cite{2011_Caux_PRL_106},  we
 have presented the result for the exact two-spinon contribution to this sum and shown that this contribution is a highly accurate truncation, saturating two independent sum rules to around 99\% at $\Delta=0.5$. 

The main purpose of the current paper is to explain the derivation of the results presented in brief in \cite{2011_Caux_PRL_106}. In particular, we show how the relevant two-spinon form factors $\dvac| S_0^{z}|\alpha\rangle$ are obtained for the massless phase of the XXZ model.
 The general method that we follow to obtain massless XXZ form-factors is usually called the vertex operator approach (VOA). The VOA for the antiferromagnetic XXZ model is described in detail in \cite{JM}, where the representation theory of the
 quantum affine algebra $U_q(\slth)$ plays an essential role. 
This theoretical framework has been exploited to offer results on dynamical correlation functions
of the Heisenberg chain both at the isotropic antiferromagnetic point, where two \cite{1996_Bougourzi_PRB_54,1997_Karbach_PRB_55} and four-spinon \cite{1997_Abada_NPB_497,2006_Caux_JSTAT_P12013} contributions have been obtained, and for the gapped antiferromagnet, where two-spinon contributions to
the transverse correlator were given \cite{1998_Bougourzi_PRB_57,2008_Caux_JSTAT_P08006}.
In order to extend this approach to deal with the massless regime we follow the strategy proposed in 
 \cite{JKM}: we use the VOA for the XYZ spin chain in the principal regime, 
map it to the XYZ disordered regime, and then take a massless 
limit to the XXZ model.

 The XYZ model Hamiltonian is given by 
 \bea H_{XYZ}=
-\frac{1}{4} \sli_{i\in \ZZ} (J_x \gs^x_i \gs^x_{i+1} + J_y \gs^y_i \gs^y_{i+1}+ J_z  \gs^z_i \gs^z_{i+1}).\lb{HXYZ}\eea
%{\bf[Note: I have absorbed the $J$ into $(J_x,J_y,J_y)$ appearing in Equation \eqref{eqn:Js}]}
The VOA for the XYZ model was developed in \cite{JMN,JKKMW}. The role of $U_q(\slth)$ in the XXZ model is taken in the more general XYZ case by the elliptic quantum group of vertex type ${\cal A}_{q,p}(\slth)$ \cite{FIJKMY,MR1356513,JKOS}. 
The VOA is directly valid in the principal regime of the XYZ model for which 
$|J_y|\leq J_x\leq -J_z$.  However, as demonstrated in \cite{BaxterBOOK}, is is possible to map the principal regime to any other region in the phase diagram of the XYZ model.
In particular, we can map to the disordered region 
for which $|J_z|\leq J_y \leq J_x$.
A transformation which achieves this is 
$H^{\tiny{disord}}_{XYZ} = G \,H^{princ}_{XYZ}\, G^\dagger$, where
\bea
 G= \cdots \ot U_1 \ot U_0 \ot U_1 \ot U_0 \ot \cdots,\quad U_0=\frac{-1}{\sqrt{2}}\bpm i&1\\i&-1 \epm ,\ws U_1=\frac{1}{\sqrt{2}}
\bpm 1&i\\-1&i\epm.
\lb{gauge}\eea
%{\bf[ Note: I have changed the order of $U_0$ and $U_1$ in this $G$ - and in sec2.tex where the posn that $U_0$ acts is more accurately defined - in order to simplify the discussion in sec2]} 
In this way, it is possible to use the VOA results in the principal regime in order to find form-factors in the disordered region. It remains only to take the limit $J_x\ra J_y$ limit, in order to obtain form-factors for the massless XXZ model.

However, things are not quite so simple: while it is true that the VOA to the XYZ model parallels that of the XXZ model, it does differ in one important respect. The explicit multiple-integral expressions for form-factors in the XXZ case are obtained by using a Bosonization technique - more precisely a free-field-representation of the quantum affine algebra $U_q(\slth)$. The problem is that such a free-field representation has not yet been found for the quantum elliptic algebra ${\cal A}_{q,p}(\slth)$ 
relevant to the XYZ model. The reason for this technical problem is ultimately linked to the absence of charge conservation around a vertex in the 8-vertex model associated with the XYZ chain. 

This problem has been considered before in the literature, and there are two ways to get around it. The first approach involves mapping the XYZ model to a solid-on-solid (SOS) model, for which a free-field realisation {\it does} exist (and for which the relevant algebraic structure is the elliptic quantum group of face type $U_{q,p}(\slth)$\cite{Konno98,JKOS99,Konno09}). This was the method developed and used by Lashkevich and Pugai to obtain expressions for both form-factors and correlation functions in the principal regime XYZ model \cite{LaP98,La02}. The method was extended to higher spin analogues of the XYZ model in \cite{KKW05}. 
The second approach is applicable specifically to the massless XXZ model; the idea here is to derive and solve a difference equation 
(a deformed KZ-equation) for correlation functions from the analogous equation for the XYZ model \cite{JM96}, or to construct a field realisation only after having already mapped to the disordered regime and taken the massless limit \cite{JKM}.

In this paper, we take the first approach, the main reason being that simplified expressions for the resulting XYZ two-particle form-factors mapped to the disordered regime are already present in the literature \cite{LT03}. Our contribution is to take the appropriate massless XXZ scaling limit of these existing results, and to use them to compute the exact two-particle contribution to the longitudinal structure factor.

In Section 2 of this paper, we describe the key components of the VOA to the XYZ model in the principal regime, the map to the disordered regime, and the limit from the existing disordered-regime XYZ form-factor results to our new 
massless XXZ form-factor expressions. In Section 3, we give the derivation of  expression \mref{eq:Szz2} for the two-spinon 
contribution to the longitudinal structure factor $S^{zz}(k,w)$. This was the key result quoted in the earlier paper \cite{2011_Caux_PRL_106}.
In Section 4, we present a detailed quantitative analysis of the structure factor, an analytic derivation of the asymptotic threshold 
behaviour close to the upper and lower $w$ limits of the two-particle $(w,k)$ continuum, and an analysis of the region of the $(w,k)$ band over which this threshold behaviour is a good practical approximation for different $\Delta$ values. We present some concluding remarks in Section 5. Finally, in Appendices A and B, we give the definitions and required properties of elliptic functions, and present an alternative derivation of the mapping of principal form factors to disordered ones.

 % Intro
\section{From XYZ to Massless XXZ}
\setcounter{equation}{0} 
A general multiple-integral expression for principal regime XYZ form-factors can be constructed by following the approach of \cite{LaP98,La02}. The case of the form-factor of the operator $\gs^z$ is considered in detail in the paper \cite{La02}, and the author demonstrates a technique that enables him to obtain an expression for this form-factor which involves no integrals. This approach is extended to $\gs^x$ and $\gs^y$ form-factors by Lukyanov and Terras in \cite{LT03}. Using the mapping mentioned in the previous section these authors present results directly in the disordered region of the XYZ model. In this section, we review these results and take the appropriate scaling limit to the massless XXZ model. This limit is different to the sine-Gordon limit considered in \cite{LT03}.

\subsection{The XYZ model in the principal regime}
The XYZ Hamiltonian is derived from the 8-vertex model elliptic $R$ matrix \cite{BaxterBOOK} given 
by 
\bea
R(u)=\rho(u)\mat{a(u)&&&d(u)\cr
                                &b(u)&c(u)&\cr
                                &c(u)&b(u)&\cr
                                d(u)&&&a(u)\cr} \label{ellR}
\ena
with
\be
a(u)=\snh(\la(1-u)),\ b(u)=\snh(\la u),\ c(u)=\snh(\la),\ d(u)=k \snh(\la(1-u))\snh(\la u)\snh(\la).\en
Here $\snh(u)=-i\sn(iu)$ is the Jacobi elliptic function with modulus $k$. The definitions, relations between and required properties of all elliptic functions used in this paper can be found in Appendix \ref{app:ellipticfn}.  Let $K, K'$ be the corresponding complete elliptic integrals given in Appendix \ref{app:ellipticfn}. We use the variables
\be
x^{2r}=e^{-\frac{\pi K'}{K} },\quad x=e^{-\frac{\pi \la}{2K}},\quad \zeta=x^u
%e^{-\frac{\pi u}{K}}
,
\en 
from which definition it follows that $\la=K'/r$. We also define $p=e^{-\frac{4\pi K}{K'}}$ and 
$\delta=\frac{\la}{K}$, and sometimes use $\xi=r-1$ when connecting to the results of \cite{LT03}. The principal regime is given by $0<x^r<x<\zeta<1$. 
We choose the scalar function $\rho(u)$ as follows.
\be
{\rho(u)}&=&x^{1-r/2}\frac{(x^{4r};x^{4r})_\infty}{(x^{2r};x^{2r})_\infty^2}\frac{\Theta_{x^{4r}}(x^{2r}x^2)\Theta_{x^{4r}}(x^{2r}\zeta^{-2})}{\Theta_{x^{4r}}(x^2\zeta^{-2})}\frac{g(\zeta^{-2})}{g(\zeta^2)},\\
g(z)&=&\frac{(x^2z;x^4,x^{2r})_\infty(x^{2r}x^2z;x^4,x^{2r})_\infty}
{(x^4z;x^4,x^{2r})_\infty(x^{2r}z;x^4,x^{2r})_\infty},\\
&&\hspace{-2cm}(z;q_1,\cdots,q_m)_\infty=\prod_{n_1,\cdots,n_m=0}^\infty(1-zq_1^{n_1}\cdots q_m^{n_m}),\\
\Theta_{q}(z)&=&(q;q)_\infty(z;q)_\infty(q/z;q)_\infty.
\en
This form of the R-matrix coincides with that used in \cite{La02} except for a minus sign in $d(u)$. However,  our notation differs slightly: most importantly, our $p$ is not equal to the $p$ of \cite{La02}. A full dictionary between our notation and that of both \cite{La02} and \cite{LT03} is given in Table \ref{table:dict}.

\begin{table}\caption{A Dictionary of Notation} \begin{center}
\begin{tabular}{|l |c | r | }
\hline
  This paper & Reference \cite{La02} & Reference \cite{LT03} \\
\hline 
 $u$ & $u$ & - \\
  $\lambda$ & - & - \\
  $r$ & $r$ & $\frac{1}{1-\eta}$ \\
$\delta=\frac{\la}{K}$ & $\frac{2\epsilon}{\pi}$ & $\delta$\\
$\xi=r-1$& $r-1$&$\xi=\frac{\eta}{1-\eta}$\\
$x=e^{ -\frac{\pi \lambda}{2K} }$ & $x=e^{-\epsilon}$ & $e^{-\pi \delta/2}$\\
$x^{2r}=e^{-\frac{\pi K'}{K} }$ & $p=x^{2r}$ & $e^{-\pi \delta(\xi+1)}$\\
$p=e^{-\frac{4\pi K}{K'}}$ & $e^{\frac{-2\pi^2}{\epsilon r}}$& $p=e^{-\frac{4\pi}{\delta(\xi+1)}}$\\
\hline
\end{tabular}\end{center}\label{table:dict}
\end{table}

For $V=\C v_+\oplus \C v_-$, we regard $R(u)$ as a linear map on $V\otimes V$ by
\be
R(u)v_{\vep_1}\otimes v_{\vep_2}=\sum_{\vep_1',\vep_2=\pm}R^{\vep_1\vep_2}_{\vep_1'\vep_2'}(u)v_{\vep_1'}\otimes v_{\vep_2'}.
\en
 Let $V_i\ (i=0,1,..,N)$ denote $N+1$ copies of $V$ and regard  
$R_{ij}(u)$ as a linear map on $V_N\otimes \cdots\otimes V_1\otimes V_0$  
acting  on the $i$-th and $j$-th tensor components as $R(u)$ and  on the other components trivially. We define the finite transfer matrix $T(u)$ by 
\be
T(u)&=&\tr_{V}R_{N0}(u)R_{N-10}(u)\cdots R_{10}(u).
\en
Then one can verify that in the infinite $N$ limit the XYZ Hamiltonian \eqref{HXYZ} is obtained as 
\bea
H_{XYZ}=-\frac{J\sn(\la,k')\cn(\la,k')}{2\la}\frac{d}{du}\ln T(u)\Bigl|_{u=0}+\hb{constant},
\lb{dLogT}
\ena
with $k'=\sqrt{1-k^2}$ and 
\bea
J_x&=&J\left(\cn^2(\la,k')+k \sn^2(\la,k')\right),\nn\\
J_y&=&J\left(\cn^2(\la,k')-k \sn^2(\la,k')\right),\lb{eqn:Js}\\
J_z&=&-J\dn(\la,k'). \nn
\eea

\subsection{The vertex operator approach to the XYZ model in the principal regime}
The vertex operator approach to the infinite-lattice massive antiferromagnetic 
XXZ model was developed in \cite{Daval} and is described in detail in the book \cite{JM}.
This approach was then extended to the principal regime XYZ model 
%in \cite{MR1225309,MR1356513,MR1299042}. 
in \cite{JMN}-\nocite{JKKMW,FIJKMY,MR1356513}\cite{JKOS}. 
The essence of the approach is to identify the transfer matrix, the space on which it acts, and local operators in terms of the representation theory of the underlying symmetry algebra. In the XXZ case, this algebra is the quantum affine algebra $\uq$; in the XYZ case it is the elliptic algebra $\Aqp(\slth)$ (beware that the $(q,p)$ indicated in the name $\Aqp(\slth)$ do not correspond directly to the notation of the current paper - in fact we have
$(q,p)=(-x,x^{2r})$). In this section and Appendix \ref{VOA}, we give a brief outline of the key features of the approach relevant to the present work. We refer the interested reader to the original articles cited above and \cite{KKW05} for further details.

The lattice transfer matrix of the infinite-size principal-regime XYZ model
acts on the infinite tensor product space $\cdots \ot V \ot V\ot V\ot V\ot \cdots$ with antiferromagnetic boundary conditions at plus and minus infinity. Let us choose to label the position of our spin-chain sites as $\cdots,2,1,0,-1,-2,\cdots$.
Then the two antiferromagnetic boundary conditions we consider are labelled by $j=0$ or $j=1$ and correspond to considering only those states which have the spin at site $i$, denote $\vep(i)$, restricted to $\bar{\vep}^{(j)}(i)=(-1)^{i+j+1}$  for $|i|>>0$. More precisely, 
we introduce the space of states $\H^{(j)}\ (j=0,1)$ as a half of the infinite tensor space with the antiferromagnetic boundary condition $j$. Namely
\bea
\cH^{(j)}
&:=&{\rm Span}_{\C}\left\{ \cdots \otimes  v_{\vep(1)} \otimes
  v_{\vep(0)}\ |\ \vep(i)=\pm,\ 
\ep(i)=\bar{\vep}^{\,(j)}(i)\hb{ for } i\gg 0
\right\},\lb{spaceH}
\ena
where $v_+=\mat{1\cr 0\cr}, v_-=\mat{0\cr 1\cr}$. 
 The starting point of the vertex operator approach is to identify the space of states $\H^{(j)}$ with the level-1 highest weight modules
 $V(\Lambda_j)$ ($j=0,1$) of the algebra ${\cal A}_{q,p}(\slth)$, where $\Lambda_j$ denotes the fundamental weight of $\slth$. Then, the total infinite tensor product space is identified with the tensor product 
\bea \cF^{(j)}:=\H^{(j)}\ot \H^{*(j)}\simeq \hb{End}(\H^{(j)}).\lb{eq:space}\eea

The transfer matrix of the XYZ model and local operators are then identified in terms of certain vertex operators that act on the space  \eqref{spaceH}. The relevant `type I' vertex operators are maps involving both $\H^{(j)}$ and a finite-dimensional $\Aqp(\slth)$ evaluation module $V_u=V\otimes \C[\z,\z^{-1}]$. They are $\Aqp(\slth)$ homomorphisms of the form
\bea
\Phi^{(1-j,j)}(u):\H^{(j)}\longrightarrow  \H^{(1-j)}\ot V_u,
\eea
whose components $\Phi^{(1-j,j)}_\pm(u)$ are defined by
\ben 
\Phi^{(1-j,j)}(u)=\sum_{\vep=\pm} \Phi^{(1-j,j)}_\vep(u)\otimes v_\vep.\een
%\Psi^{*(1-j,j)}(u):V_u\otimes \H^{(j)}\longrightarrow \H^{(1-j)},
%\qquad
%\Psi^{*(1-j,j)}_\vep(u)=\Psi^{*(1-j,j)}(u)\left(v_\vep\otimes\cdot\right).
%\end{eqnarray*}
The transfer matrix of the XYZ model is identified with the map $T(u):\cF^{(j)}\ra \cF^{(1-j)}$ defined by
\bea
T(u)=\sum_{\vep=\pm}\Phi^{(1-j,j)}_\vep(u)\otimes \Phi^{(j,1-j)}_{-\vep}(u)^t, \lb{TPhiPhi}
\ena
where $t$ denotes transpose. The XYZ Hamiltonian is then identified as
\bea
-\frac{J\sn(\la,k')\cn(\la,k')}{2\la}\frac{d}{du}\ln T(u)\Bigl|_{u=0}.
\lb{dLogT2}
\ena

Now we consider the $2\times2$ matrix $E_{\ep,\ep'}$ that acts as 
$E_{\ep,\ep'}v_{\alpha}=\delta_{\alpha,\ep'} v_\ep$ at site $0$ of the lattice. This local operator is realised as an operator on $\H^{(j)}$ in the vertex operator approach as 
\bea
\cO(E_{\vep\vep'})^{(j)}=\Phi_{-\vep}^{(j,1-j)}(u-1)\Phi^{(1-j,j)}_{\vep'}(u)\Big|_{u=0}.\lb{Eee}
\ena 
Then the spin operator $\sigma^x$ for example is realised as $
\cO(\sigma^x)^{(j)}=\cO(E_{+-})^{(j)}+\cO(E_{-+})^{(j)}$.

The vacuum eigenvector of the transfer matrix $T(u)$ is denoted by $\pvac^{(j)}\in \cF^{(j)}$ and defined by
\bea
T(u)\pvac^{(j)}=\pvac^{(1-j)}.\lb{Tvac}
\ena 
This eigenvector\footnote{We are using the term eigenvector loosely:
 $\pvac^{(j)}$ is a true eigenvector only of $T^2(u)$.}
has a very simple form in the vertex operator picture, and is constructed in terms of a certain grading operator $H^{(j)}$ that acts on $\H^{(j)}$. More precisely, we define $H^{(j)}= -\frac{1}{2}\rho + \frac{j}{4}$, where $\rho=\gL_0+\gL_1$, and
 identify
\ben  
\pvac^{(j)}= \frac{1}{   (  Z^{(j)}   )^\half   } x^{2H^{(j)}},
\een
where we are regarding $\cF^{(j)}\simeq \hbox{End}(\H^{(j)})$. 
Namely, $T(u)$ acts on $f\in \hbox{End}(\H^{(j)})$ as 
\bea
T(u)f=\sum_{\vep=\pm}\Phi^{(1-j,j)}_\vep(u)\circ f \circ \Phi^{(j,1-j)}_{-\vep}(u),
\lb{Tonf}
\ena
see \cite{JM}. 
The normalisation is defined by the $\slth$ principal character
\bea 
Z^{(j)}=\hbox{Tr}_{\H^{(j)}}(x^{4H^{(j)}})=\frac{1}{(x^2;x^4)_\infty}, 
\lb{eqn:char}
\eea 
and is chosen such that $ {^{(j)}} \langle\hbox{vac};pr\pvac^{(j)}=1$.
 Here the inner product of two elements $f,g\in \hbox{End}(\H^{(j)})$ is defined by $(f,g)= \hbox{Tr}_{ \H^{(j)} }(f \circ g)$. We denote a vector in $\F^{(j)}=\H^{(j)}\otimes \H^{*(j)}$ by a ket vector in this section, but it should be understood that it is identified with an operator in $\End(\H^{(j)})$ whenever one considers an action of the vertex operators on it.   
In what follows, we refer to arguments based on the identification $\F^{(j)}\cong
\End(\H^{(j)})$ as the `vertex operator picture'.   

The reader may at this point be thinking that the vertex operator approach is wholly algebraic and formal, but in fact $H^{(j)}$, $\Phi_\vep(u)$ and $Z^{(j)}$ have a direct lattice interpretation in terms of the 8-vertex model: the operator $H^{(j)}$ is identified with Baxter's 
corner-transfer-matrix Hamiltonian, $\Phi_\vep(u)$ with the half-transfer matrix,  and $Z^{(j)}$ with the partition function. In fact, it was Baxter's observation \cite{BaxterBOOK} that it it possible to express the partition function in terms of the corner-transfer-matrix  Hamiltonian as in Equation \mref{eqn:char}, and the subsequent observation that this partition function was related to the $\slth$ principal character, that were the starting points for the development of the vertex operator approach.

In order to construct other eigenstates of the operator \mref{TPhiPhi} it is necessary to introduce a new `type II' vertex operator $\Psi^{*(1-j,j)}(u)$, defined as the map
\begin{eqnarray*}
\Psi^{*(1-j,j)}(u):V_u\otimes \H^{(j)}\longrightarrow \H^{({1-j})},
\end{eqnarray*}
with components $\Psi^{*(1-j,j)}_\pm(u)$ specified by
\begin{eqnarray*}
\Psi^{*(1-j,j)}_\vep(u)=\Psi^{*(1-j,j)}(u)\left(v_\vep\otimes\cdot\right).
\end{eqnarray*}
The full list of properties of both type I and type II vertex operators can be found in \cite{La02}.  One property that we require in the current paper is the commutation relation of type I and type II vertex operators:
\bea 
\Phi^{(j,1-j)}_{\vep_1}(u_1)\Psi^{*(1-j,j)}_{\vep_2}(u_2)
=
\tau(u_1-u_2)\Psi^{*(j,1-j)}_{\vep_2}(u_2)\Phi^{(1-j,j)}_{\vep_1}(u_1),\label{eqn:PhiPsi}
\eea
where the function $\tau(u)$ is given by \cite{La02}
\bea
\tau(u)=i\frac{\vartheta_1\left(\frac{1}{4}-\frac{u}{2},
 p^{\frac{r}{4}}\right)}{\vartheta_1\left(\frac{1}{4}+\frac{u}{2},
 p^{\frac{r}{4}}\right)}.\lb{tau}
\ena
Our convention for theta functions is given in Appendix \ref{app:ellipticfn}. 

Let us consider a state 
defined by
\be
\ket{\theta_1,\theta_{2};pr}^{(j)}_{\vep_1,\vep_{2}}=
\Psi^{*(j,1-j)}_{\vep_2}({i\gt_{2}}/{\pi})\Psi^{*(1-j,j)}_{\vep_1}({i\gt_{1}}/{\pi})
\pvac^{(j)}.
%\frac{1}{   (  Z^{(j)}   )^\half   } x^{2H^{(j)}}.
\en
It then follows immediately from \eqref{eqn:PhiPsi}, \eqref{TPhiPhi} and \eqref{Tonf} that we have 
\bea 
T(u) \ket{\theta_1,\theta_{2};pr}^{(j)}_{\vep_1,\vep_{2}} = \tau(u-i\gt_1/\pi)\tau(u-i\gt_2/\pi)
\ket{\theta_1,\theta_{2};pr}^{(1-j)}_{\vep_1,\vep_{2}}.\lb{Ton2spinon}
\ena
Hence, we may create a new eigenstate of $T(u)$, i.e., an excited state with an 
eigenvalue $\tau(u-i\gt_1/\pi)\tau(u-i\gt_2/\pi)$, by acting on the vacuum with the type II vertex operators. More precisely,
the type II vertex operators $\Psi^{*(1-j,j)}_\vep(u)$ are
 identified with the creation operators of quasi-particle (spinon) excitations 
with rapidity $\theta=-i\pi u$ and spin $\vep$. The $2n$-spinon state 
 (spinons are always excited in pairs) is given by
\bea
\ket{\theta_1,\cdots, \theta_{2n};pr}^{(j)}_{\vep_1,\cdots,\vep_{2n}}=
\Psi^{*(j,1-j)}_{\vep_{2n}}\left({i\gt_{2n}}/{\pi}\right)\cdots \Psi^{*(1-j,j)}_{\vep_1}\left({i\gt_1}/{\pi}\right)\pvac^{(j)}.
\lb{Nspinon}
\ena
From \eqref{Ton2spinon}, one finds that the 
eigenvalue of $T(u)$ per spinon is $\tau(u-i\gt/\pi)$.     
Then from \eqref{dLogT2}, one can deduce that
the pseudomomentum $k(\gt)$ and energy $\omega(\gt)$ of a spinon state in the principal XYZ model are 
\ben
e^{ik(\gt)}&=&\tau(-i\gt/\pi)
,\\
\omega(\gt)&=&\frac{J\sn(\la,k')\cn(\la,k')}{2\la}\frac{\pd}{\pd u}
\ln \tau(u-i\gt/\pi)\Bigl|_{u=0}. 
\een
Hence we obtain
\bea
k(\gt)&=&\am\left(\frac{2I'\gt}{\pi},k_I\right)+\frac{\pi}{2},\label{eqn:spinonk}\\
\omega(\gt)&=&\frac{J I'\sn(\la,k')\cn(\la,k')}{\la}\dn\left(\frac{2I'\gt}{\pi},k_I\right)\nn\\
&=&\frac{J I'\sn(\la,k')\cn(\la,k')}{\la}\sqrt{1-k_I^2\cos^2(k(\gt))}.\lb{eqn:spinonw}
\eea
Here we have introduced new complete elliptic integrals $I, I'$ by 
\ben 
x=e^{-\frac{\pi I'}{I}}
\een
and denote by $k_I$, $k_I'$ the corresponding moduli (that is, we now consider elliptic function with nome $x$ as opposed to the original functions involved in the R-matrix which had nome $x^{2r}$).   
The symbols $\sn(u,k_I), \dn(u,k_I), \am(u,k_I)$ denote Jacobi's elliptic function with modulus $k_I$. In deriving \mref{eqn:spinonk} and \mref{eqn:spinonw}, we have used 
the identity between elliptic functions of different nomes given in Equation \mref{hitid}.
% the following formulas.
% \be
% &&i\frac{\vartheta_1\left(\frac{1}{4}-\frac{i\gt}{\pi}, 
% p^{\frac{r}{4}}\right)}{\vartheta_1\left(\frac{1}{4}
% +\frac{i\gt}{\pi},p^{\frac{r}{4}}\right)}=sn\left(\frac{2I'\gt}{\pi},k_I\right)+i\cn\left(\frac{2I'\gt}{\pi},k_I\right),\\
% &&\sn(u,k)=\sin(\am(u,k)),\quad \cn(u,k)=\cos(\am(u,k)),\quad \dn(u,k)=\sqrt{1-k^2\sin^2(\am(u,k))}. 
% \en
Expressions \eqref{eqn:spinonk} and \eqref{eqn:spinonw} for spinon pseudomomentum and energy are consistent with the results of \cite{JKM73}.

The 2n-spinon form factor of the local operator $E_{\ep,\ep'}$ can now be expressed in the vertex operator picture as the following trace:
\bea
&&{}^{(j)}\dpvac| E_{\ep,\ep'}
\ket{\gt_1,\cdots,\gt_{2n};pr}^{(j)}_{\vep_1,\cdots,\vep_{2n}}\nn\\
&&=\frac{1}{Z}{\tr_{{\cal H}^{(j)}}\Big(x^{4 H^{(j)}}\cO(E_{\ep,\ep'})^{(j)}
\Psi^{*(j,1-j)}_{\vep_{2n}}(i\gt_{2n}/\pi)\Psi^{*(1-j,j)}_{\vep_{2n-1}}(i\gt_{2n-1}/\pi)\cdots
\Psi^{*(1-j,j)}_{\vep_{1}}(i\gt_{1}/\pi)\Big)}
.\label{eqn:tr}
\ena

The massive, antiferromagnetic XXZ model corresponds to the $r\ra \infty \ (k\to 0)$ limit of the above picture. In this limit the above trace can be computed directly in terms of the free-field realisation of $\uq$ \cite{JM}. However, there is no known free-field realisation for the general elliptic case. This problem was overcome in \cite{LaP98,La02} by mapping the 8-vertex model to the SOS model using Baxter's intertwiners. A free field realisation {\it does} exist for the SOS model \cite{MR1403536,MR1436155,Konno98,JKOS99} and this was used to produced an integral expression for  \eqref{eqn:tr} which may be found in \cite{La02}.

\subsection{The map to the disordered regime}
Any regime of the XYZ model can be obtained from the principal regime by a suitable gauge transformation \cite{BaxterBOOK}.
% The principal regime of the XYZ Hamiltonian \eqref{HXYZ} corresponds to the restriction $|J_y|\leq J_x\leq -J_z$ \cite{BaxterBOOK}, while the disordered regime  corresponds to $|J_z|\leq J_y\leq J_x$. 
%The latter can be obtained from the former by making use of a gause transformation that involves the following matrices:
%Let us consider a gauge transformation of the  XYZ Hamiltonian \eqref{HXYZ} specified in terms of the following matrices:
In this section, we construct such a transformation in terms of the following matrices
\ben 
U_0=\frac{-1}{\sqrt{2}}\bpm i&1\\i&-1 \epm ,\ws U_1=\frac{1}{\sqrt{2}}
\bpm 1&i\\-1&i\epm.\een
The adjoint action of these matrices on Pauli matrices is given by
\bea
U_0 (\sigma^x,\sigma^y,\sigma^z) U_0^{-1}=  (\sigma^y,\sigma^z,\sigma^x),\quad
U_1 (\sigma^x,\sigma^y,\sigma^z) U_1^{-1}=  (\sigma^y,-\sigma^z,-\sigma^x).\lb{adact}\eea
We consider the following gauge transformations.
\bea
\widetilde{R}(u)=(U_1\otimes U_0)R(u)(U_0^{-1}\otimes U_1^{-1})=
(U_0\otimes U_1)R(u)(U_1^{-1}\otimes U_0^{-1}).\lb{URU}
\ena
Note that a similar gauge transformation has been discussed in \cite{JKM}.
 The difference is due to the shift $\la\to \la-2iK$ made in sec.2.4 of \cite{JKM}.  

Now define the infinite tensor product $G_j\ (j=0,1)$ by
\bea
 G_j= \cdots \ot U_{1-j}\ot U_j \ot U_{1-j} \ot U_j \ot \cdots.\lb{gaugeG}
\ena
where $U_j$ acts at even sites of our infinite product space. 
Then it follows that with $H_{XYZ}$ given by \eqref{HXYZ}, we have
\ben &&H^{dis}_{XYZ}=G_j H_{XYZ} G_j^{-1} =
-\frac{1}{4} \sli_{i\in \ZZ} (J^{dis}_x \gs^x_i \gs^x_{i+1} + J^{dis}_y \gs^y_i \gs^y_{i+1}+J^{dis}_z  \gs^z_i \gs^z_{i+1}),\\
&&\hbox{where}\ws  J^{dis}_x=-J_z,\ws
J^{dis}_y=J_x,\ws J^{dis}_z=-J_y.\een 
With the parametrisation \mref{eqn:Js}, we have  $|J_y|\leq J_x\leq -J_z$ which  corresponds to the principle regime. 
Hence, we have $|J^{dis}_z|\leq J^{dis}_y\leq J^{dis}_x$ which corresponds to the disordered regime \cite{BaxterBOOK}.

In order to apply the gauge transformation to the space of states 
$\H^{(\ell)}\ (\ell=0,1)$, let us devide $G_j$ into two parts in the following way.
\be
G_j=\G_j\otimes \widetilde{\G}_{1-j}
\en
with 
\be
&&\G_j=\cdots \otimes U_j\otimes U_{1-j} \otimes U_j,\quad \widetilde{\G_j}=U_j\otimes U_{1-j}\otimes U_j \otimes \cdots
\en
Here we assume the rightmost $U_j$ of $\G_j$ acts on the $0$-th site 
of our infinite product space. 

By transforming $\H^{(\ell)}$ by $\G_j$, one finds the following  
two spaces.
\be
&&\H^{(0)}_{dis}:={\rm Span}_\C\{\cdots \otimes w_0\otimes w_0\otimes w_0\otimes \cdots \otimes w_{j_1}\otimes w_{j_0}\ |\ j_0, j_1,\cdots \in\{0, 1\}\ \},\\
&&\H^{(1)}_{dis}:={\rm Span}_\C\{\cdots \otimes w_1\otimes w_1\otimes w_1\otimes \cdots \otimes w_{j_1}\otimes w_{j_0}\ |\ j_0, j_1,\cdots \in\{0, 1\}\ \},
\en
where $w_0, w_1$ denote the eigenvectors of $\sigma^x$ given by
\ben 
w_0 =\frac{1}{\sqrt{2}} \bpm 1\\ 1\epm,\quad w_1 =\frac{1}{\sqrt{2}} \bpm 1\\ -1\epm.
\een
Namely we have 
\bea
&&\H^{(j)}_{dis}=\G_{j+\ell}\H^{(1-\ell)}\quad (\ell=0,1).
%=\G_1\H^{(1)},\\
%&&\H^{(1)}_{dis}=\G_0\H^{(1)}=\G_1\H^{(0)}.
\lb{Hdis}
\ena
Here and hereafter index $j+\ell$ should be understood in mod 2. 
We regard $\H^{(j)}_{dis} \ (j=0,1)$ as the spaces of states in the 
disordered regime. We also set $\H^{*(j)}_{dis}=\widetilde{\G}_{1+j+\ell}\H^{*(1-\ell)}$ 
and define the total space $\F^{(j)}_{dis}=G_{j+\ell}\F^{(1-\ell)}=\H^{(j)}_{dis}\otimes 
\H^{*(j)}_{dis}$.

Accordingly, eigenstates of $H^{dis}_{XYZ}$ are obtained by acting with $G_j$ on the eigenstates of $H_{XYZ}$ and have the same
energy. 
Noting the duplication \eqref{Hdis}, we have a 
new vacuum vector $\vac^{(j)}$ in $ \F^{(j)}_{dis}$ expressed in two ways  
as 
\be
\vac^{(j)}
=G_{j+\ell}\pvac^{(1-\ell)}
\en
with $\ell=0,1$.  
%and will be states with boundary conditions corresponding to
%\ben  &\cdots \ot w_- \ot w_- \ot w_- \ot w_-\ot \cdots ,\quad &\hbox{for}
% \ws j=0,\\
%\hb{and}\ws &\cdots \ot w_+ \ot w_+ \ot w_+ \ot w_+\ot \cdots ,\quad &\hbox{for}
% \ws j=1.\een

\subsection{The massless XXZ limit}\label{subsec:limit}
The massless XXZ Hamiltonian is obtained by taking the limit $x\to 1\ (K\ra +\infty)$ of the disordered Hamiltonian $H^{dis}_{XYZ}$ while keep
$r$ fixed. This corresponds to the following limits of the various elliptic parameters:
\ben \ws K'\ra \frac{\pi}{2},\ws k\ra 1,\ws k'\ra 0, \ws \delta\ra 0_+, \ws\gl\ra \frac{\pi}{2r}.\een 
Corresponding to this limit, we have 
\ben J_x\ra J,\quad J_y\ra J \cos\left(\frac{\pi}{r}\right),\quad J_z\ra -J.\een
Defining $\Delta=\cos\left(\frac{\pi}{r}\right)$, we then have 
\ben 
H^{dis}_{XYZ}\ra -\frac{J}{4} \sli_{i\in \ZZ} (\gs^x_i \gs^x_{i+1} + \gs^y_i \gs^y_{i+1}-\Delta \gs^z_i \gs^z_{i+1}),\een
which is the Hamiltonian of the massless XXZ model.

In order to obtain the dispersion relation for the massless XXZ model, 
let us set $\gt_+=\beta$ and $\gt_-=\beta-\frac{2\pi}{\delta}$. One should note $p^{\frac{r}{4}}=e^{-\frac{\pi}{\delta}}$. Then it follows from \eqref{eqn:spinonk} and \eqref{eqn:spinonw} that we have
\ben
k(\gt_\pm)&=&\am\left(\frac{2I'\beta}{\pi},k_I\right)\pm\frac{\pi}{2},\\
\omega(\gt_\pm)&=&\frac{J I'\sn(\la,k')\cn(\la,k')}{\la}\dn\left(\frac{2I'\beta}{\pi},k_I\right).
\een
The above massless XXZ limit implies 
%may be expressed in terms of the new modulus and elliptic integrals as
\ben  I\ra +\infty,\ws I'\ra \frac{\pi}{2},\ws k_I\ra 1,\ws k'_I\ra 0.\een
 Defining $\kappa(\beta)$ by \ben \kappa(\beta):= 2\arctan (e^\beta),\een
 we find the pseudomomentum $k(\gt_\pm)$ and energy $\omega(\gt_\pm)$ of a spinon
 of the massless XXZ model are given by
\ben
\lim k(\gt_+)&=&\kappa(\beta),\quad \lim k(\gt_-)=\kappa(\beta)-\pi\\ 
\lim \omega(\gt_\pm)&=&\frac{v_F}{\cosh(\gb)}=v_F |\sin \kappa(\beta)|.  
\een 
Here $v_F$ denotes the Fermi velocity given by
\bea
v_F=\frac{Jr}{2}\sin\left(\frac{\pi}{r}\right)=\frac{\pi J}{2} \frac{\sqrt{1-\Delta^2}}{\arccos(\Delta)}.\lb{eq:vF}
\ena
In deriving these limits we have made use of the conjugate modulus transformation for $\dn$ given by Equation \eqref{cmtrans} and of the limits of elliptic functions given  by Equations \eqref{eqn:am0limit} and \eqref{eqn:jac1limit}.

For $\beta$ real, the range of $2\arctan(e^\beta)$ is $(0,\pi)$, and so the + parametrisation gives us right-moving spinons occupying half the first Brillouin zone, and the - parametrisation gives us left-moving spinons occupying the other half. However, as we shall discuss in Section 3, spinons come only in pairs,
and right-moving spinons are alone sufficient to span the complete Hilbert space of the quantum spin chain.
% It is also worth to note the formula
% \be
% \lim \omega(\gt_\pm)&=&v_F\lim_{k_I'\to 0}\frac{\dn\left(\frac{-2iI'\beta}{\pi},k_I'\right)}{\cn\left(\frac{-2iI'\beta}{\pi},k_I'\right)}
% =\frac{v_F}{\cosh(\gb)}.
% \en
% \vspace{2mm}
% \noindent

\subsubsection{ The sine-Gordon Limit}
In \cite{LT03}, the sine-Gordon theory is discussed by taking a similar  $x\to 1$ scaling 
limit of the XYZ model. It is interesting to compare our massless XXZ limit with this relativistic field theory limit.
The approach of \cite{LT03} involves shifting the rapidity in a different way: as $\gt=\vartheta-\frac{\pi}{\delta} $ (we here use $\vartheta$ to indicate the parameter denoted by $\theta$ in \cite{LT03}). Then, with a slightly different normalisation of the Hamiltonian, the momentum $p$ and the excitation energy 
$\vep$ of the quantised soliton are given in \cite{LT03} by
\ben
e^{i p(\vartheta)}&=&\tau(-i\vartheta/\pi+i/\delta)=\frac{\vartheta_4\left(\frac{1}{4}+\frac{i\vartheta}{2\pi}, p^{\frac{r}{4}}\right)}{\vartheta_4\left(\frac{1}{4}-\frac{i\vartheta}{2\pi}, p^{\frac{r}{4}}\right)},
\\
\vep(\vartheta)&=&\frac{\pd}{\pd \vartheta}p(\gt).  
\een 
Then in the same scaling limit $x\to 1$ as above together with 
 the limit that the lattice spacing $\epsilon \to 0$, the 
dispersion relation for massive relativistic particles is given in \cite{LT03} as
\ben
\lim \frac{p(\vartheta)}{\epsilon}&=&M\sinh \vartheta,\\
\lim \frac{\vep(\vartheta)}{\epsilon}&=&M\cosh \vartheta,
\een
where the mass $M$ is given by
\be
M=\lim \frac{2e^{-\frac{\pi}{\delta}}}{\epsilon}.
\en
This sine-Gordon limit is different to our massless XXZ limit and was used in \cite{LT03} in order to connect lattice and
field theory operators. 

\subsection{Form factors in the disordered regime}
In the same way that vacuum vectors in the disordered regime were constructed in Section 1.3, excited states of $H^{dis}_{XYZ}$ are given by  
\ben  
\ket{\theta_1,\cdots, \theta_{2n}}^{(j)}_{\vep_1,\cdots,\vep_{2n}} =
G_{j+\ell} 
\ket{\theta_1,\cdots, \theta_{2n};pr}^{(1-\ell)}_{\vep_1,\cdots,\vep_{2n}}
\een
in $\F^{(j)}_{dis}$ with $\ell=0,1$. 
Hence a form factor of a local operator acting on the site 0 for the disordered regime is given by 
\bea
{}^{(j)}\!\bra{{\rm vac}} E_{\ep,\ep'}
\ket{\gt_1,\cdots,\gt_{2n}}^{(j)}_{\vep_1,\cdots,\vep_{2n}}
={}^{(1-\ell)}\!\bra{{\rm vac};pr} U_{j+\ell}^{-1}E_{\ep,\ep'}U_{j+\ell}
\ket{\gt_1,\cdots,\gt_{2n};pr}^{(1-\ell)}_{\vep_1,\cdots,\vep_{2n}}.
\lb{eqn:fftrans}\eea
The correspondingly gauge transformed Pauli operators acting at
site 0 of the lattice are given by
\ben 
U_{j+\ell}^{-1}(\sigma^{x},\sigma^{y},\sigma^{z})U_{j+\ell}=((-)^{j+\ell}\sigma^{z},\sigma^x,(-)^{j+\ell}\sigma^y).
\een
A derivation of the formula \eqref{eqn:fftrans} in the vertex operator picture is given in Appendix \ref{VOA}.  

% The evaluation of the trace in \eqref{disFF} has been carried out by using the free 
% field realisation of the type I and II vertex operators, the corner Hamiltonian $H^{(j)}$ and the space of states $\cH^{(j)}$ by way of the vertex-face correspondence\cite{LaPu,La}. 
As we mentioned above, a general integral formula for the principal form factors appearing on the right-hand-side of Equation \mref{eqn:fftrans} can be found in \cite{La02}.
Furthermore, this integral was there performed explicitly in the case when the local operator was $\sigma^z$ - see (3.14)-(3.16) of \cite{La02}. This same method was then used to compute the integrals associated with the other Pauli operators in \cite{LT03} and a summary of all cases can be found in Appendix A of \cite{LT03} (where the results are presented directly in the disordered regime). 
Let us define the function
\bea
f(z_1,z_2)^{a,b}_{c,d}:=
\frac{
F_0 \, \overline{G}(z_1-z_2,p)\, \vartheta_a(0,p^\half) \vartheta_b(\frac{z_1+z_2}{2\pi i},p^{\frac{\xi+1}{2}})
}{
 \vartheta_c(\frac{z_1}{2\pi i}-\frac{1}{4},p^{\frac{\xi+1}{4}}) 
 \vartheta_c(\frac{z_2}{2\pi i}-\frac{1}{4},p^{\frac{\xi+1}{4}}) 
\vartheta_d(\frac{z_1-z_2+i\pi}{2\pi i\xi },p^{\frac{\xi+1}{2\xi}})
}
\eea
where $F_0$ and $ \overline{G}(z,p)$ are defined in Appendix A of \cite{LT03}, and we now use the notation $r=\xi+1$.
The results of Appendix A of \cite{LT03} may be
expressed succinctly as 
\bea 
^{(j)}\dvac| \gs^x |\gt_1,\gt_2 \rangle^{(j)}_{\pm,\mp}&=&
 (-1)^j f(\gt_1+\frac{\pi}{\delta},\gt_2+\frac{\pi}{\delta})^{4,4}_{4,1} \pm  f(\gt_1+\frac{\pi}{\delta},\gt_2+\frac{\pi}{\delta})^{4,1}_{4,4},\nn\\[2mm]
^{(j)}\dvac| \gs^y | \gt_1,\gt_2 \rangle^{(j)}_{\pm,\pm}&=& - f(\gt_1+\frac{\pi}{\delta},\gt_2+\frac{\pi}{\delta})^{3,3}_{4,2} \pm (-1)^{j+1}  f(\gt_1+\frac{\pi}{\delta},\gt_2+\frac{\pi}{\delta})^{3,2}_{4,3},\lb{2ptff}\\[2mm]
^{(j)}\dvac| \gs^z | \gt_1,\gt_2\rangle^{(j)}_{\pm,\pm}&=&  i (-1)^jf(\gt_1+\frac{\pi}{\delta},\gt_2+\frac{\pi}{\delta})^{2,2}_{4,2} \pm  if(\gt_1+\frac{\pi}{\delta},\gt_2+\frac{\pi}{\delta})^{2,3}_{4,3}.\nn
\eea
All other components, for example $ ^{(j)}\dvac| \gs^z | \gt_1,\gt_2\rangle^{(j)}_{+,-}$, are zero. Note that $\ep=\pm$ labels on the form-factors \mref{2ptff} are inherited from spin labels in the principal regime but no longer have this interpretation in the disordered phase.  
Note also that the apparent $\frac{\pi}{\delta}$ shifts in the arguments of the functions $f$ relative to \cite{LT03} are again due to the fact that
that our $\theta$ and the corresponding
symbol in \cite{LT03}, which we here denote $\vartheta$, are related by $\theta=\vartheta-\frac{\pi}{\delta}$. 
We can remove these shifts by using the half-period property of theta functions given by Equation \mref{halfperiod}, from which it follows that
% % Note however that these results are given in a 
% % parametrization relevant to taking the sine-Gordon limit. In order to get 
% % two-spinon form factor relevant to the massless XXZ limit, we need to re-parametrize them  by shifting the rapidity as $\gt\pm\frac{\pi}{\delta}$ according as considering the right or left mover. 
% % In order to consider XXZ form-factors 
% %it is necessary to consider the function $f^{a,b}_{c,d}(\gt_1,\gt_2)$ with 
% %arguments shifted by $\pm \frac{\pi}{\delta}$. 
% To this end we note the half-period property of theta functions
% \bea &&\vartheta_a(u\pm \frac{\tau}{2},q)=(\pm i)^{g_a} e^{-i\pi \tau/4} e^{\mp i \pi u} \vartheta_{\bar{a}}(u,q),\ws\nn\\
% &&\hb{where}\ws q=e^{i\pi \tau}\ws and\ws
% \bar{1}:=4,\, \bar{2}:=3, \,\bar{3}:=2, \,\bar{4}:=1,\quad g_1=g_4=1, \,g_2=g_3=0,\lb{halfperiod}\eea
% from which it follows that 
\ben f(z_1+ \frac{\pi}{\delta},z_2+\frac{\pi}{\delta})^{a,b}_{c,d} =
-i(-1)^{g_c} (-i)^{g_b} f(z_1,z_2)^{a,\bar{b}}_{\bar{c},d}.
%,\quad 
%f(\gb_1\pm \frac{\pi}{\delta},\gb_2\mp \frac{\pi}{\delta})^{a,b}_{c,d} =
%  i^{\pm(g_d+1)} f(\gb_1,\gb_2)^{a,b}_{\bar{c},\bar{d}} 
\een
Using this property leads to the following expressions for the two-spinon XYZ form-factors in the disordered regime:
\bea ^{(j)}\dvac| \gs^x |\gt_1,\gt_2\rangle^{(j)}_{\pm,\mp}&=& (-1)^{j}  f(\gt_1,\gt_2)^{4,1}_{1,1} \pm  f(\gt_1,\gt_2)^{4,4}_{1,4},\nn\\[2mm]
^{(j)}\dvac| \gs^y |\gt_1,\gt_2 \rangle^{(j)}_{\pm,\pm}&=& -if(\gt_1,\gt_2)^{3,2}_{1,2} \pm (-1)^{j+1}i  f(\gt_1,\gt_2)^{3,3}_{1,3},\lb{eqn:6ff}\\[2mm]
^{(j)}\dvac| \gs^z |\gt_1,\gt_2 \rangle^{(j)}_{\pm,\pm}&=&  (-1)^{j+1}f(\gt_1,\gt_2)^{2,3}_{1,2} \mp  f(\gt_1,\gt_2)^{2,2}_{1,3}.
\nn\eea
%Here we set 
%\bea
% \ket{\gb_1,\cdots,\gb_{2n}}^{(j)}_{\ep_1,\cdots,\ep_{2n}} = 
%\ket{B(\gb_1+\frac{\pi}{\delta}),\cdots,B(\gb_{2n}+\frac{\pi}{\delta})}^{(j)}_{\ep_1,\cdots,\ep_{2n}}\lb{xxzstate}
%\eea
%(and by an abuse of notation, we continue to use $\vac^{(j)}$ to refer 
%to the $\delta\ra 0$ limit of the vacuum state).
% The left-left two spinon form-factors $^{(j)}\dvac| \gs^a |\gb_1-\frac{\pi}{\delta},\gb_2- \frac{\pi}{\delta}\rangle^{(j)}_{\pm,\mp}$ are given by the same expressions except that $\sigma^x$ form-factor has an extra minus sign.

\subsection{Massless XXZ form factors} 
We now consider the disordered XYZ form factors in the $x\ra 1$, $r$ fixed, limit discussed in Section \ref{subsec:limit}.
For right-moving spinons, we identify $\theta=\beta$ and take the $p\ra 0$ limit of the function $f(\beta_1,\beta_2)^{a,b}_{c,d}$. 
This limit is obtained from that of the theta functions and from the following limits: 
\ben \lim \overline{G}(\beta, p) =\widetilde{G}(\beta):=\exp\left( -\int_0^\infty \frac{dt}{t} 
\frac{\sinh((\xi+1)t)\, \sinh^2((1+\frac{\gb}{\pi i})t)}
{\sinh(\xi t) \sinh(2t) \cosh(t)}\right),\quad \lim F_0 = 2(1+\xi^{-1}) p^{\frac{\xi^2+\xi+1}{8\xi}}.\een
We find that only one of the six $f(\beta_1,\beta_2)^{a,b}_{c,d}$ appearing in \eqref{eqn:6ff}, namely $f(\gb_1,\gb_2)^{2,3}_{1,2}$, is non-zero in this limit. As a result, the only non-zero, two-spinon massless XXZ form-factor is given by 
\ben 
\lim \:   ^{(j)}\!\dvac |\gs^z \ket{\gb_1,\gb_2}^{(j)}_{\ep_1,\ep_2} =  (-1)^{j+1} \delta_{\ep_1,\ep_2}\frac{  (1+\xi^{-1}) \widetilde{G}(\gb_1-\gb_2)}
{2\, \sin\left( \frac{\gb_1}{2i} -\frac{\pi}{4}\right)\sin\left( \frac{\gb_2}{2i} -\frac{\pi}{4}\right)
 \cos\left( \frac{\gb_1-\gb_2+i\pi}{2i\xi}\right) }.
\een
By an abuse of notation, we continue to use $^{(j)}\!\dvac |\gs^z \ket{\gb_1,\gb_2}^{(j)}_{\ep_1,\ep_2}$ to refer to the massless XXZ limit of this form factor. 
With $\gb_1$, $\gb_2$ real, we have
\bea F(\gb_1,\gb_2,\xi):=|^{(j)}\!\dvac |\gs^z \ket{\gb_1,\gb_2}^{(j)}_{\ep,\ep}|^2 
 = 
\frac{
2(1+\xi^{-1})^2 e^{-I_\xi(\frac{\gb_1-\gb_2}{2\pi})}
}{\cosh(\gb_1)\cosh(\gb_2) (\cos(\frac{\pi}{\xi}) + \cosh(\frac{\gb_1-\gb_2}{\xi}) )}, \lb{ffsq}\lb{FFS}\eea
where the integral $I_\xi(z)$ is defined by
\bea I_\xi(z)= \int_0^\infty \frac{dt}{t}\frac{\sinh((\xi+1)t}{\sinh(\xi t)} \frac{(\cosh(2t) \cos(4t z) -1)}
{\sinh(2 t)\cosh(t)}.\label{keyint}\eea
Note that the expression \mref{ffsq} is symmetric with respect to exchange of $(\gb_1,\gb_2)$, and independent of both $\ep$ and $j$.

 % XYZ to Massless XXZ
\section{The Longitudinal Structure Factor}
\setcounter{equation}{0} 
The longitudinal structure factor has already been defined by Equation (\ref{lsf}).
%as
%\bea S^{zz}(k,\omega)= \sli_{j\in \ZZ} e^{-ikj} \int_{-\infty}^{\infty} dt e^{iwt} \dvac| S_j^z(t) S_0^z(0)\vac\lb{lsf}\eea 
%where $S^z=\half \gs^z$. 
In order to be able to compute the form-factor expansion of this object we need to know the resolution of the identity in terms of a basis of states. We conjecture that 
\bea \id = \sli_{j=0,1}\sli_{n\geq 0} \sli_{\ep_1,\cdots,\ep_{2n}}\frac{1}{(2n)!} \int_{-\infty}^\infty \frac{d\beta_1}{2\pi}
\cdots \int_{-\infty}^\infty \frac{d\beta_{2n}}{2\pi} 
%%%%
\ket{\gb_1,\cdots,\gb_{2n}}^{(j)}_{\ep_1,\cdots,\ep_{2n}} \ws
 _{\ep_1,\cdots,\ep_{2n}} ^{\quad\quad(j)}\langle\gb_1,\cdots,\gb_{2n}|.
\lb{resol}\eea 
This conjecture is an analogue of the conjecture made in the antiferromagnetic
 regime of the XXZ model in \cite{JM}.
 Note however, that there is a slight but important difference to \cite{JM}. In considering XXZ single spinons in Section 2.2, we have characterised right-moving spinons as having momentum in the range $(0,\pi)$ and left-moving spinons as having momentum in the range $(-\pi,0)$. In \mref{resol} however, we only include even spinon states consisting of right-moving spinons. The reason for this are two-fold: firstly, that right-left moving pairs are simply absent from the Bethe Ansatz states, and secondly that a right-right pair already spans the entire $(0,2\pi)$ Brillouin zone - to include the left-left pair would be to double count. This point is discussed in some detail in \cite{2008_Caux_JSTAT_P08006} for the massive antiferromagnetic phase of the XXZ model. Independent  numerical justification that we have made the correct choice of normalisation is given by the sum rule calculations in Section 4.

In writing \mref{lsf} we have not specified to which vacuum we are referring. However, the 
result is the same whether we choose to use $\vac^{(0)}$, $\vac^{(1)}$ or the linear combinations $\vac_\pm=\frac{1}{\sqrt{2}}( \vac^{(0)}\pm \vac^{(1)})$ considered in \cite{LT03}. For notational convenience, let us specify that $\vac=\vac^{(0)}$. Then inserting the resolution \mref{resol} we have the 2-particle contribution
\ben
S^{zz}_2(k,\omega)\!=\! \frac{1}{8}\!\sli_{\ep}\int_{-\infty}^\infty\int_{-\infty}^\infty \frac{d\beta_1}{2\pi}  \frac{d\beta_2}{2\pi}
\sli_{j\in \ZZ} e^{-i(k-K(\gb_1)-K(\gb_2))j} \!\!\!\int_{-\infty}^{\infty} \!\!\!\!\!\! dt e^{i(\omega-W(\gb_1)-W(\gb_2))t} \,
|^{(0)}\!\dvac  |\gs^z |\gb_1,\gb_2 \rangle^{(0)}_{\ep,\ep}|^2,
\een
where the spinon momentum and energy are defined by
\ben K(\gb)=2\arctan(e^\gb),\quad W(\gb)=\frac{v_F}{\cosh(\beta)}.
\een
We then write both the $j$ sum and $w$ integral in terms of delta functions to give
\bea
S^{zz}_2(k,\omega)= \frac{1}{4}\int_{-\infty}^\infty\int_{-\infty}^\infty \!\! d\beta_1 d\beta_1  \,\delta\big(k-K(\gb_1)-K(\gb_2)\big) \, \delta\big(\omega-W(\gb_1)-W(\gb_2)\big)\,
F(\gb_1,\gb_2,\xi),
\lb{penul}\eea
where $F(\gb_1,\gb_2,\xi)$ is defined by \mref{FFS}.

We now recall that if we have a suitably smooth function $g:\RR^2\ra \RR^2$ with a finite number of zeroes $(x_1^{(j)},x_2^{(j)})$ at which $\det g'(x_1^{(j)},x_2^{(j)})\neq 0$ , and a function $f:\RR^2\ra \RR$, then we have
\ben 
\int_{-\infty}^\infty\int_{-\infty}^\infty dx_1 dx_2
\, f(x_1,x_2) \delta(g(x_1,x_2)) = \sli_{j} \frac{ f(x_1^{(j)},x_2^{(j)})}{| \det g'(x_1^{(j)},x_2^{(j)})|}. \een
To use this fact on \mref{penul}, we make use of the determinant
\ben \bdet{ll}  W'(\gb_1) & W'(\gb_2)\\ K'(\gb_1) & K'(\gb_2)\edet=
\frac{v_F(\tanh(\gb_2)-\tanh(\gb_1))}{\cosh(\gb_1)\cosh(\gb_2)}. \een
For each choice of $k$ and $\omega$ in the two-spinon band there is a unique (up to exchange of 
$\tilde{\beta}_1$ and $\tilde{\beta}_2$) pair $(\tilde{\beta}_1,\tilde{\beta}_2)$ 
 satisfying the combined conditions $k=K(\tilde{\beta}_1)+K(\tilde{\beta}_2)$ and 
$\omega=W(\tilde{\beta}_1)+W(\tilde{\beta}_2)$.
Thus, \mref{penul} becomes
 \ben
S^{zz}_2(k,\omega)= \frac{1}{2} \frac{\cosh(\tilde{\beta}_1)\cosh(\tilde{\beta}_2) 
F(\tilde{\beta}_1,\tilde{\beta}_2,\xi)
}{v_F|\tanh(\tilde{\beta}_1)-\tanh(\tilde{\beta}_2)|}.\een

Let us denote the upper two-spinon energy threshold and and lower two-particle energy threshold by 
$\omega_{2,u}(k)$ and $\omega_{2,l}(k)$. They are given by the following expressions: 
\bea  \omega_{2,u}(k)=2v_F\sin(k/2),\quad 
\omega_{2,l}(k)=v_F|\sin(k)|\label{def:2ptthresh},\eea 
where the Fermi velocity $v_F$ is given by Equation \eqref{eq:vF}.
There is a useful identity
\ben 
\sqrt{\omega_{2,u}^2(k)-\omega^2}=v_F|\tanh(\tilde{\beta}_1)-\tanh(\tilde{\beta}_2)|.\een
Using this identify, and substituting the expression \mref{ffsq} for the modulus squared of the form-factor, we arrive at the expression

\begin{equation}
 S_2^{zz}(k,\omega)=\frac{\Theta(\omega_{2,u}(k)-\omega)\Theta(\omega-\omega_{2,l}(k))(1+\xi^{-1})^2 e^{-I_\xi(\frac{\gb}{2\pi})}}{
\sqrt{\omega_{2,u}^2(k)-\omega^2 } \, \left(\cos(\frac{\pi}{\xi}) + \cosh(\frac{\gb}{\xi})\right) },
\label{eq:Szz2}
\end{equation}
where $\gb:=\tilde{\beta}_1-\tilde{\beta}_2$ and $k=K(\tilde{\beta}_1)+K(\tilde{\beta}_2)$, $\omega=W(\tilde{\beta}_1)+W(\tilde{\beta}_2)$.
 % DSF
\section{Results}
\setcounter{equation}{0} 
In this section, we offer quantitative results and plots for the longitudinal structure factor (LSF).  Here, for convenience of comparison with previous results in the literature, 
we use a slightly different convention for the Hamiltonian (corresponding to the ones we
used in \cite{2011_Caux_PRL_106}), writing it as
\begin{equation}
H = J \sum_{i \in \ZZ} \left(S^x_i S^x_{i+1} + S^y_i S^y_{i+1} + \Delta S^z_i S^z_{i+1}\right)
\label{eq:HXXZ} \end{equation}
where $S_i^\alpha=\half \sigma_i^\alpha$, $J>0$. 
% and $0<\bD=-\Delta=\cos\left(\frac{\pi}{\xi+1}\right)\leq 1$. 
We have a gauge equivalence $H= O H_{XXZ} O^{\dagger}$, where $O=\cdots \otimes 1\otimes  \sigma^z\otimes 1\otimes \sigma^z\otimes \cdots$. 
% The only significant change to the discussion of the previous section is that the dispersion relation of the spinon states is now of the form 
% \begin{eqnarray*} \omega(k)=v_F |\sin(k)|,\end{eqnarray*}
% where the Fermi velocity is given in \eqref{eq:vF}.
% %\begin{eqnarray} 
% %v_F =\frac{\pi J}{2} \frac{\sqrt{1-\Delta^2}}{\arccos(\Delta)}=\frac{J(\xi+1)}{2}
% % \sin\left( \frac{\pi}{\xi+1}\right).
% %\label{eq:vF}
% %\end{eqnarray}
The two-spinon part of the LSF is given by Equation (\ref{eq:Szz2})
% ), with the upper and lower two-particle thresholds modified to
% \begin{eqnarray} \omega_{2,u} (k)=2 v_F \sin(\frac{k}{2}),\quad \omega_{2,l} (k)=v_F|\sin(k)|,
% \label{eq:boundaries}
% \end{eqnarray}
% and 
in which the
parameter $\beta (k, \omega) =: 2\pi \rho (k, \omega)$ is obtained from the constraint 
\begin{equation}
\cosh (\pi \rho(k, \omega)) = 
\sqrt{ \frac{\omega_{2,u}^2(k) - \omega_{2,l}^2(k)}{\omega^2 - \omega_{2,l}^2(k)} }.
\label{rhodef}
\end{equation}
% \section{Results}
% In this section, we offer quantitative results and plots for the longitudinal structure factor.
% Here, for convenience of comparison with previous results in the literature, 
% we use a slightly different convention for the Hamiltonian (corresponding to the ones we
% used in \cite{XXZ_h0_Szz_2spinons_Paper1}), writing it as
% \begin{equation}
% H = J \sum_{j=1}^N \left(S^x_j S^x_{j+1} + S^y_j S^y_{j+1} + \Delta S^z_j S^z_{j+1}\right).
% \label{eq:HXXZ}
% \end{equation}
% The two-spinon part of the longitudinal DSF is given by (\ref{eq:Szz2}), in which the
% parameter $\beta (k, \omega) := 2\pi \rho (k, \omega)$ is obtained from the constraint 
% \begin{equation}
% \cosh (\pi \rho(k, \omega)) = \sqrt{\frac{\omega_{2,u}^2(k) - \omega_{2,l}^2(k)}{\omega^2 - \omega_{2,l}^2(k)}}.
% \label{rhodef}
% \end{equation}

%\myinsert{Modification of above for \begin{eqnarray*} H_{XXZ}=
%J'  \sli_{i\in \ZZ (\gs^x_i \gs^x_{i+1} + \gs^y_i \gs^y_{i+1}+ \Delta'  \gs^z_i \gs^z_{i+1})
%,\quad \Delta'=+\cos(\pi/(\xi+1))\end{eqnarray*} 
%as used in Paper 1.}\\
%\myinsert{Graphs}\\

\subsection{Full Brillouin zone}
\begin{figure}
\begin{tabular}{cc}
\includegraphics[width=80mm]{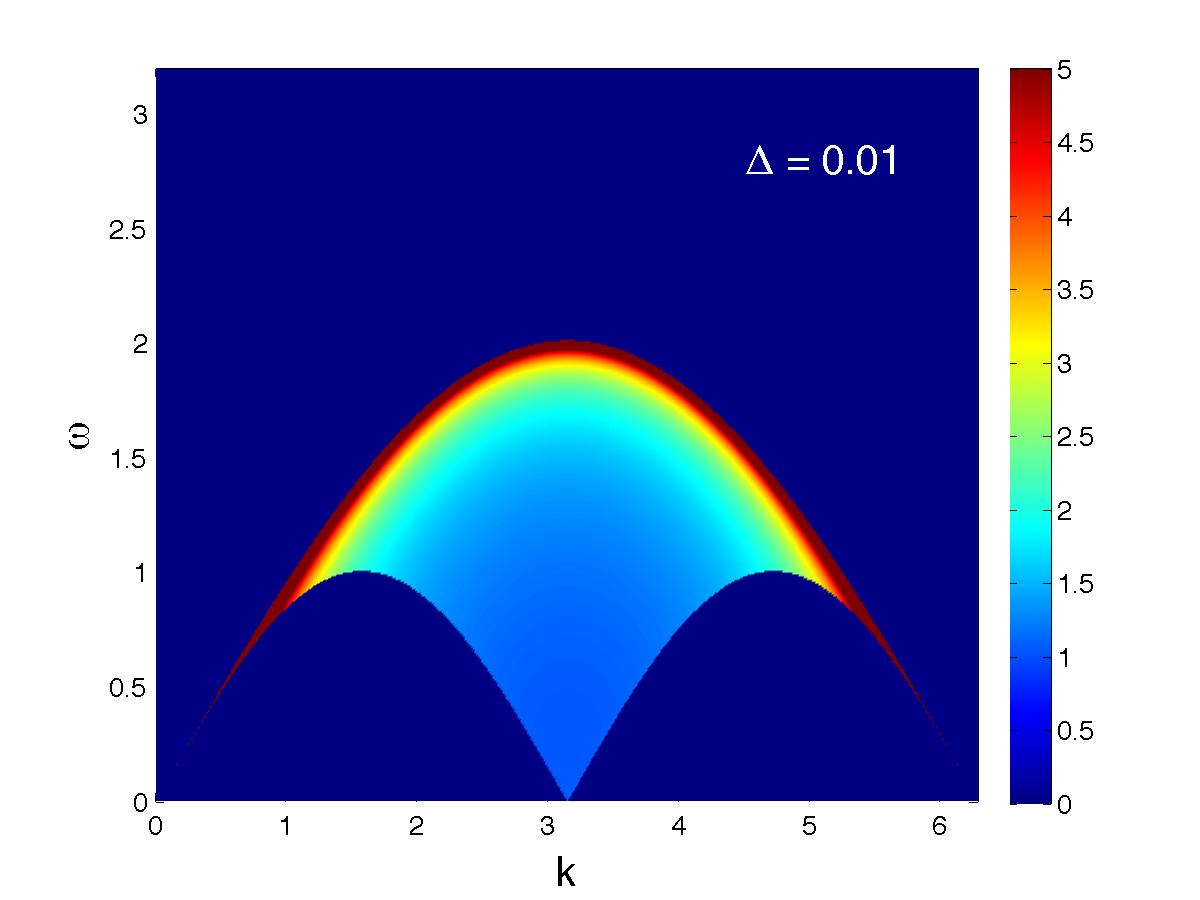}
&
\includegraphics[width=80mm]{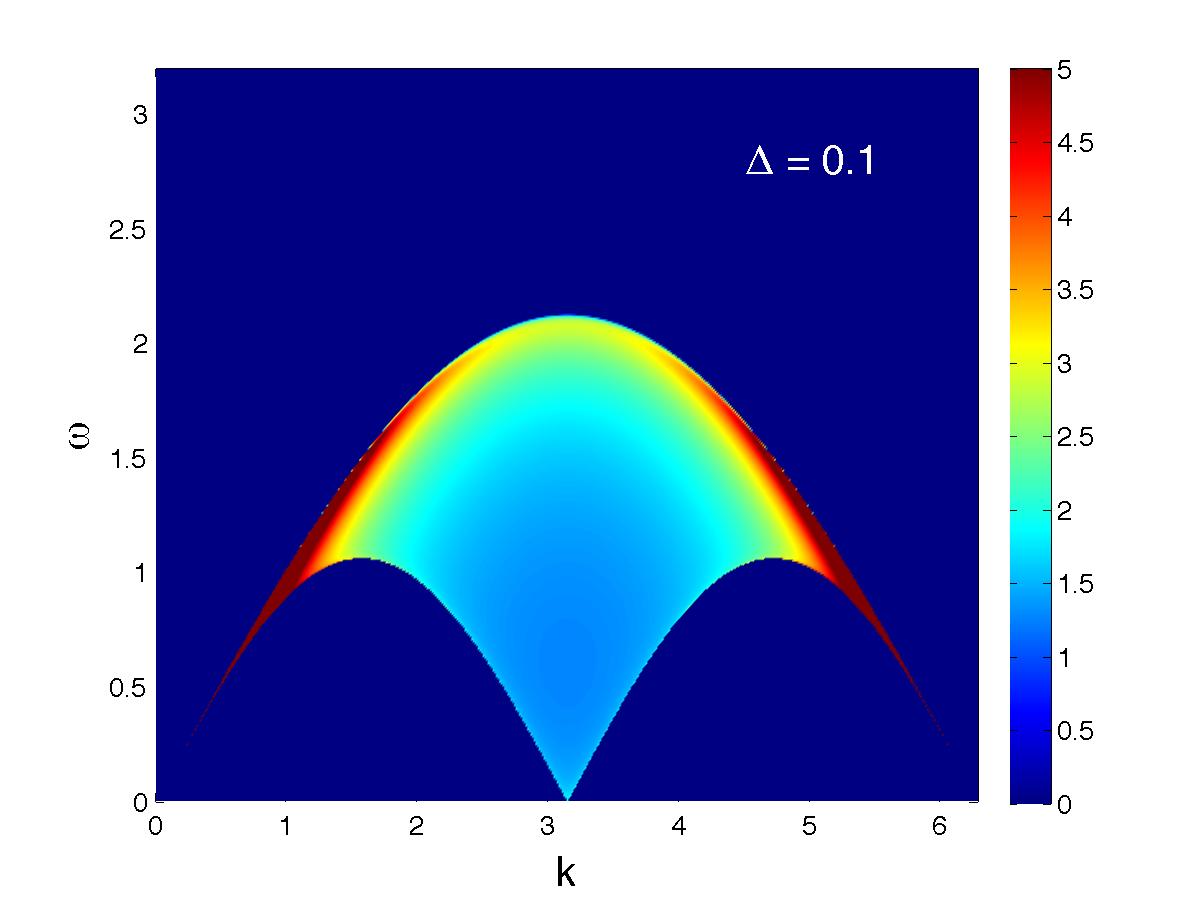}
\\
\includegraphics[width=80mm]{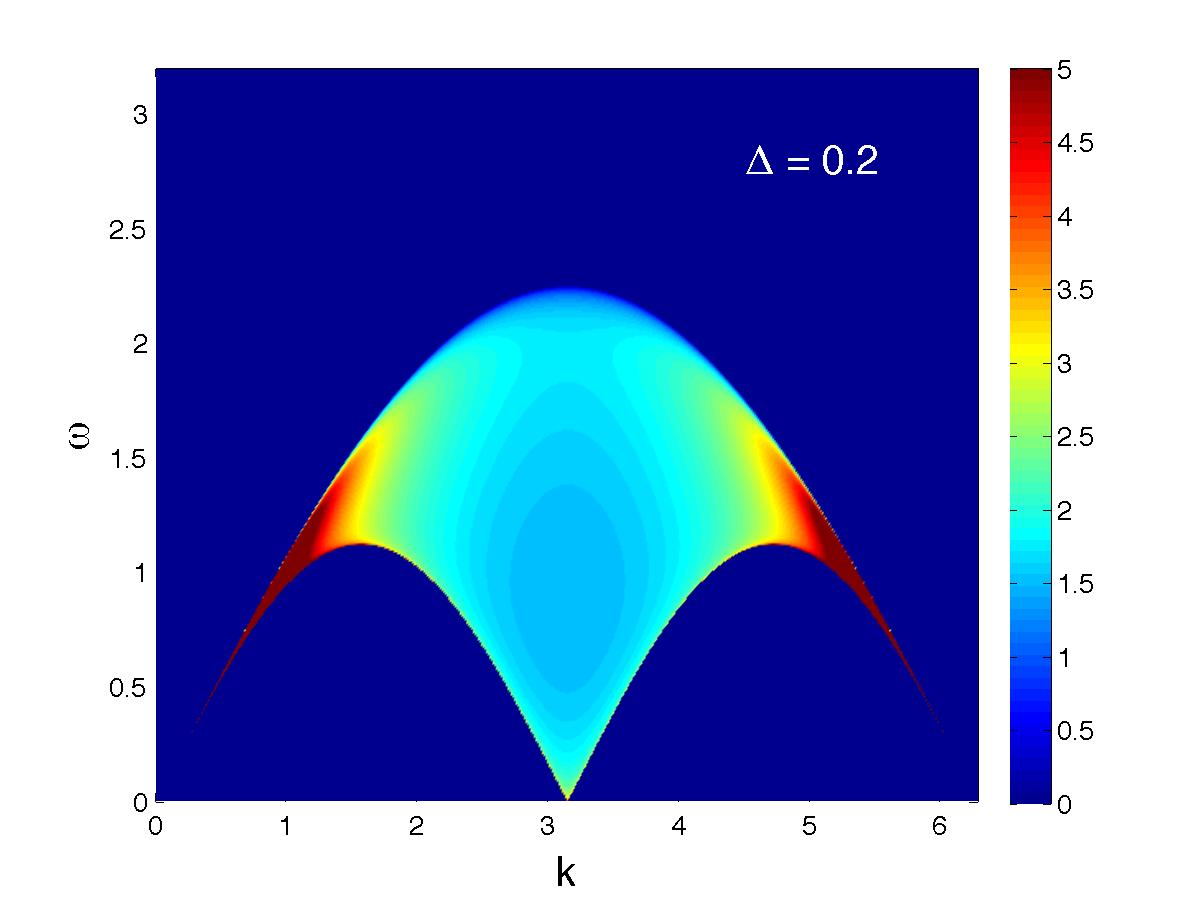}
&
\includegraphics[width=80mm]{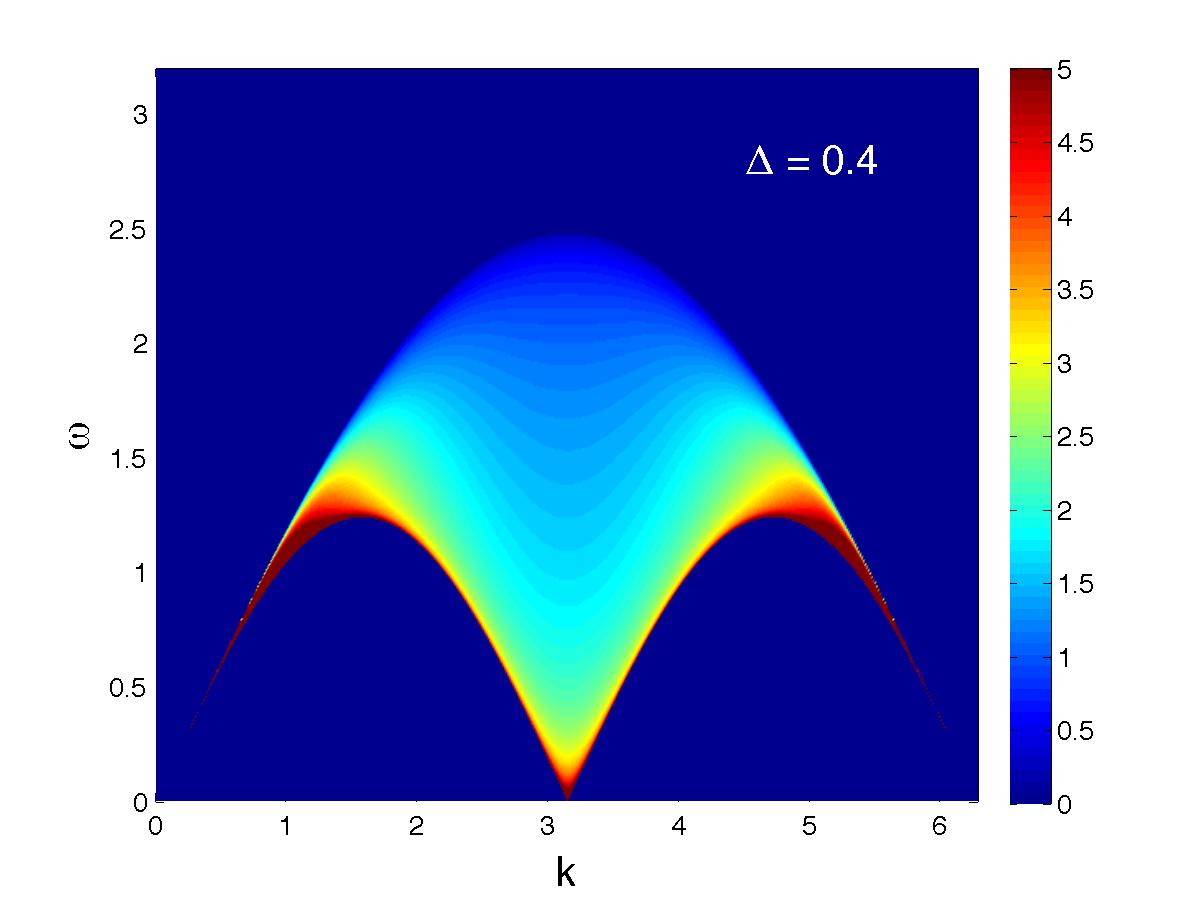}
\end{tabular}
\caption{Two-spinon part of the longitudinal structure factor of the infinite Heisenberg chain, 
for different values of the anisotropy parameter $\Delta$ (see also Figure \ref{fig:LSF2}).  For $\Delta \rightarrow 0$, the
correlation follows the density of states, and has a square root singularity at the upper threshold for all values of momenta.}
\label{fig:LSF1}
\end{figure}

\begin{figure}
\begin{tabular}{cc}
\includegraphics[width=80mm]{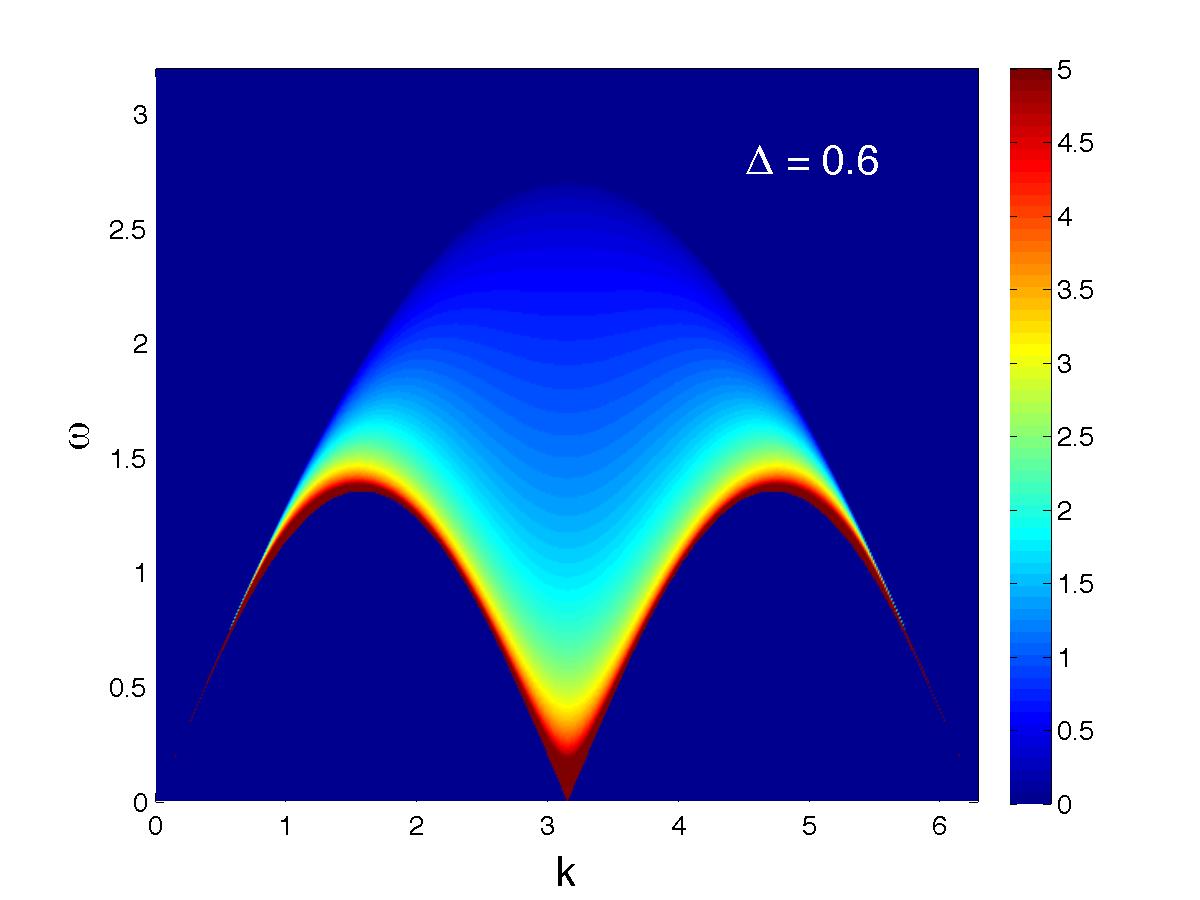}
&
\includegraphics[width=80mm]{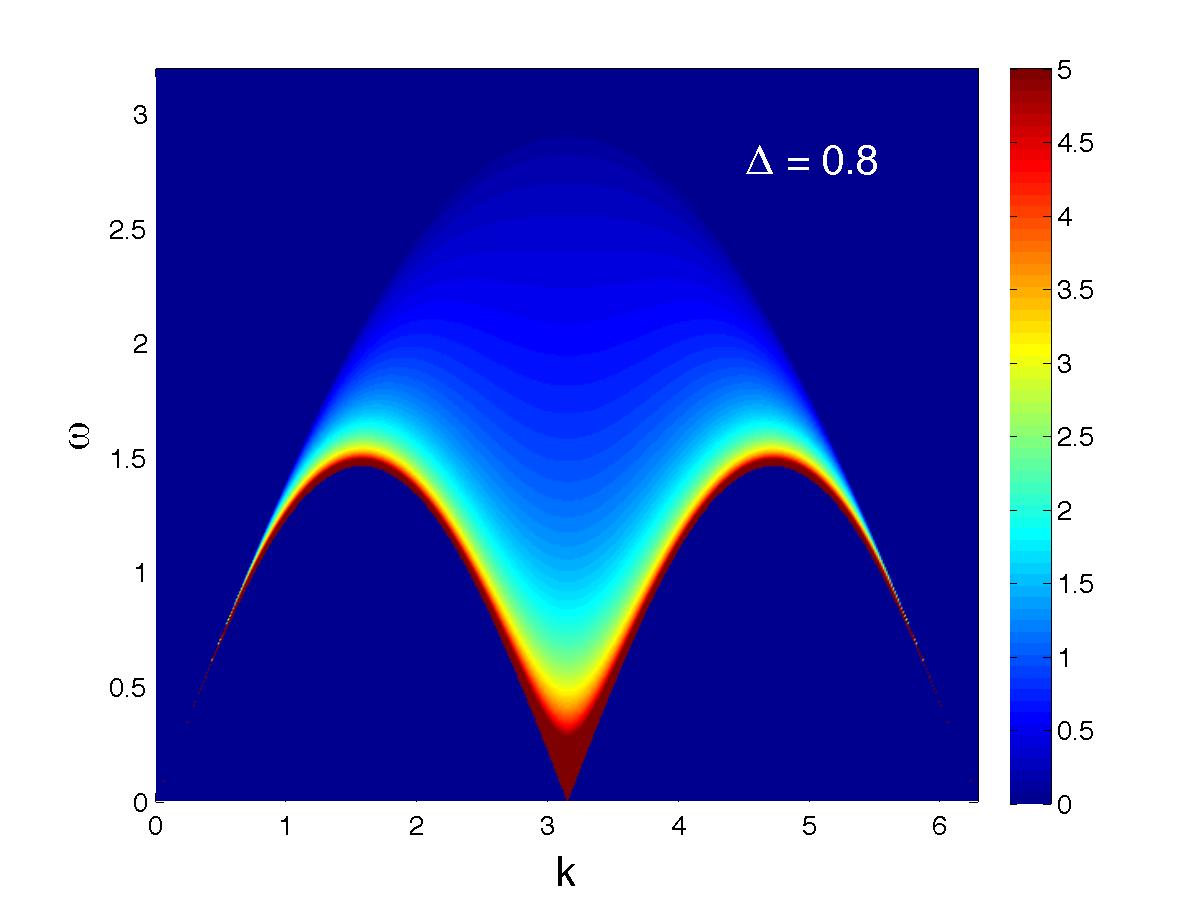}
\\
\includegraphics[width=80mm]{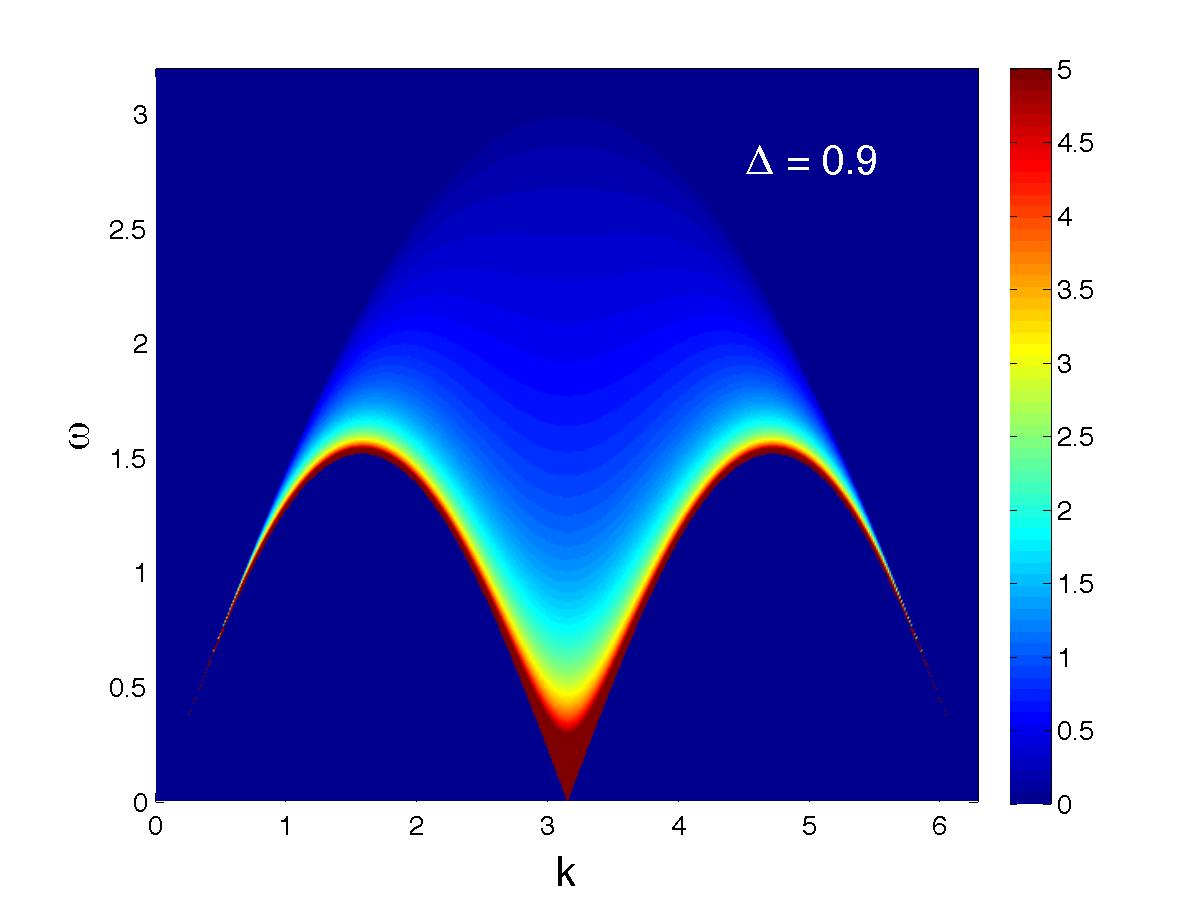}
&
\includegraphics[width=80mm]{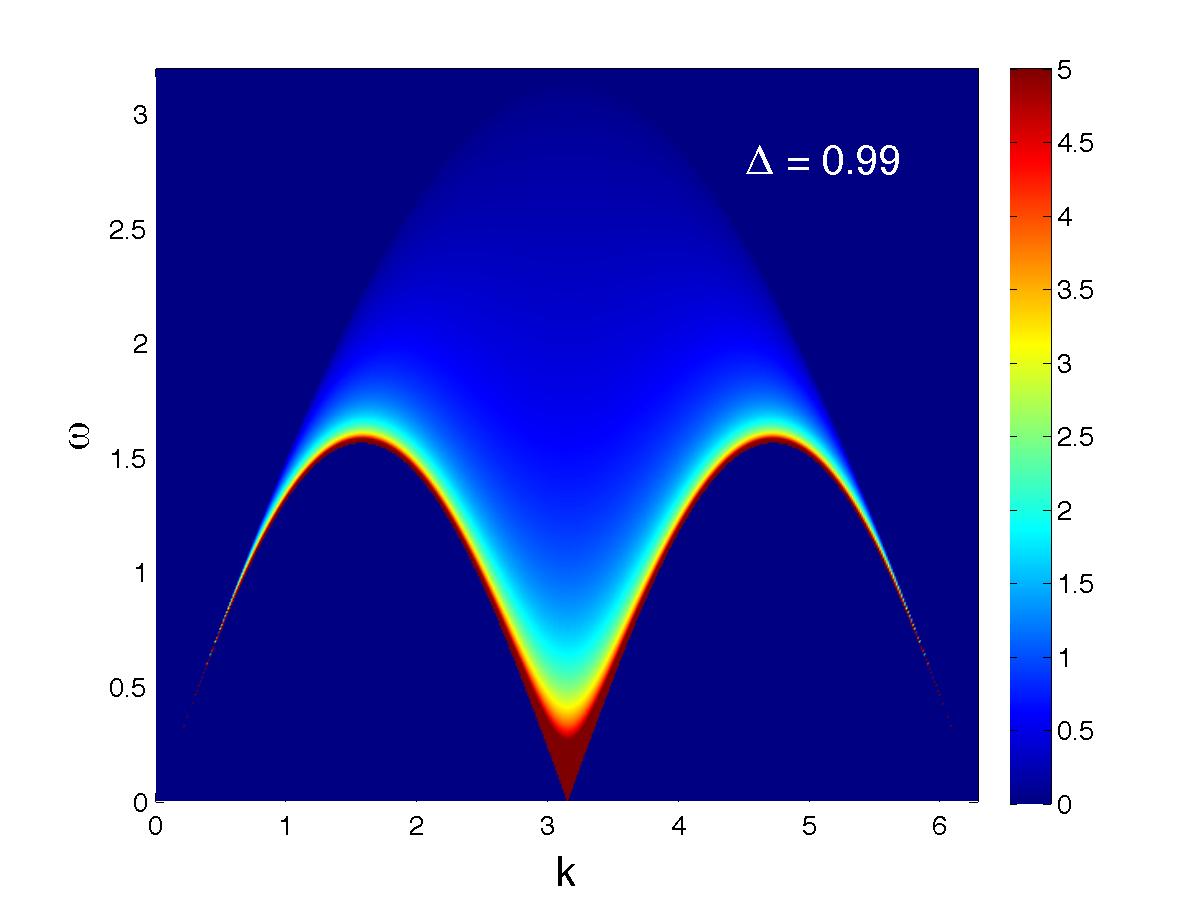}
\end{tabular}
\caption{Two-spinon part of the longitudinal structure factor of the infinite Heisenberg chain, 
for different values of the anisotropy parameter $\Delta$ (see also Figure \ref{fig:LSF2}). 
Increasing the anisotropy shifts the weight progressively towards the lower boundary. 
The lower boundary becomes increasingly sharp as the $\Delta \rightarrow 1$ limit is approached.}
\label{fig:LSF2}
\end{figure}

In Figures \ref{fig:LSF1} and \ref{fig:LSF2}, we present plots of the two-spinon longitudinal
structure factor for different values of anisotropy, starting at the $XX$ limit and going up to the isotropic
antiferromagnetic point $\Delta = 1$. All plots cover a full Brillouin zone and clearly show the continuum
over which the two-spinon correlation is non-vanishing. 
This continuum in the $k$-$\omega$ plane is located between the lower and upper boundaries
%\begin{equation}
%\omega_{2,l} (k) = v_F |\sin k|, \hspace{1cm} \omega_{2,u} (k) = 2 v_F \sin (k/2).
%\label{eq:boundaries}
%\end{equation}
%(\ref{eq:boundaries}).
(\ref{def:2ptthresh}).
In fact, for zero magnetic field, the full LSF
vanishes beneath the lower boundary,  i.e., for
$\omega < \omega_{2,l} (k)$. All  $2, 4, 6, ...$ spinon states share the line $\omega = \omega_{2,l} (k)$ 
as their lower boundary, and the LSF
is strictly positive for all $\omega > \omega_{2,l} (k)$.
Above the upper two-spinon boundary $\omega > \omega_{2,u} (k)$,
the two-spinon contribution of course vanishes, but higher-spinon states can contribute.

Starting at small $\Delta$, the top left panel of Figure \ref{fig:LSF1} clearly illustrates the
fact that the LSF diverges at the upper boundary (we will quantify all threshold behaviour
in Section \ref{subsec:threshold}), and tends to a constant at the lower one. This is easily
understood \cite{1967_Niemeijer_P_36,1970_Katsura_P_46} by mapping to a system of free fermions using the
Jordan-Wigner transformation, under which the $S^z$ operator becomes
a fermionic density operator. Since the fermionic density operator in Fourier space takes the form of a convolution 
product of creation-annihilation operators, $\rho_k = \sum_{q} \psi^{\dagger}_{k+q} \psi_k$, 
and since the ground state is a simple Fermi sea of the Jordan-Wigner fermions, all form factors
of the $S^z$ operator are energy independent, and vanish for all but the two-spinon states.
The LSF is thus simply a representation of the two-spinon density of states, which
has a square-root divergence at the upper threshold (see, e.g., \cite{1981_Mueller_JPC_14}
for details) and is constant at the lower one.

Turning the anisotropy up leads rapidly to the loss of the divergence
at the upper threshold, starting from the region $k \simeq \pi$, as can be seen in the
top right panel of Figure \ref{fig:LSF1}. The presence of a finite $\Delta$ is most importantly
felt in the form factors: while the density of states still diverges, the form factors now
decrease sufficiently rapidly to kill off this divergence. Once $\Delta$ has attained
values of around $0.4$, the remains of the divergence at the upper threshold have been
completely erased for all values of momentum away from the zone boundary, and the lower
threshold starts to feel the effects of the antiferromagnetic correlations, the peak at
$k = \pi$, $\omega = 0$ starting to develop.

As $\Delta$ is increased further (see Figure \ref{fig:LSF2}, 
the cascade of correlation weight towards the low-energy
sector continues until the isotropic limit is attained, at which point which most of the signal
is concentrated in the immediate vicinity of the lower threshold. 

Throughout this series of plots (which are presented in a uniform $\omega$ and intensity scale
for convenience of comparison), the slow increase of the Fermi velocity and of the reach of the
two-spinon continuum can be seen.


\subsection{Fixed momentum}
\begin{figure}
\begin{tabular}{cc}
\includegraphics[width=80mm]{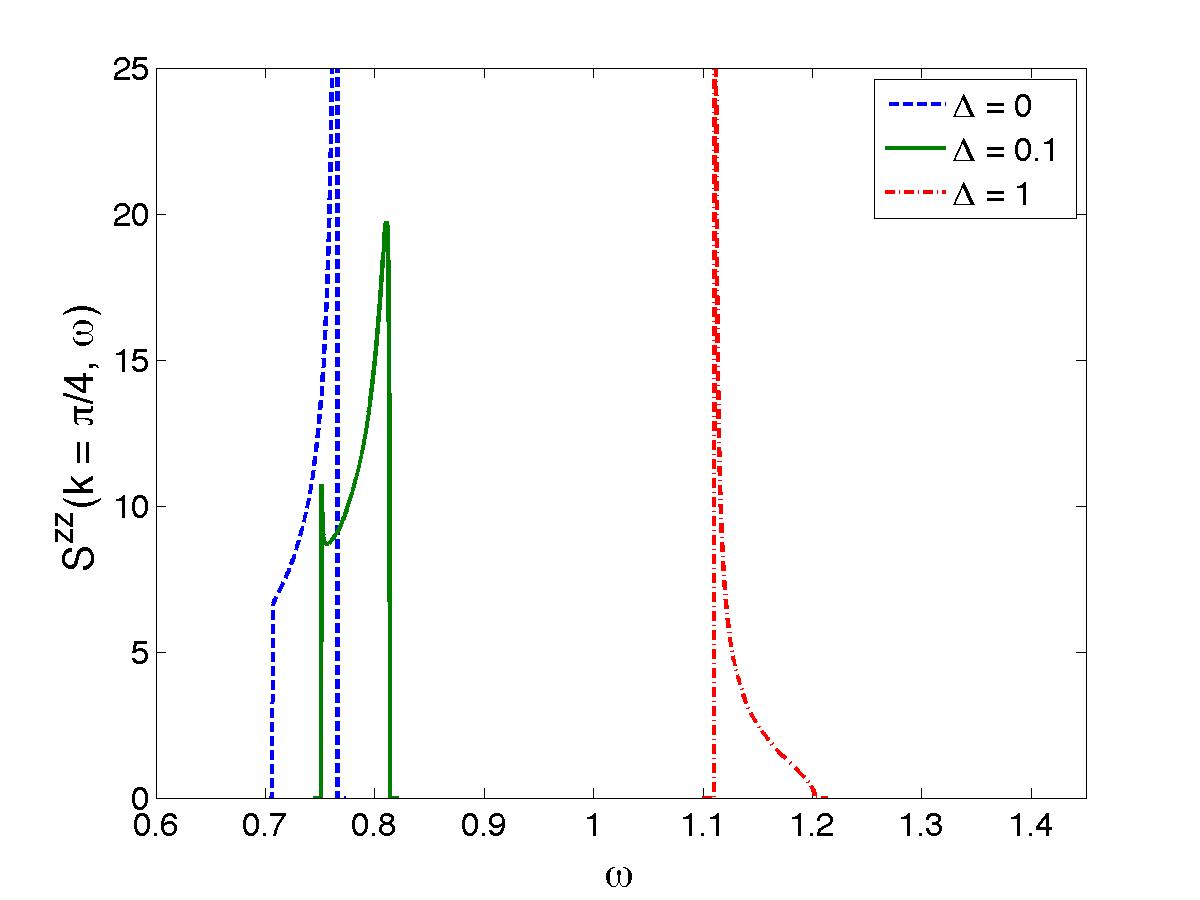}
&
\includegraphics[width=80mm]{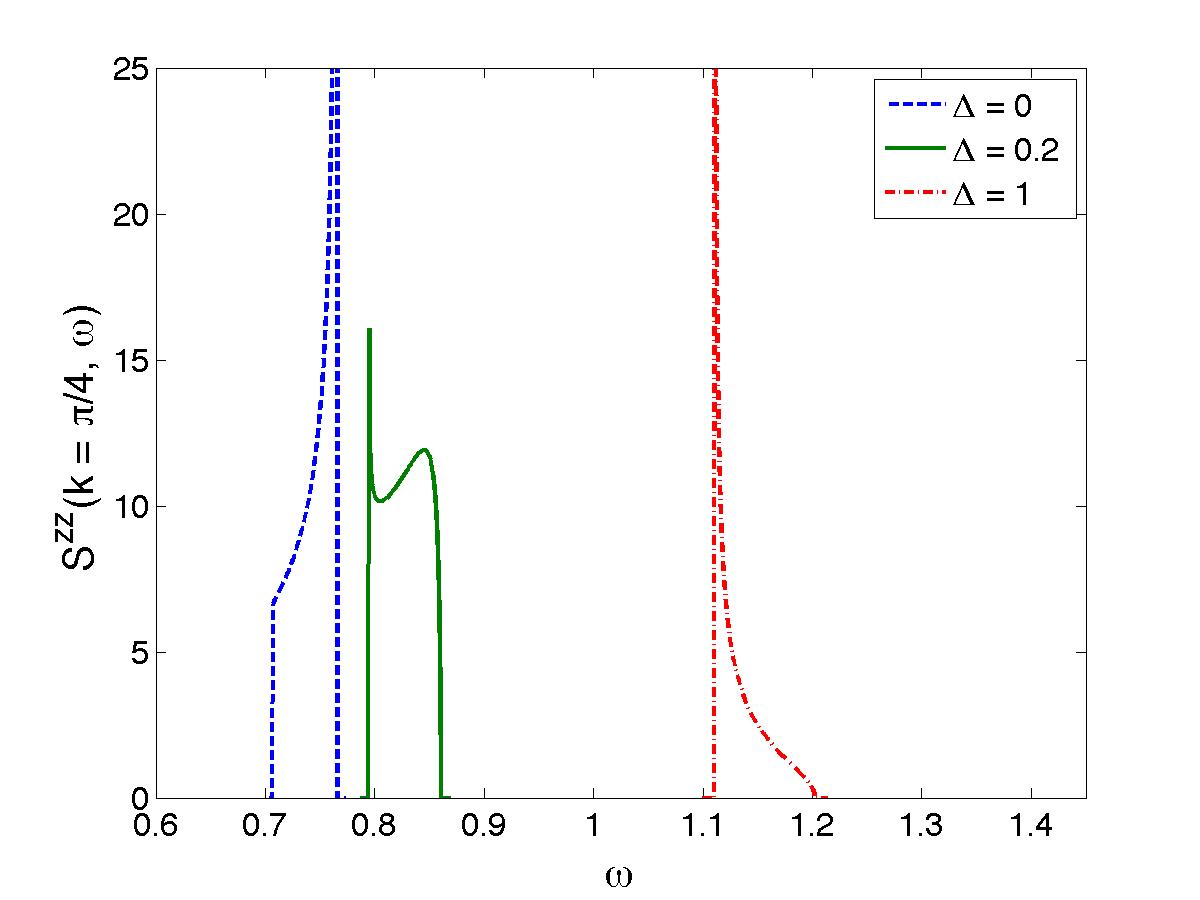}
\\
\includegraphics[width=80mm]{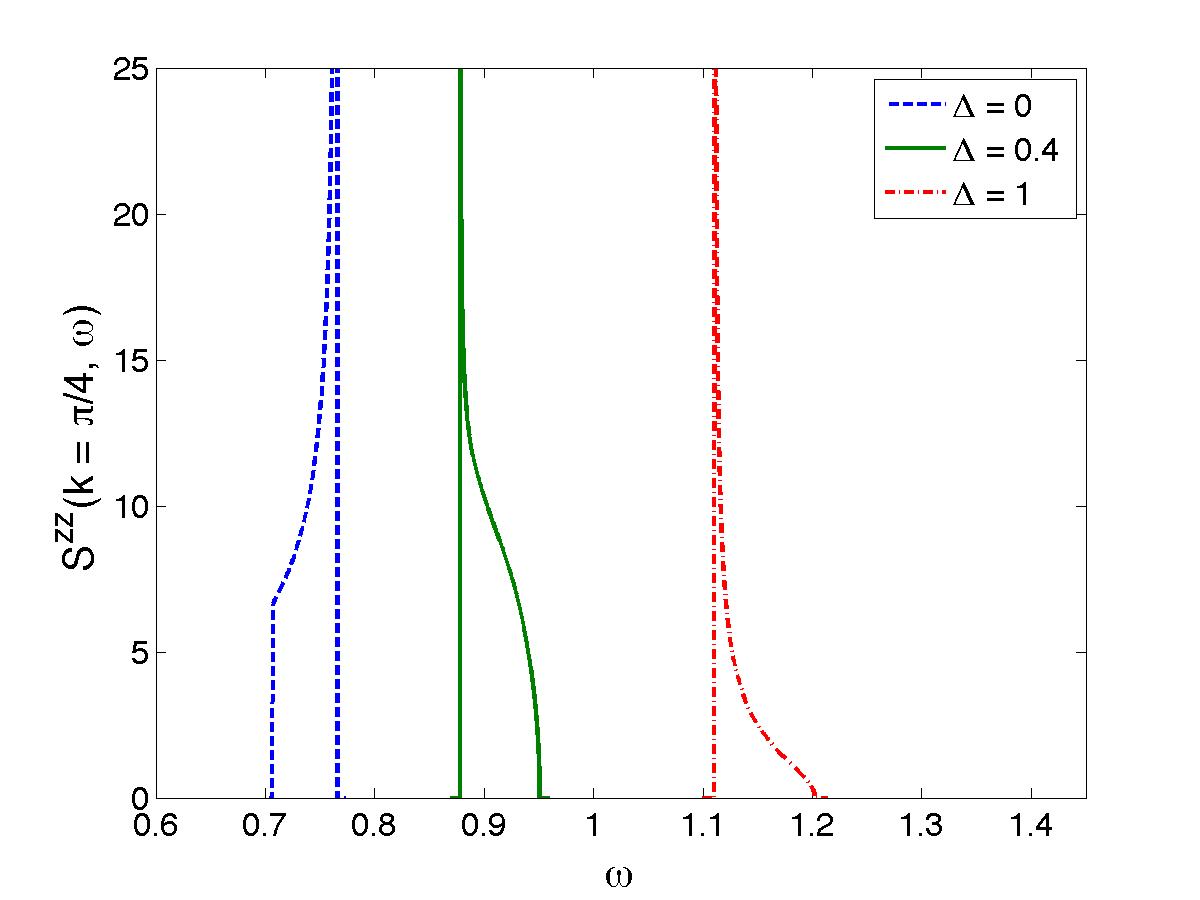}
&
\includegraphics[width=80mm]{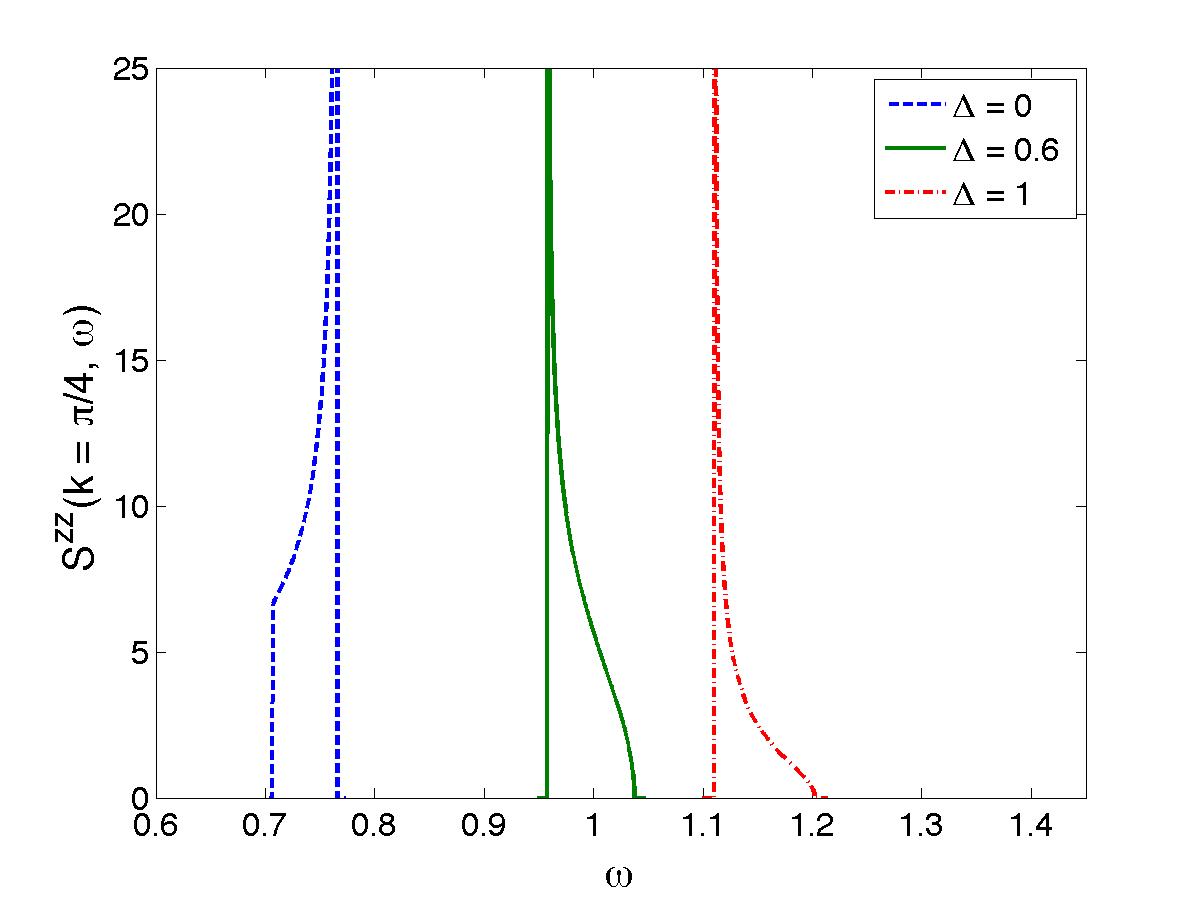}
\\
\includegraphics[width=80mm]{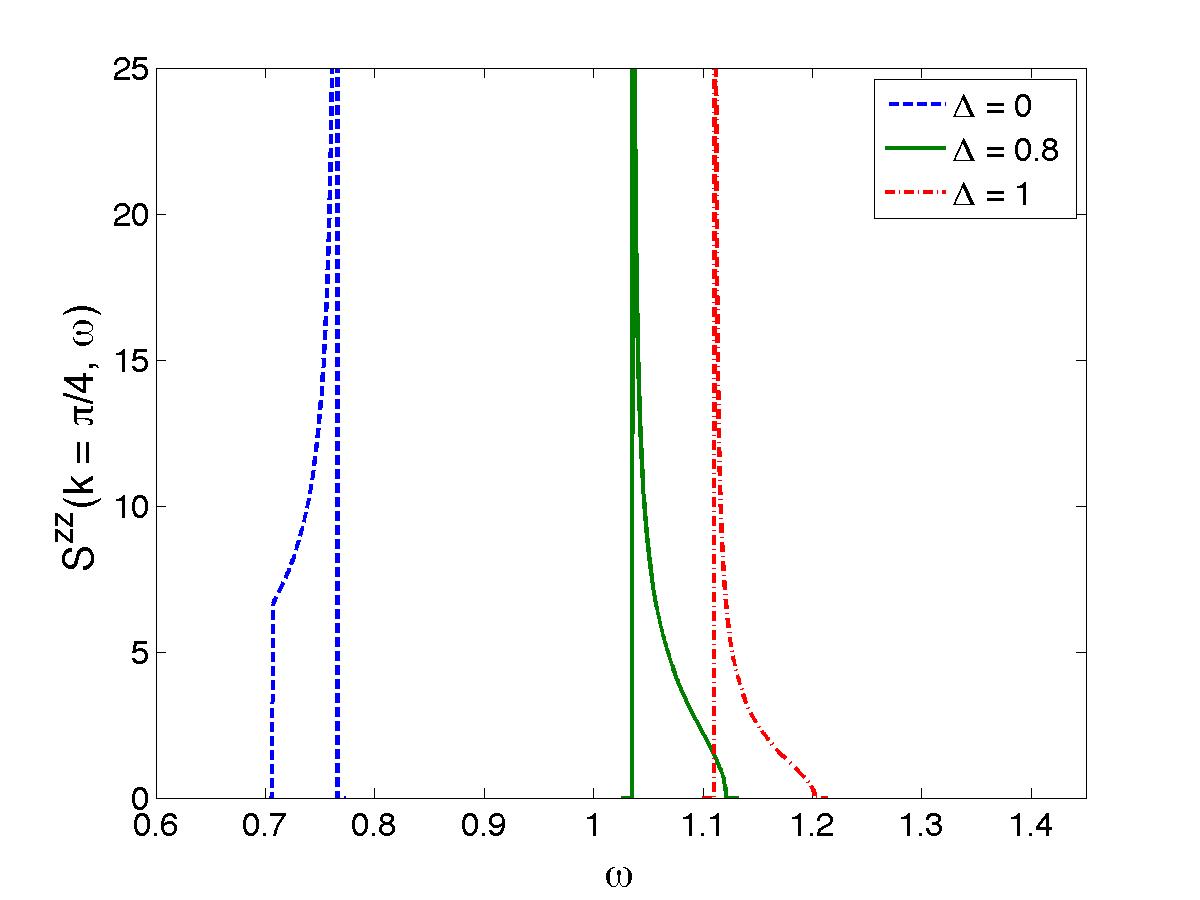}
&
\includegraphics[width=80mm]{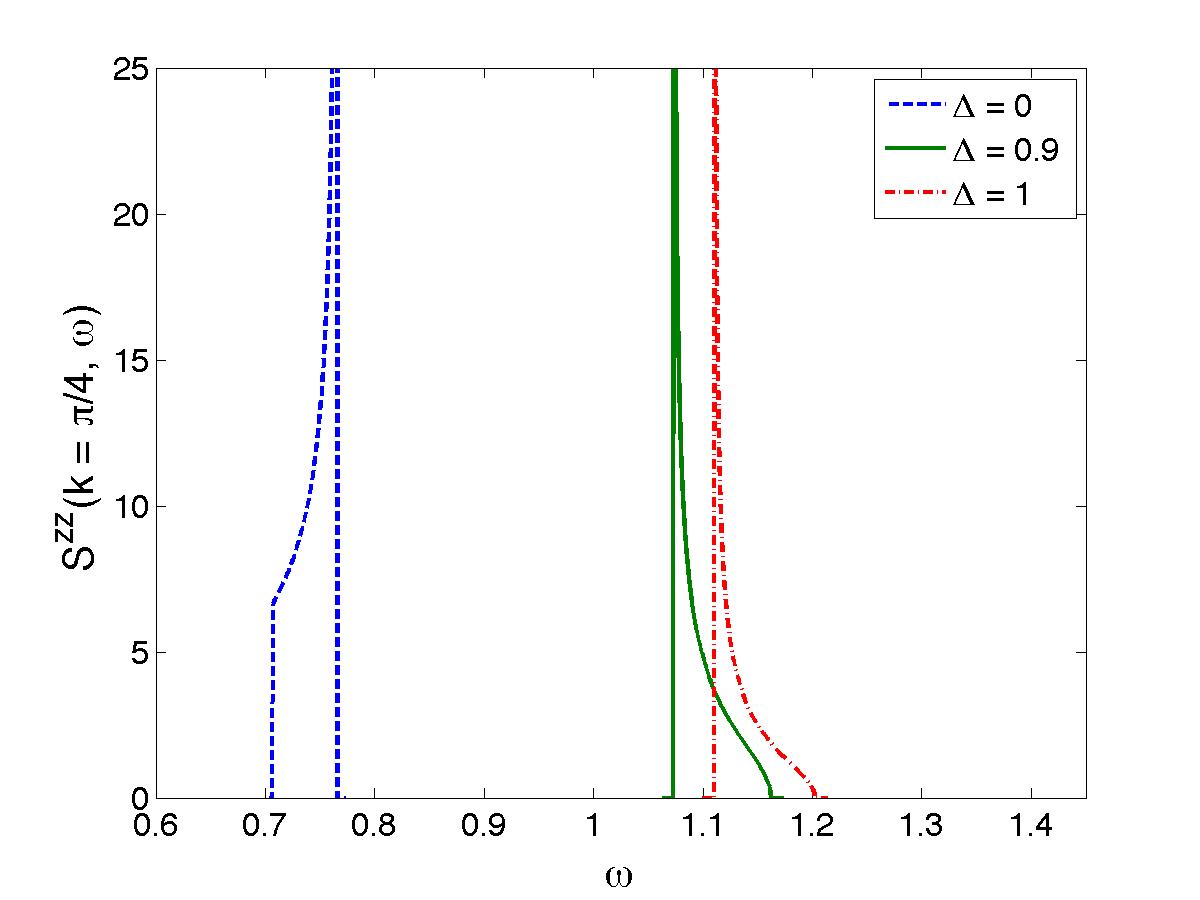}
\end{tabular}
\caption{Fixed momentum cuts at $k = \pi/4$ of the two-spinon part of the longitudinal structure 
factor of the infinite Heisenberg chain, for different values of the anisotropy parameter $\Delta$.
The $\Delta = 0$ and $\Delta = 1$ limits are displayed in all plots for comparison.}
\label{fig:LSF_K_pio4}
\end{figure}

\begin{figure}
\begin{tabular}{cc}
\includegraphics[width=80mm]{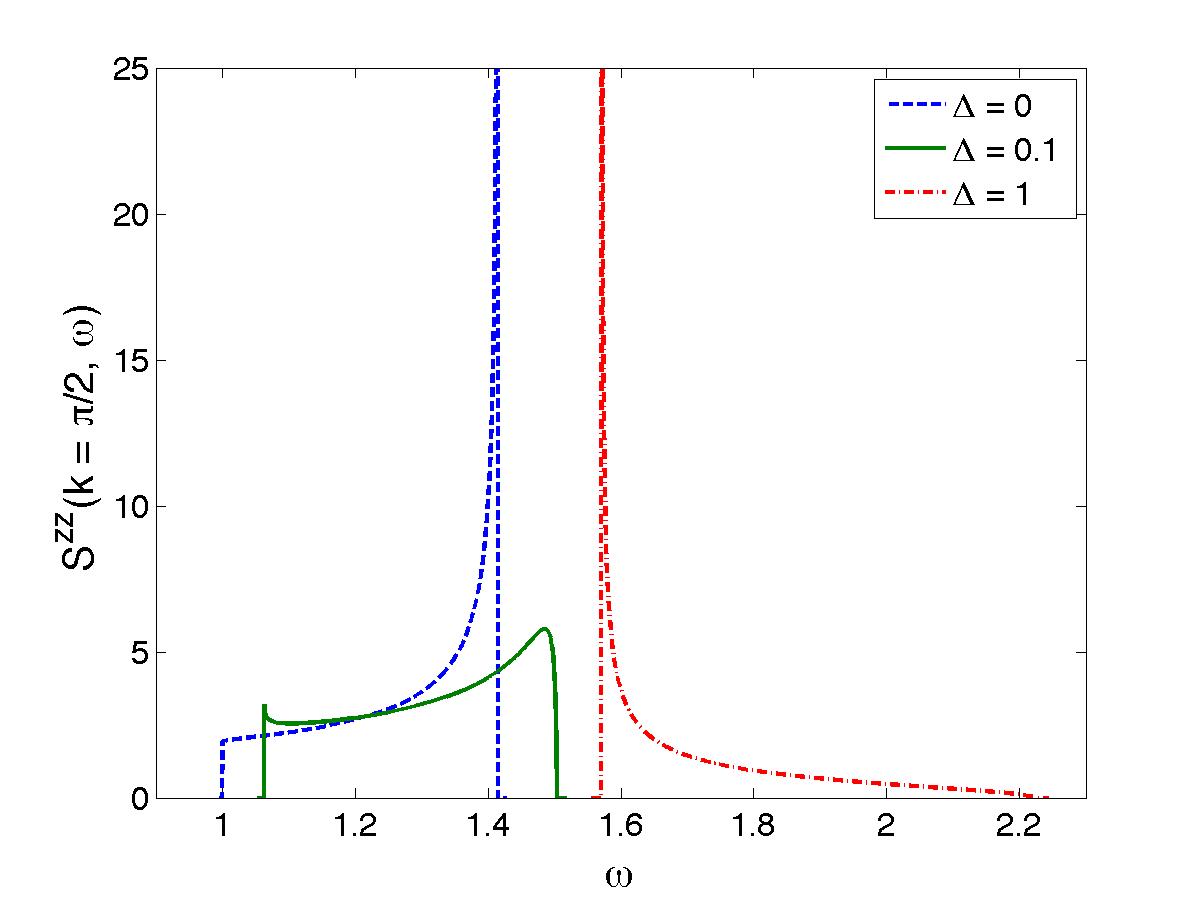}
&
\includegraphics[width=80mm]{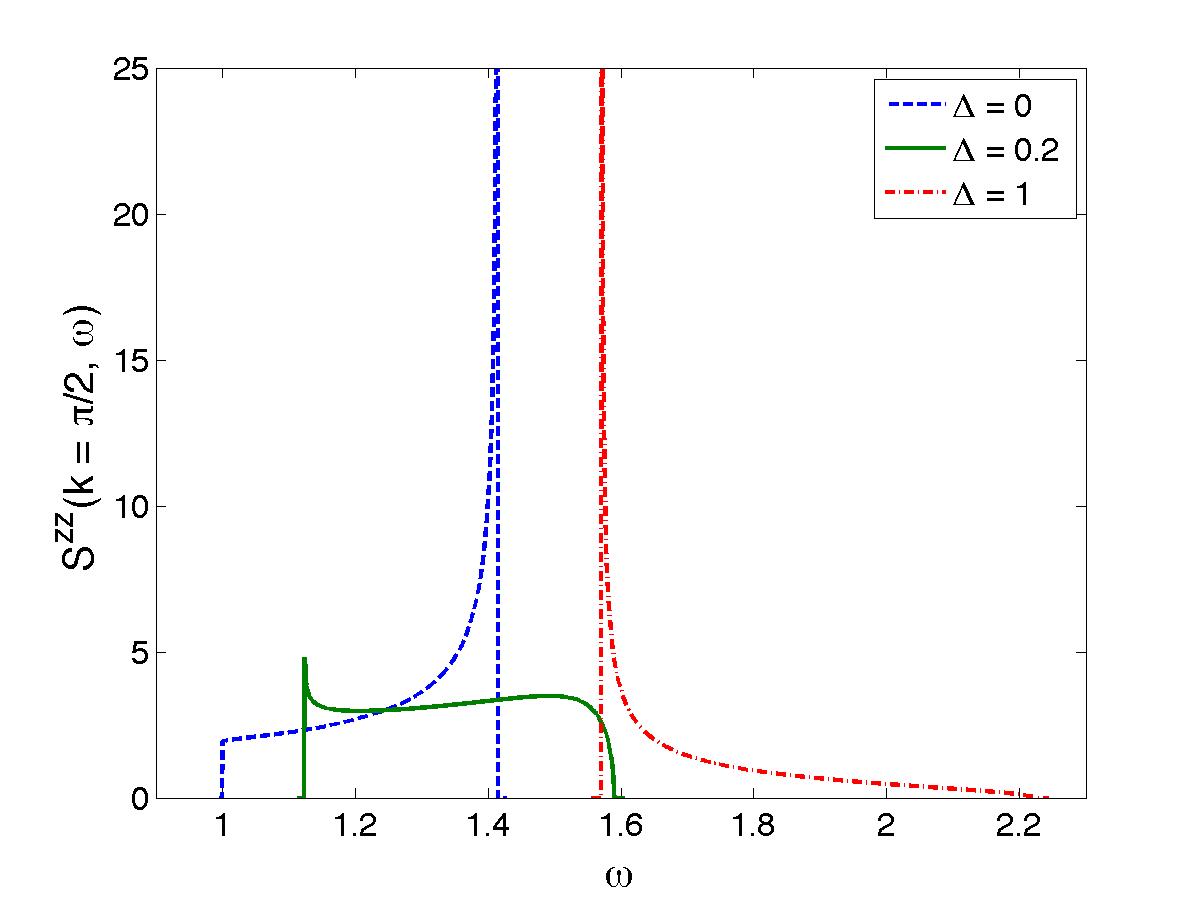}
\\
\includegraphics[width=80mm]{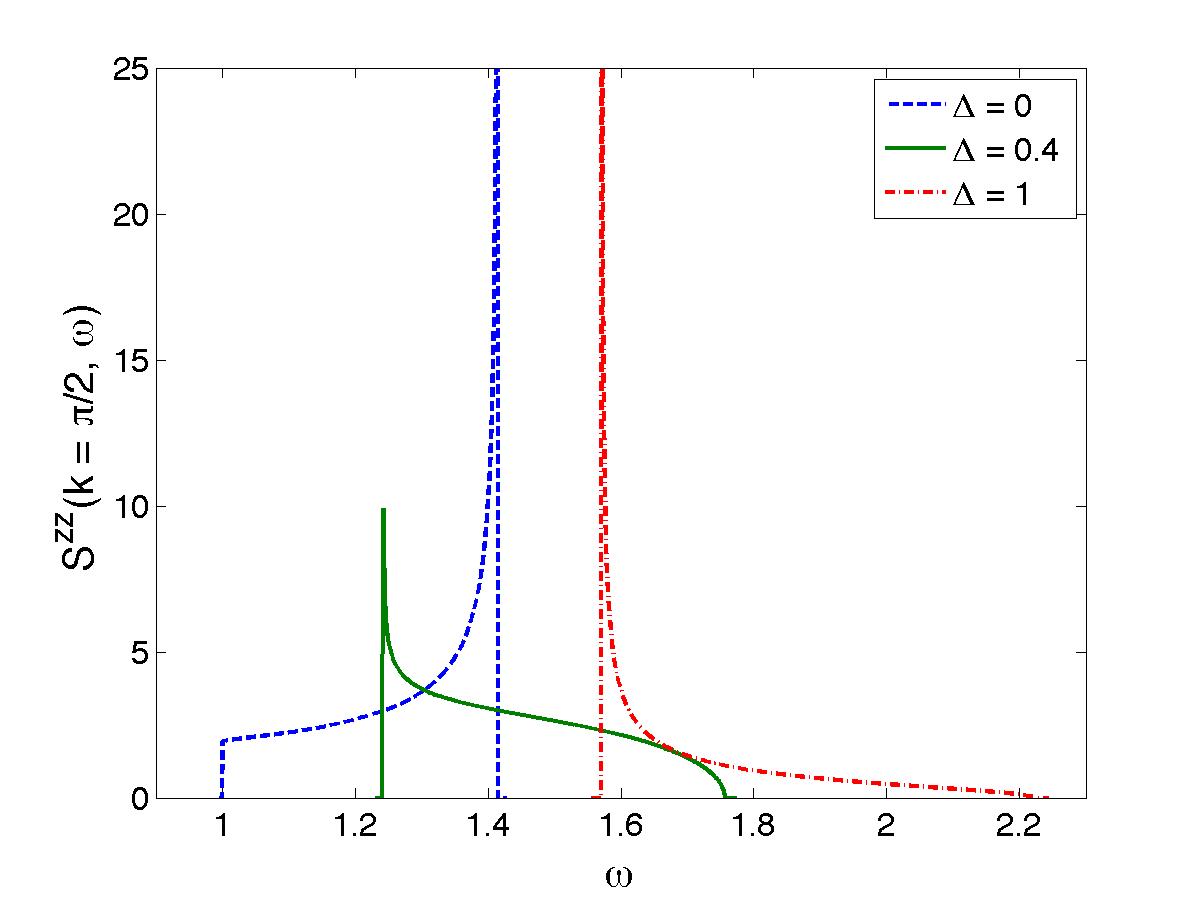}
&
\includegraphics[width=80mm]{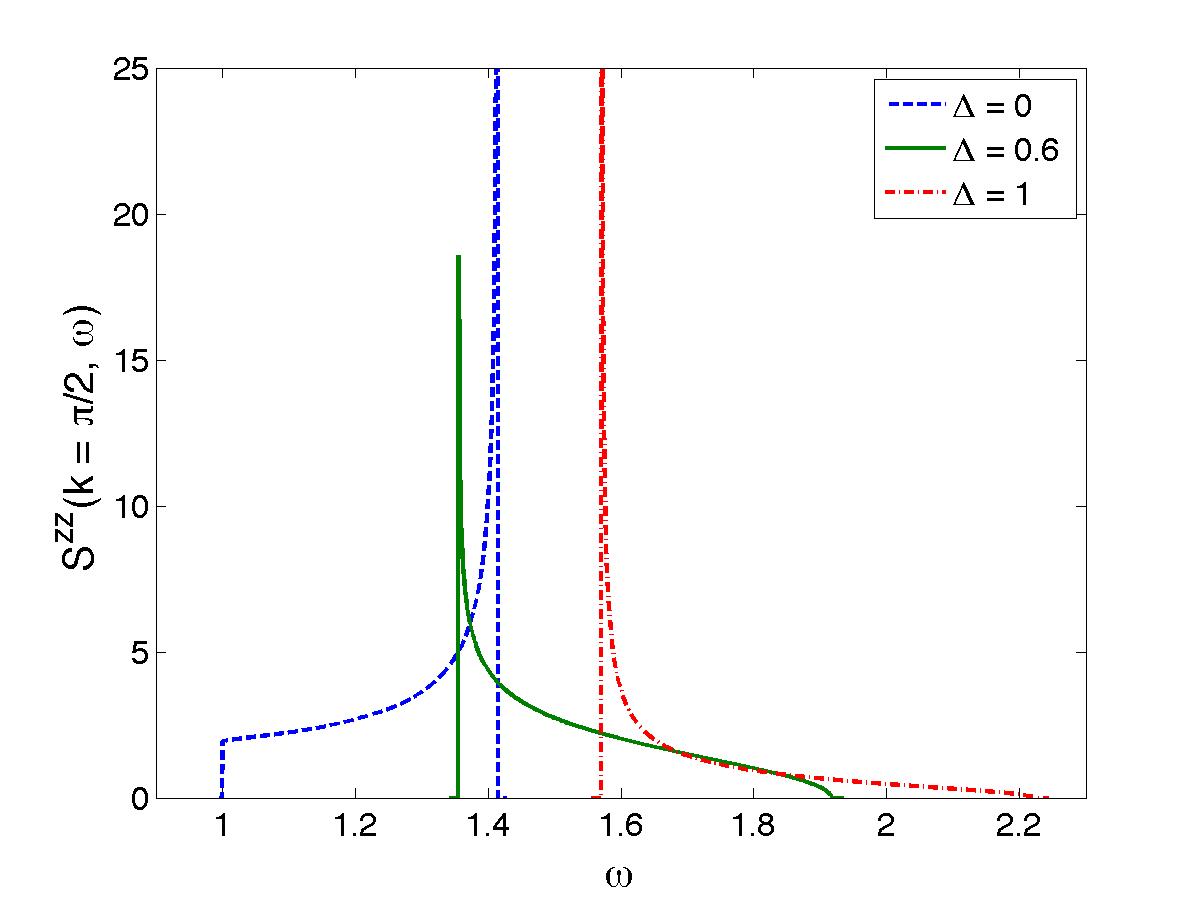}
\\
\includegraphics[width=80mm]{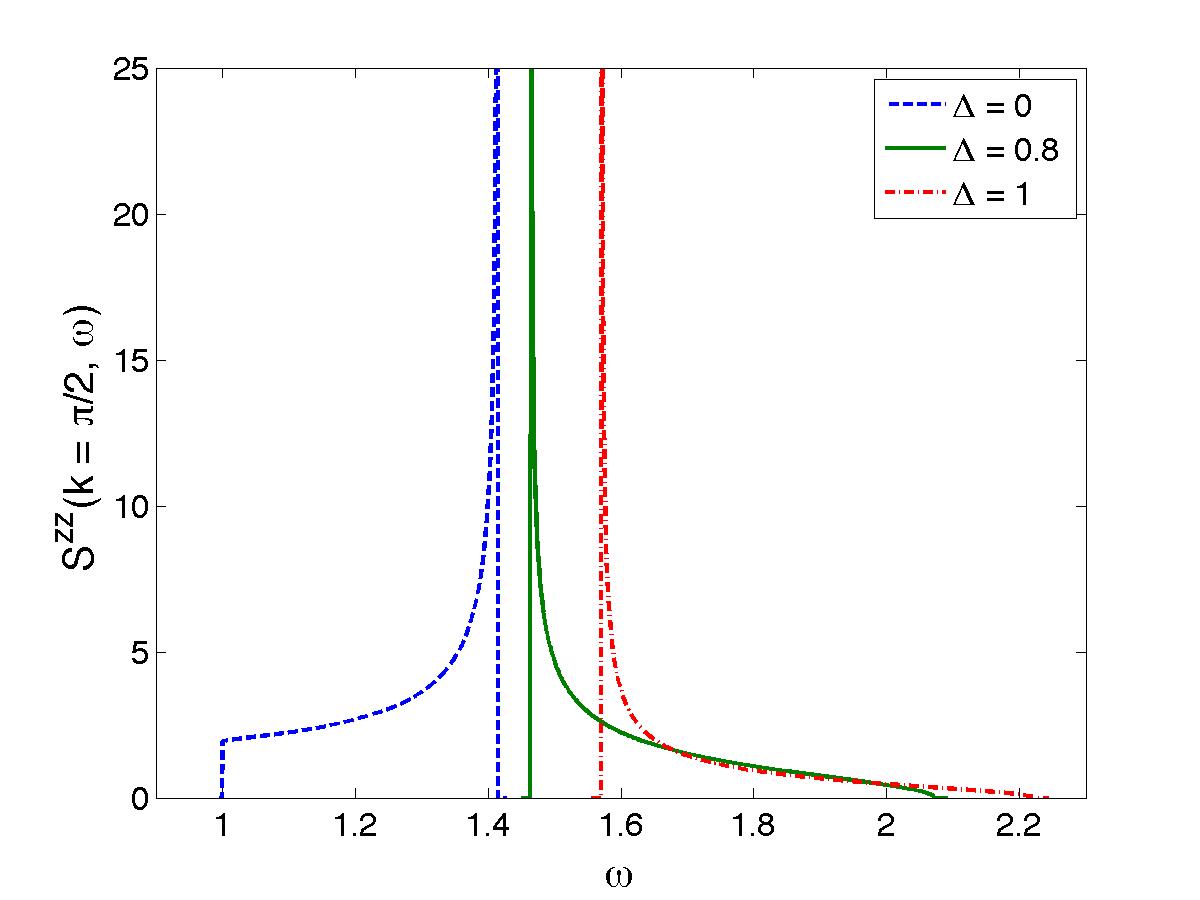}
&
\includegraphics[width=80mm]{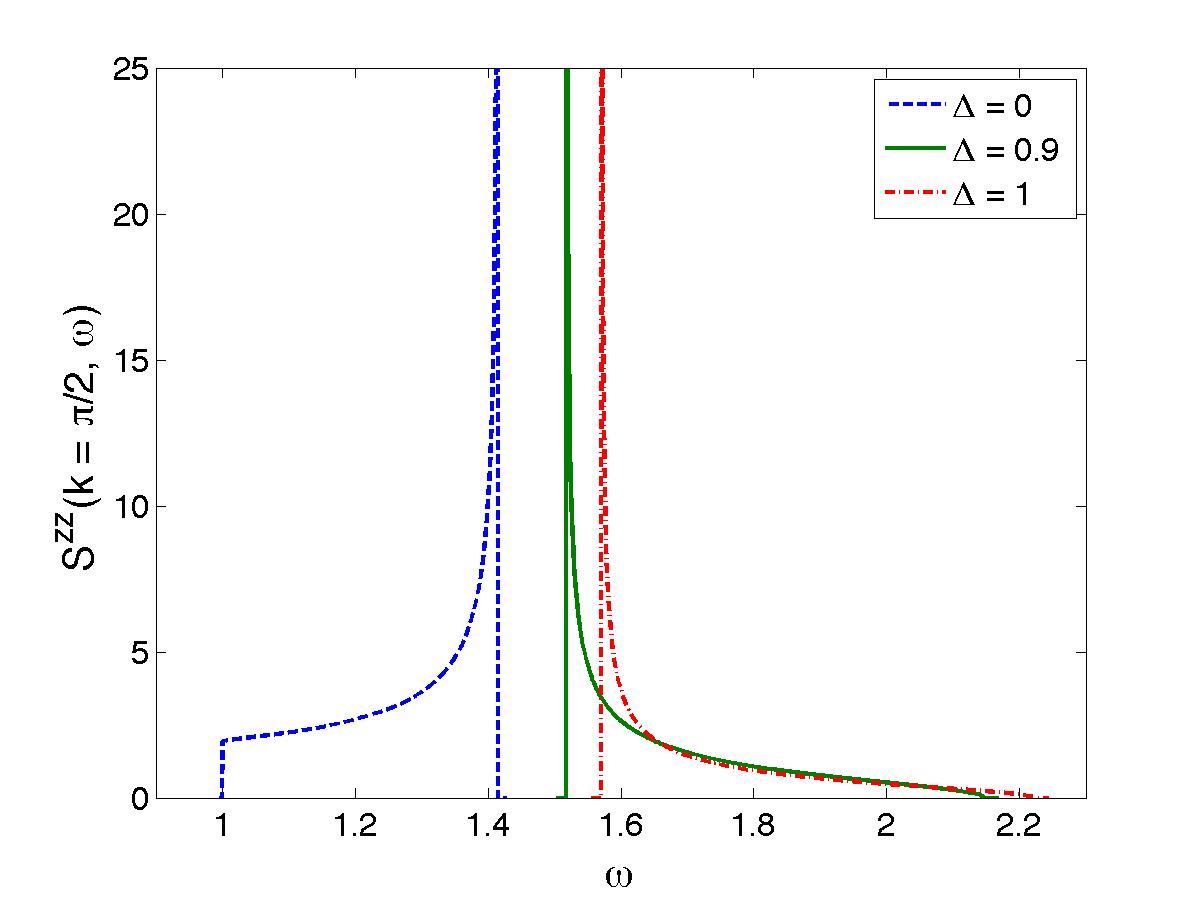}
\end{tabular}
\caption{Fixed momentum cuts at $k = \pi/2$ of the two-spinon part of the longitudinal structure 
factor of the infinite Heisenberg chain, for different values of the anisotropy parameter $\Delta$.
The $\Delta = 0$ and $\Delta = 1$ limits are displayed in all plots for comparison.}
\label{fig:LSF_K_pio2}
\end{figure}

\begin{figure}
\begin{tabular}{cc}
\includegraphics[width=80mm]{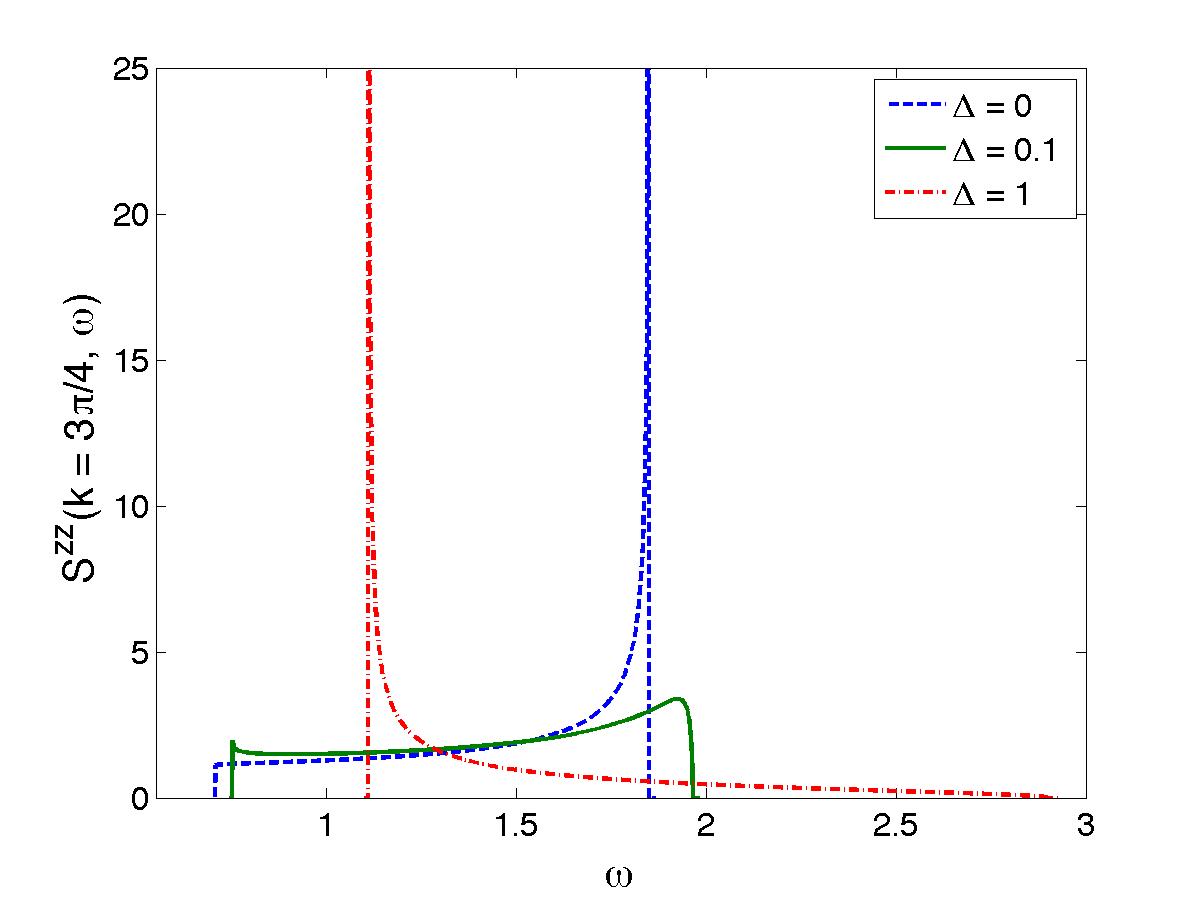}
&
\includegraphics[width=80mm]{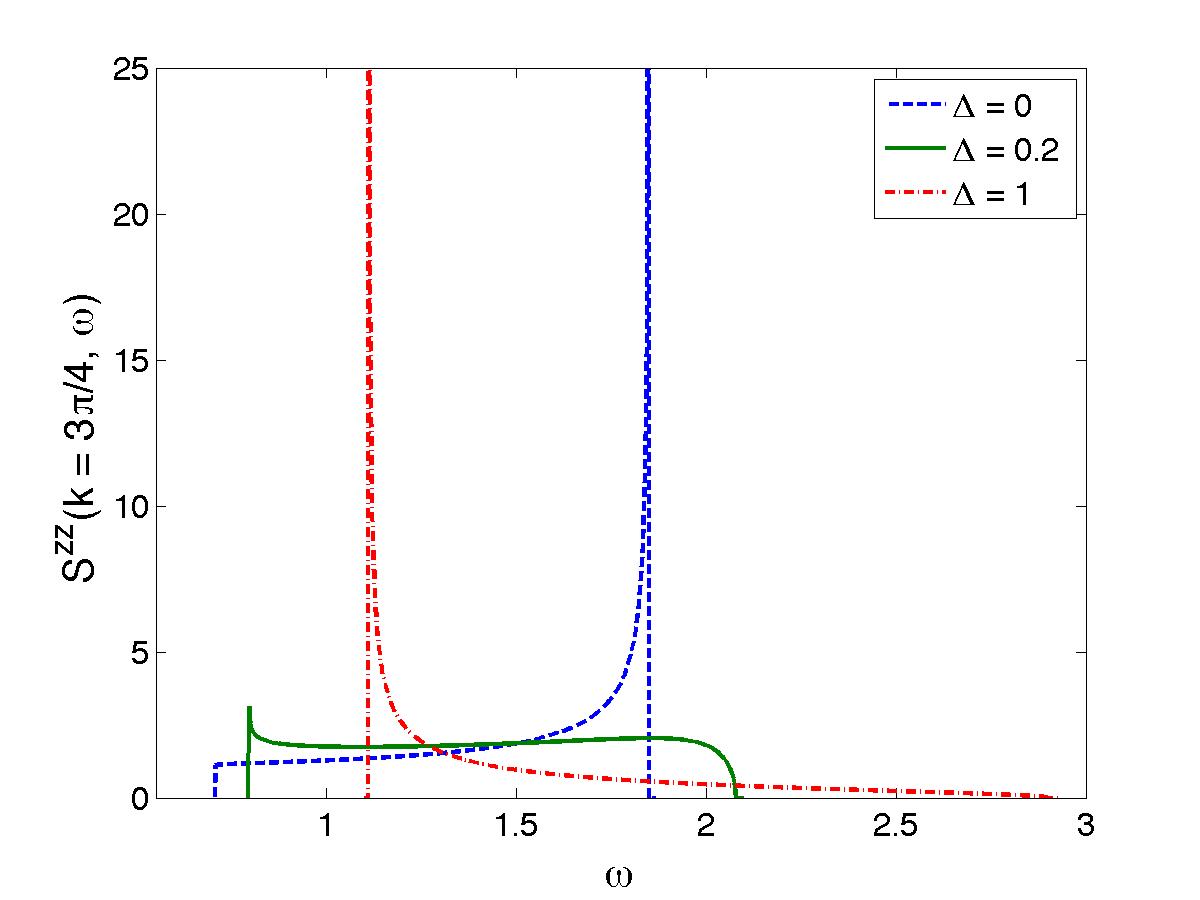}
\\
\includegraphics[width=80mm]{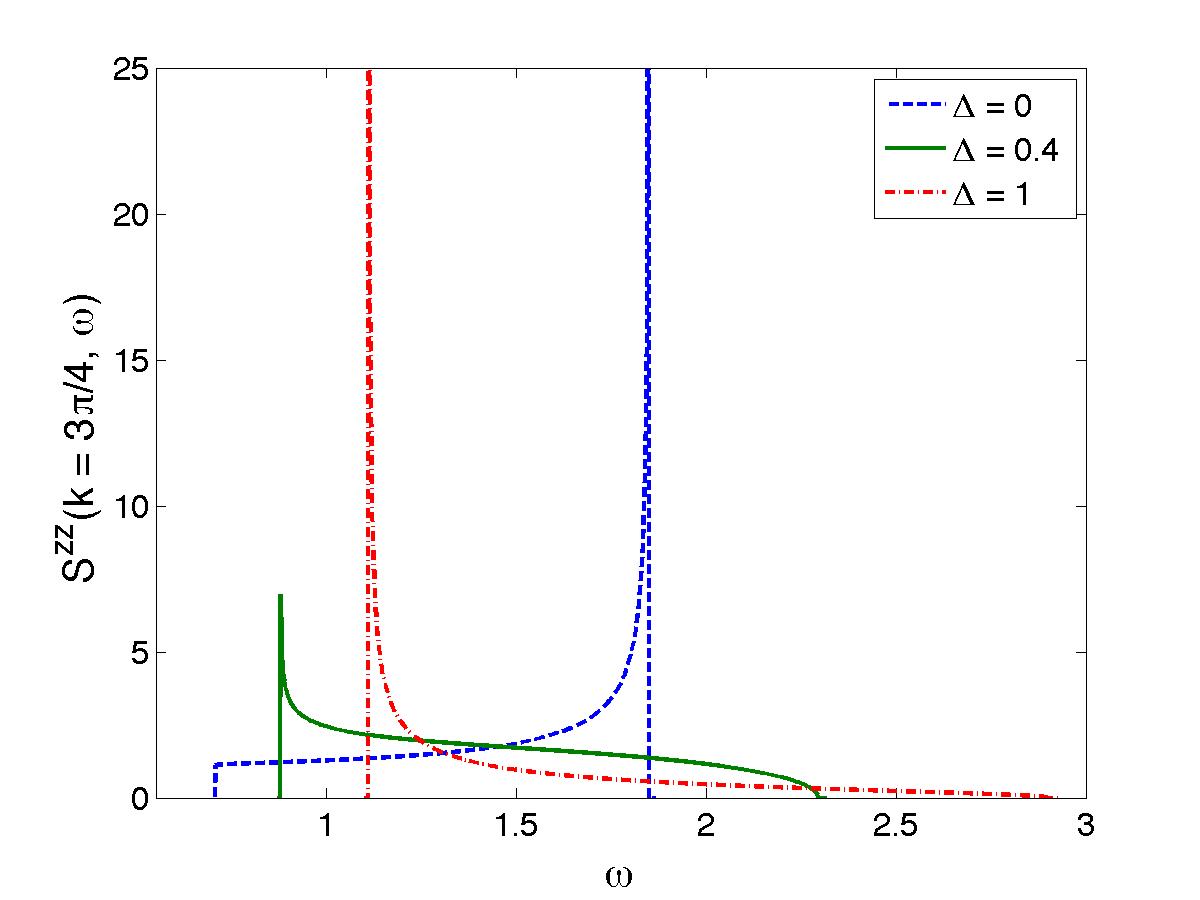}
&
\includegraphics[width=80mm]{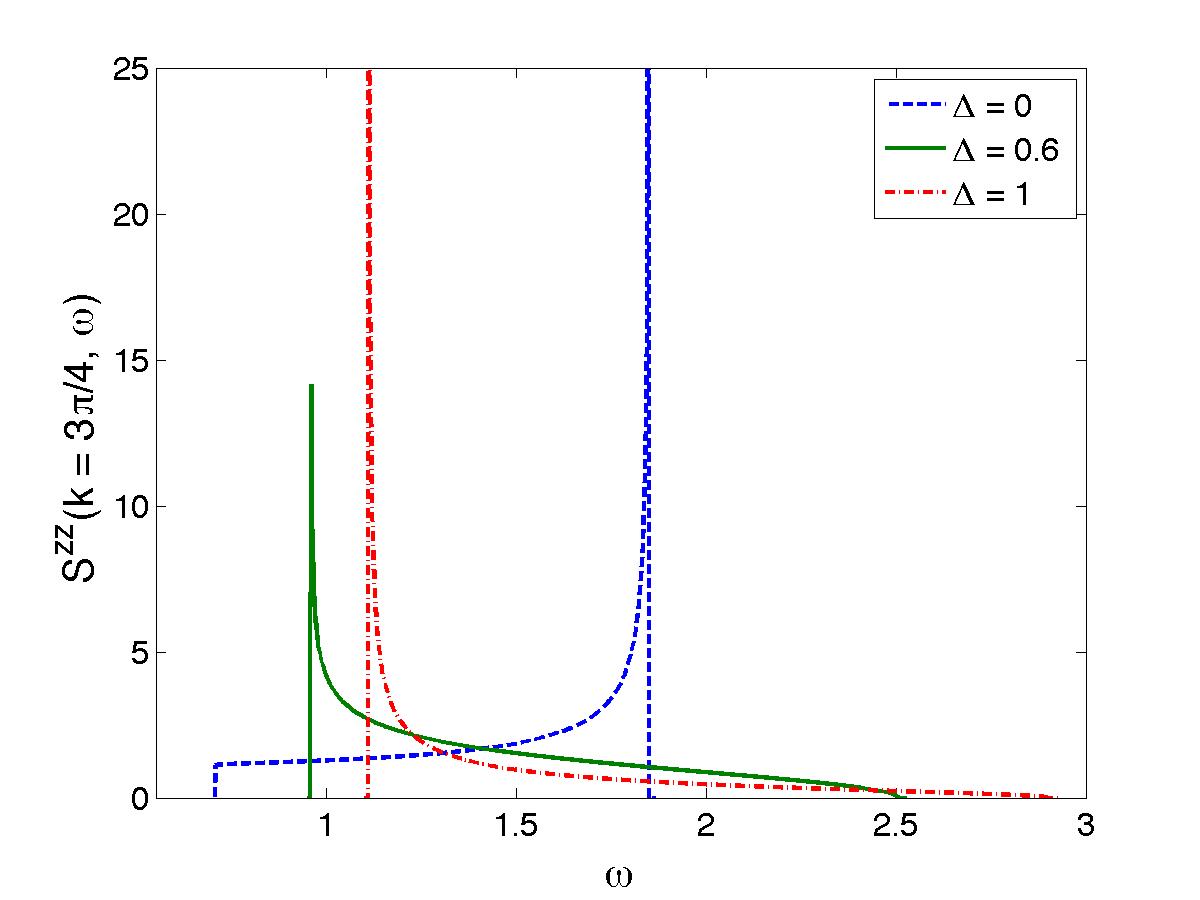}
\\
\includegraphics[width=80mm]{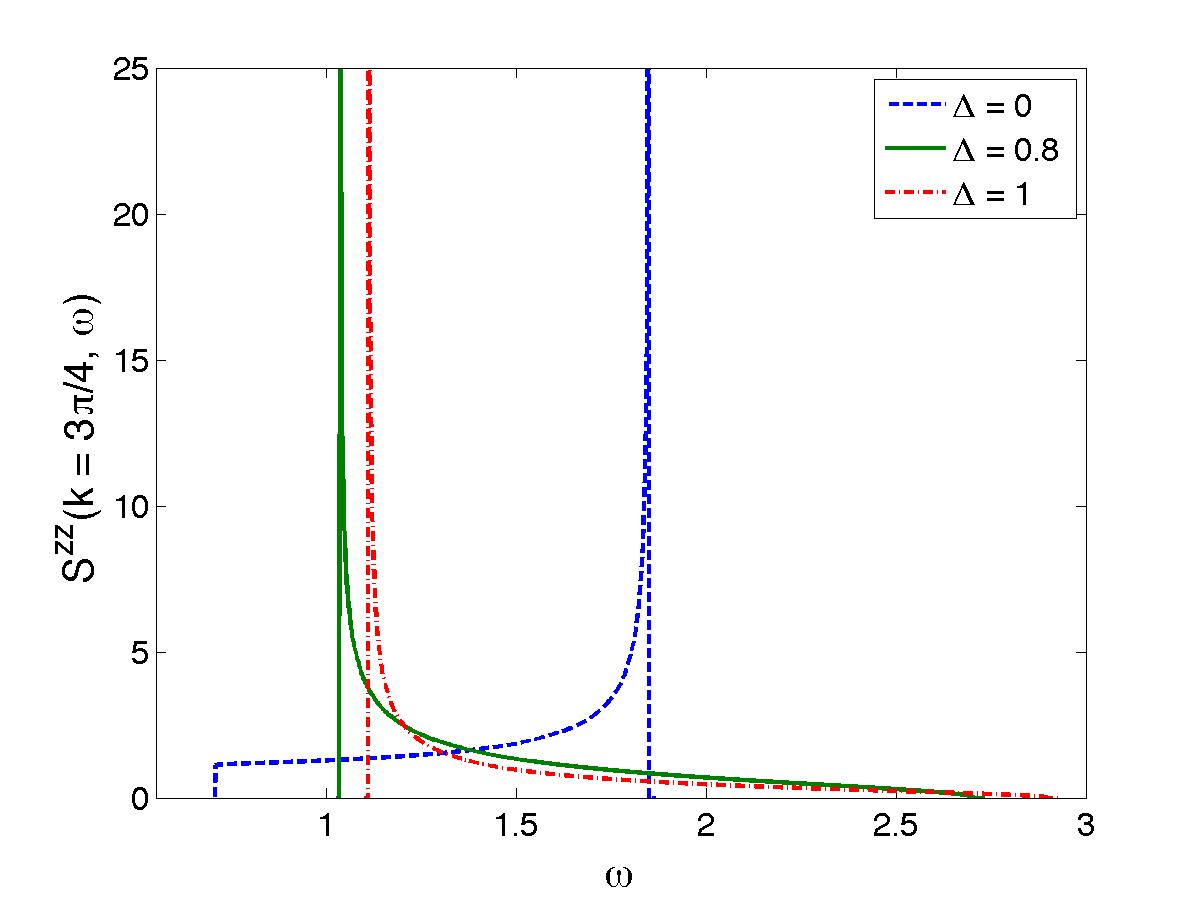}
&
\includegraphics[width=80mm]{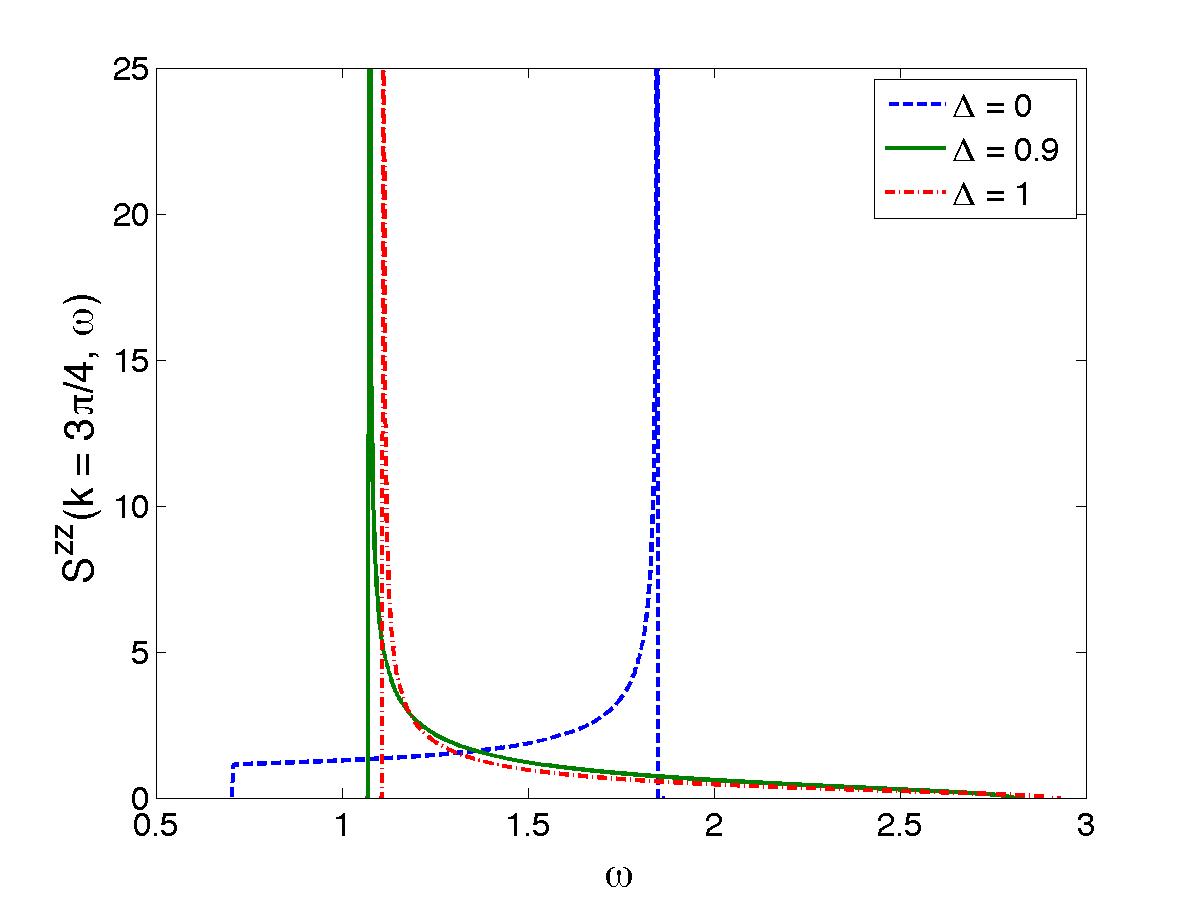}
\end{tabular}
\caption{Fixed momentum cuts at $k = 3\pi/4$ of the two-spinon part of the longitudinal structure 
factor of the infinite Heisenberg chain, for different values of the anisotropy parameter $\Delta$.
The $\Delta = 0$ and $\Delta = 1$ limits are displayed in all plots for comparison.}
\label{fig:LSF_K_3pio4}
\end{figure}

\begin{figure}
\begin{tabular}{cc}
\includegraphics[width=80mm]{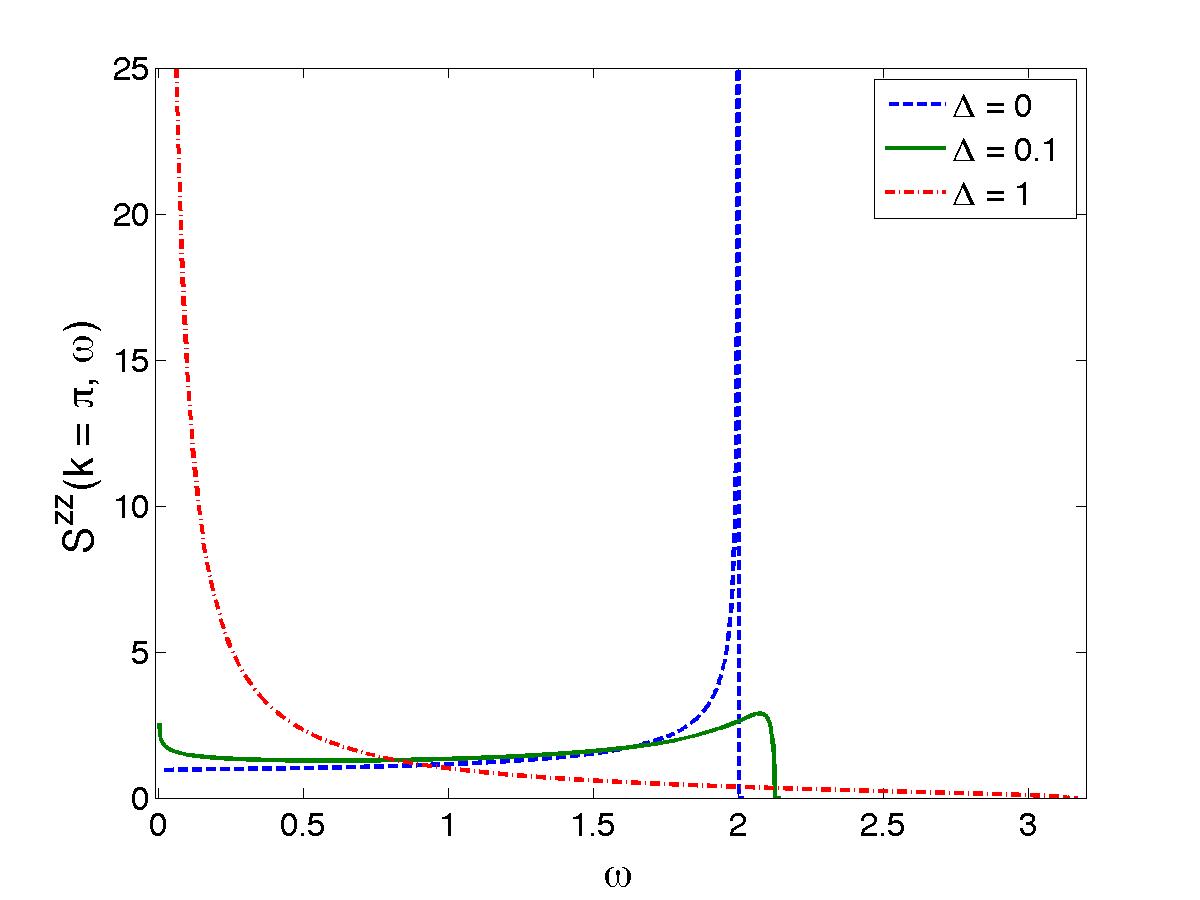}
&
\includegraphics[width=80mm]{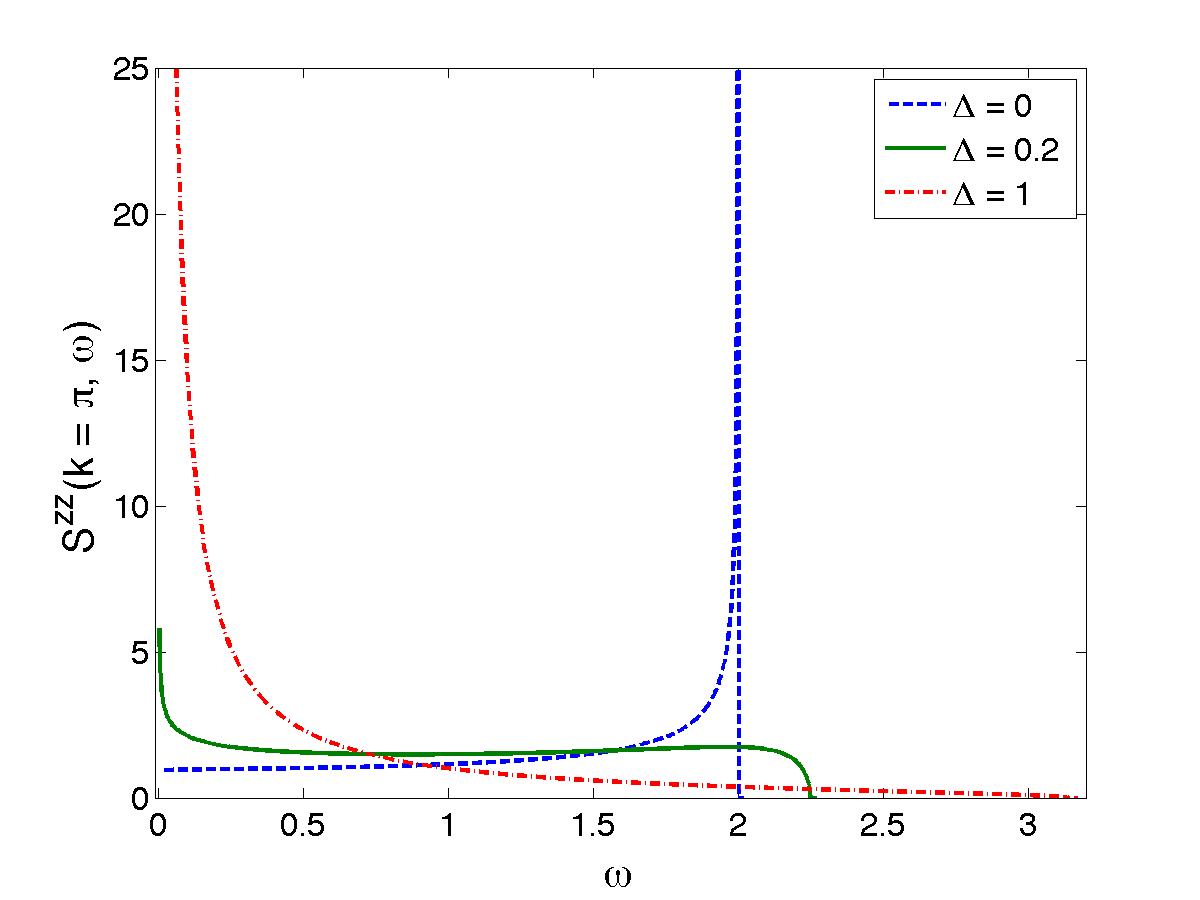}
\\
\includegraphics[width=80mm]{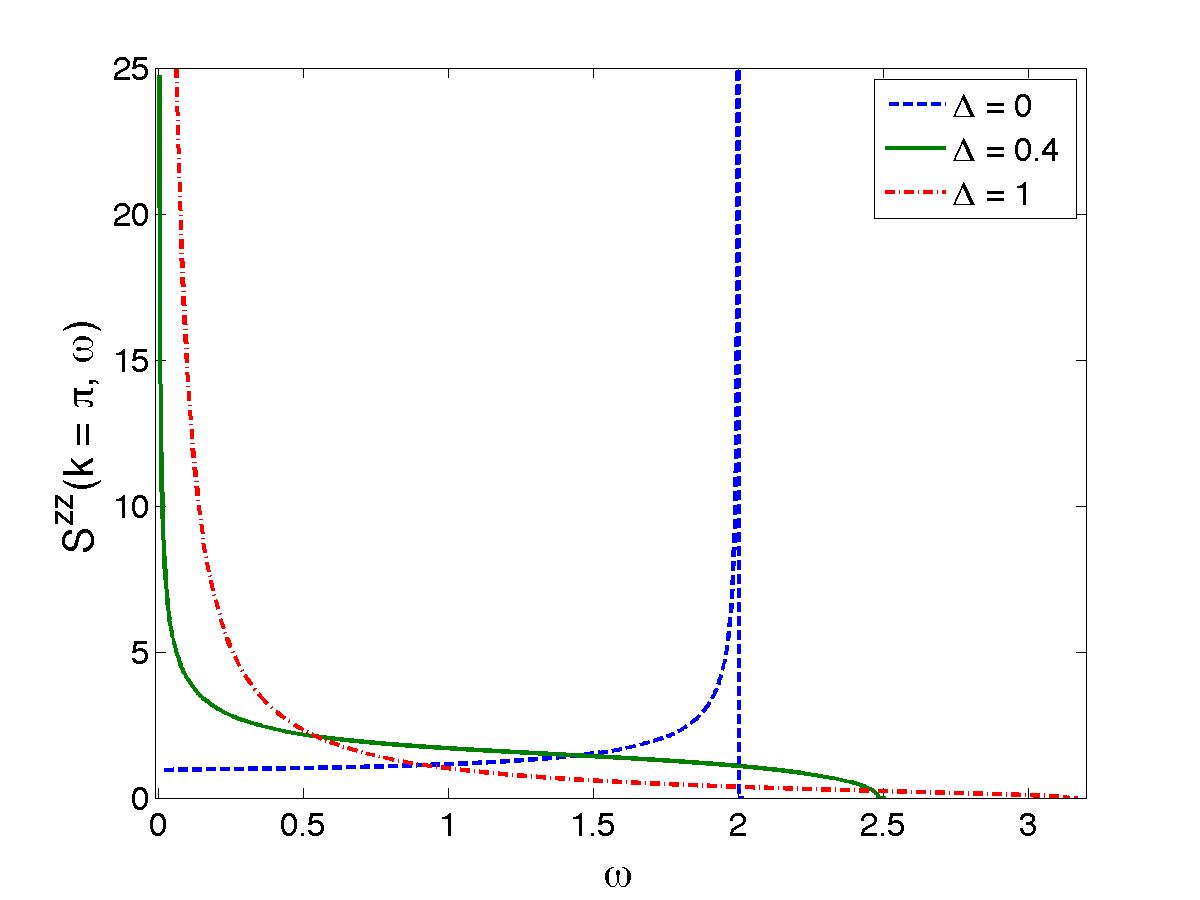}
&
\includegraphics[width=80mm]{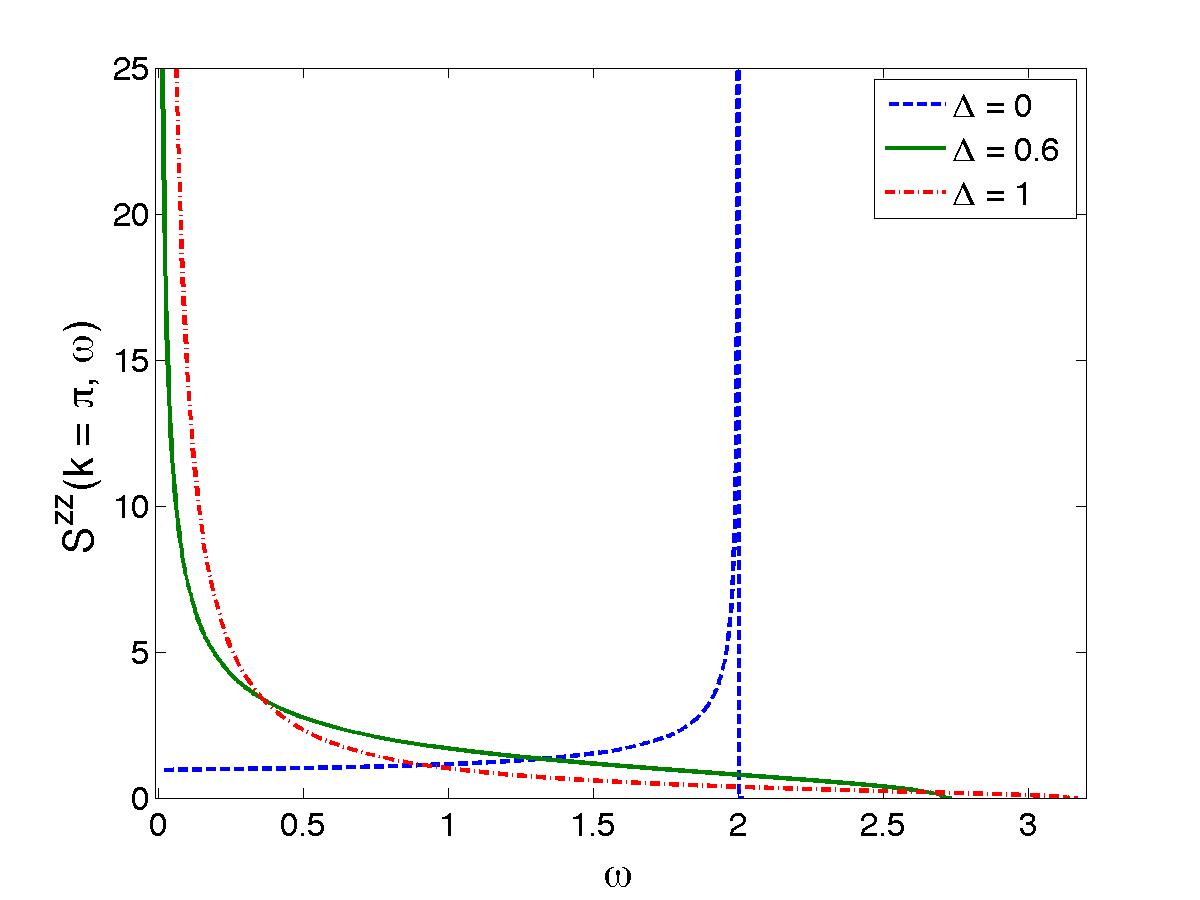}
\\
\includegraphics[width=80mm]{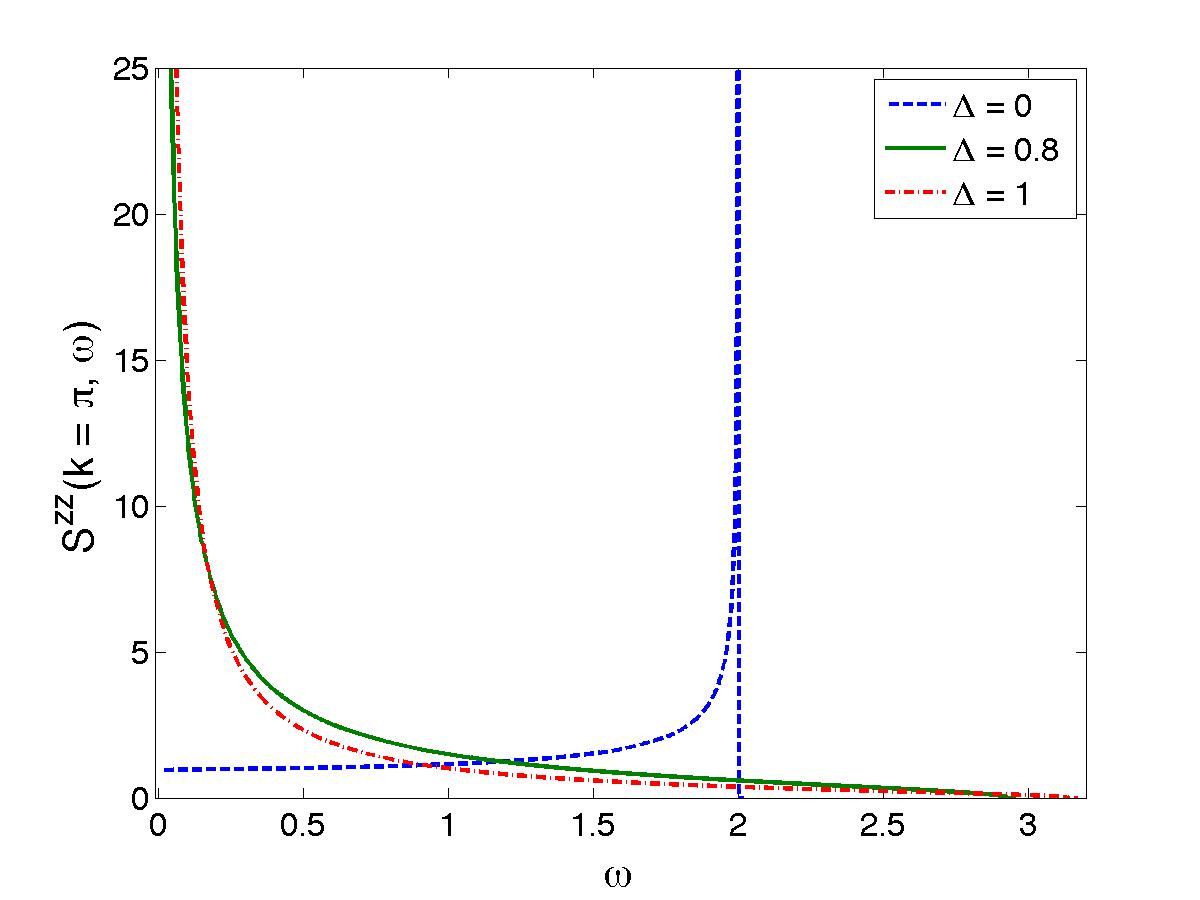}
&
\includegraphics[width=80mm]{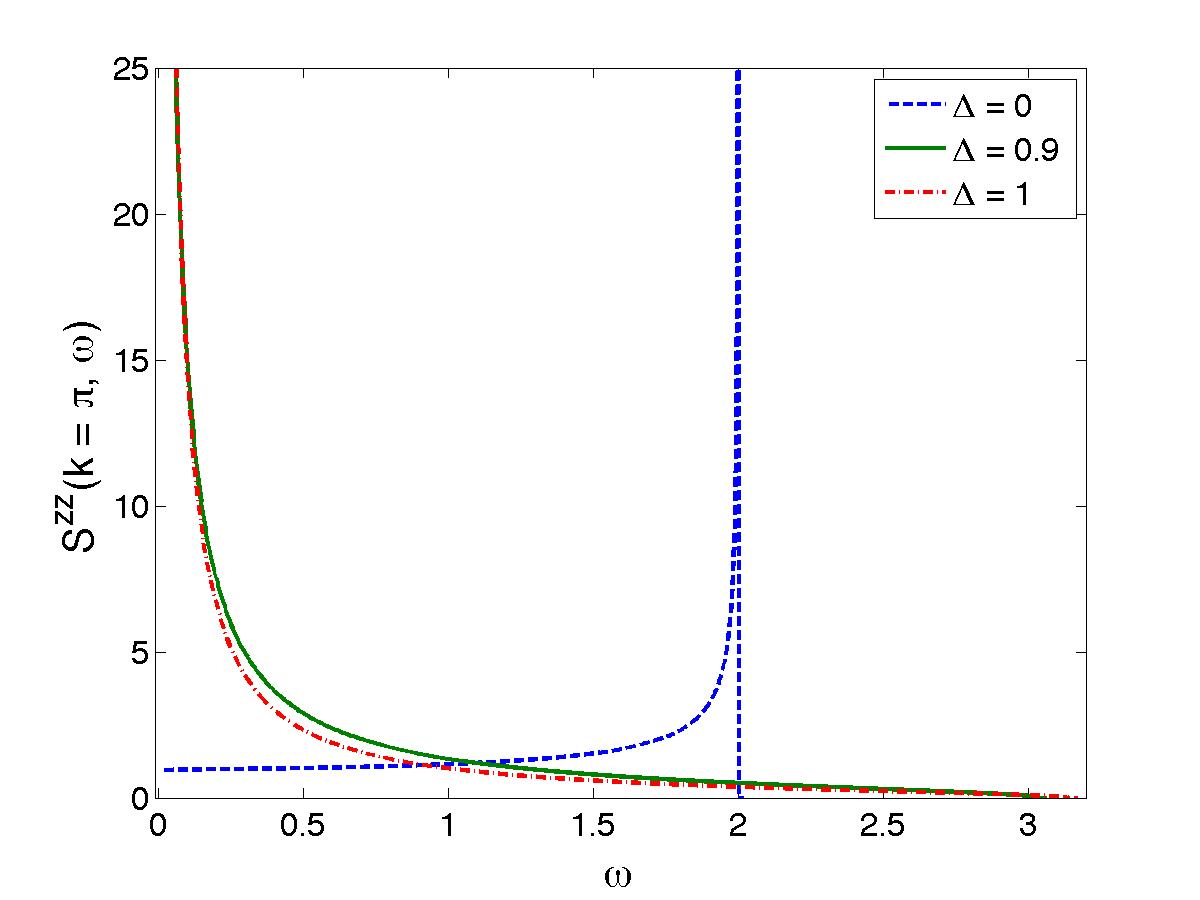}
\end{tabular}
\caption{Fixed momentum cuts at $k = \pi$ of the two-spinon part of the longitudinal structure 
factor of the infinite Heisenberg chain, for different values of the anisotropy parameter $\Delta$.
The $\Delta = 0$ and $\Delta = 1$ limits are displayed in all plots for comparison.}
\label{fig:LSF_K_pi}
\end{figure}

In Figures \ref{fig:LSF_K_pio4} - \ref{fig:LSF_K_pi}, we provide a number of fixed-momentum
cuts of the two-spinon part of the longitudinal structure factor. We organise the plots in each
figure by increasing anisotropy at fixed momentum. Each individual plot also gives the 
$\Delta = 0$ and $\Delta = 1$ curves as a reference; the effects of tuning the anisotropy
continuously between these two values can thus be easily visualised.

At $k = \pi/4$, as shown in Figure \ref{fig:LSF_K_pio4}, the two-spinon continuum is quite
narrow but its position is steadily increasing in energy as the anisotropy is turned up,
since the Fermi velocity $v_F$ (\ref{eq:vF}) is steadily increasing with increasing anisotropy.
The upper-threshold singularity at $\Delta = 0$ disappears quickly as a function of $\Delta$,
being replaced by a square-root cusp (see the discussion in Section \ref{subsec:threshold}).
The singularity at the lower threshold appears immediately but takes on a significant
weight only for around $\Delta \sim 0.3$ and above. 
The picture is very similar for the other momenta presented, namely $k = \pi/2$
(Figure \ref{fig:LSF_K_pio2}), $k = 3\pi/4$ (Figure \ref{fig:LSF_K_3pio4}) 
and $\pi$ (Figure \ref{fig:LSF_K_pi}). 

\subsection{Sum rules}
The full Hilbert space of the model in the zero magnetisation sector contains many more states
than the simple two-spinon states we have considered. Two-spinon states in fact represent
only a vanishingly small fraction of the total number of states when the system size goes to
infinity. It is thus a remarkable fact that these simple states can carry a non-vanishing
fraction of any correlation function.

To quantify the importance of the two-spinon contribution to the longitudinal structure factor, 
we consider two sum rules. First of all, we use the integrated intensity
\begin{equation}
I^{zz} = \int_0^{2\pi} \frac{dk}{2\pi} \int_0^{\infty} \frac{d\omega}{2\pi} S(k,\omega) = \frac{1}{4},
\label{eq:sr}
\end{equation}
obtained from the simple real space-time correlator $\langle S^z_{j' = j} (t = 0) S^z_j (0) \rangle = \frac{1}{4}$.
A less trivial sum rule comes from considering the integrated first frequency moment of the structure factor,
giving the so-called f-sumrule (at fixed momentum) \cite{1974_Hohenberg_PRB_10},
\begin{equation}
I^{zz}_1 (k) = \int_0^{2\pi} \frac{d\omega}{2\pi} \omega S (k,\omega) 
= -\frac{1}{2} \langle \left[ \left[ H, S^z_q \right], S^z_{-q} \right] \rangle = - 2 X^x (1 - \cos k),
\label{eq:fsumrule}
\end{equation}
where $S^z_q := \frac{1}{\sqrt{N}} \sum_j e^{i q j} S^z_j$ is the Fourier transform of the local magnetisation
operator and $X^x := \langle S^x_j S^x_{j+1} \rangle$ is the ground state expectation value of the in-plane exchange term.  
The explicit value of the right-hand side of this identity can be obtained 
from the ground-state energy density $e_0$ \cite{1966_Yang_PR_150_2} and its derivative
using the Feynman-Hellman theorem, namely $X^x = \frac{1}{2J} (1 - \Delta \frac{\partial}{\partial \Delta}) e_0$, with
\begin{equation}
e_0 = \frac{-J (\xi + 1)}{2\pi} \sin \left[\frac{\pi}{\xi \!+\! 1}\right] \int_{-\infty}^\infty \!\!\!dt \left( 1 - \frac{\tanh t}{\tanh [(\xi + 1) t]} \right).
\end{equation}

The level of saturation of the two sum rules coming from two-spinon intermediate states is inevitably anisotropy dependent.
We provide the explicit values of the sum rule saturations coming from two-spinon contributions
in Table \ref{tbl:SR} (for the f-sumrule, the saturation turns out to be exactly the same at all momenta).
Two-spinon states carry the totality of the correlation at $\Delta = 0$, a result which can be understood
by considering the mapping to free fermions using the Jordan-Wigner transformation mentioned previously. 
Two-spinon states are the only intermediate
states contributing to the longitudinal structure factor, and the sum rules are saturated to 100\%. Our results
are of course consistent with this fact.

A more remarkable fact is that the subset of two-spinon states continues to play such a determinantal role
in carrying the longitudinal structure factor even when the anisotropy has been turned on to significant values.
As our results show, two-spinon states carry essentially all the correlation weight 
up to surprisingly large values of interactions $\Delta \sim 0.8$, above
which four, six, ... spinon states become harder to neglect. This is quite surprising since,  reasoning again 
in the fermionic language obtained from the Jordan-Wigner transformation, the interaction should be able to create multiple 
particle-hole states quite easily, and arbitrarily complicated higher-spinon states should therefore 
participate in the correlation, leaving two-spinon states only negligible contribution. While true in the generic finite
magnetic field case, it turns out that for zero magnetic field, the available particle phase-space shrinks
to zero, and only the hole (spinon) part can disperse. The longitudinal structure factor thus possesses
a finite two-spinon contribution, which is not true for example of the transverse structure factor.

\begin{table}[ht]
\begin{center}
\begin{tabular}{|c|c|c|c|c|c|c|}
\hline
$\Delta$ & $I^{zz}_{2sp}/I^{zz}$ & $I^{zz}_{1,2sp}/I^{zz}_1$ &  & $\Delta$ & $I^{zz}_{2sp}/I^{zz}$ & $I^{zz}_{1,2sp}/I^{zz}_1$ \\
\hline
0 & 1 & 1 & 
& 0.6 & 0.9778 & 0.9743 \\
0.1 & 0.9997 & 0.9997 &
& 0.7 & 0.9637 & 0.9578 \\
0.2 & 0.9986 & 0.9984 &
& 0.8 & 0.9406 & 0.9314 \\
0.3 & 0.9964 & 0.9959 &
& 0.9 & 0.8980 & 0.8844 \\
0.4 & 0.9927 & 0.9917 &
& 0.99 & 0.7918 & 0.7748 \\
0.5 & 0.9869 & 0.9849 &
& 0.999 & 0.7494 & 0.7331 \\
\hline
\end{tabular}
\end{center}
\caption{Sum rule saturations as a function of anisotropy:  two-spinon contribution to
the integrated intensity $I^{zz}$ (\ref{eq:sr}) and first frequency moment $I^{zz}_1$ (\ref{eq:fsumrule}).}
\label{tbl:SR}
\end{table}

\subsection{Correspondence with finite size results}
\label{subsec:finitesize}
The longitudinal structure factor can also be obtained at finite size using explicit summations over 
intermediate states, as performed in \cite{2005_Caux_PRL_95,2005_Caux_JSTAT_P09003}.
At finite size (i.e., when the chain is made of $N$ sites; we take $N$ even to ensure a non-degenerate ground state at zero field), two-spinon states in zero field are those eigenstates obtained using $N/2 - 1$ real rapidities and a single negative-parity one-string (those readers unfamiliar with this nomenclature are referred to the explanations in \cite{2005_Caux_JSTAT_P09003}). The total number of such states then corresponds to the number of ways of choosing two hole quantum numbers from $N/2 + 1$ available ones, and thus equals $N(N+2)/8$. We can thus, at a fixed size $N$, sum over the contributions to (\ref{lsf}) coming from these two-spinon states, and compare with our infinite-size result.

\begin{figure}
\begin{tabular}{cc}
\includegraphics[width=80mm]{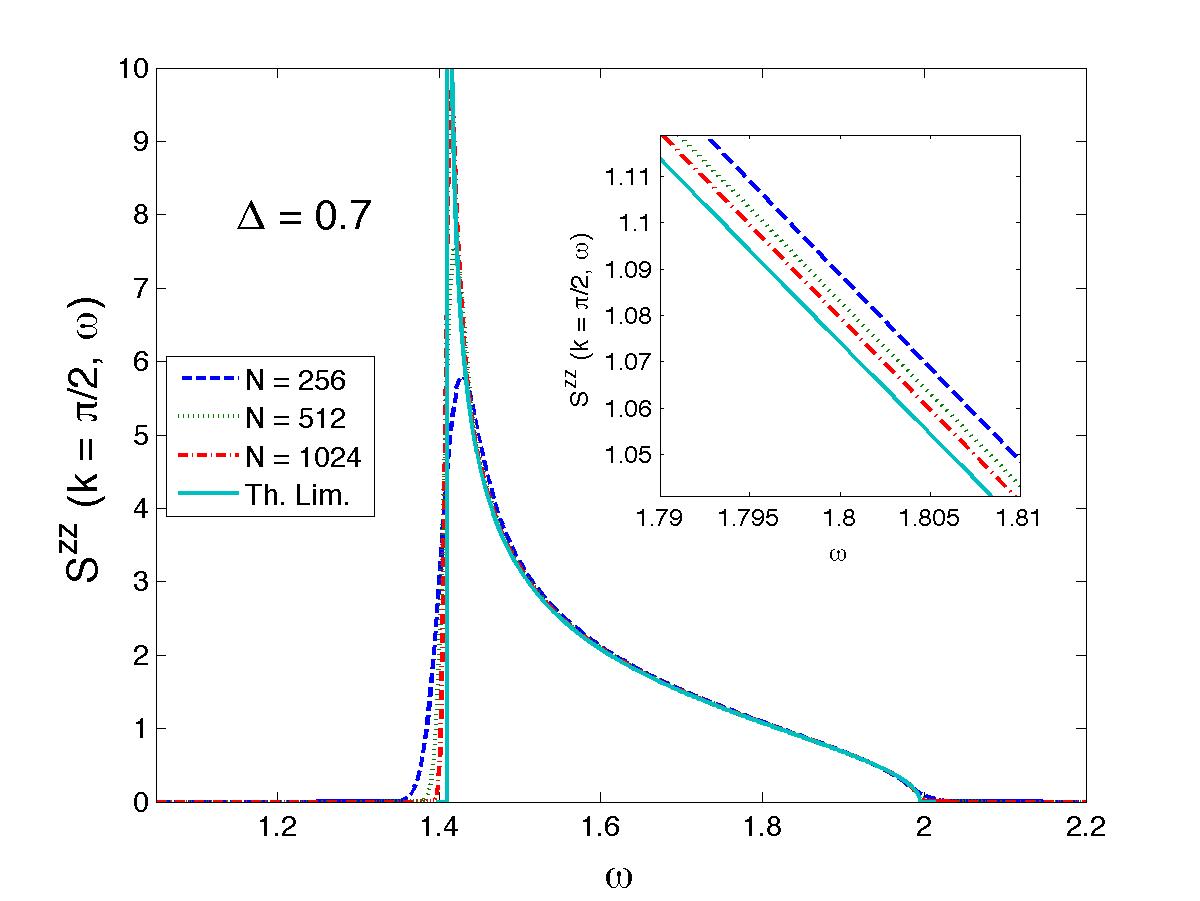}
&
\includegraphics[width=80mm]{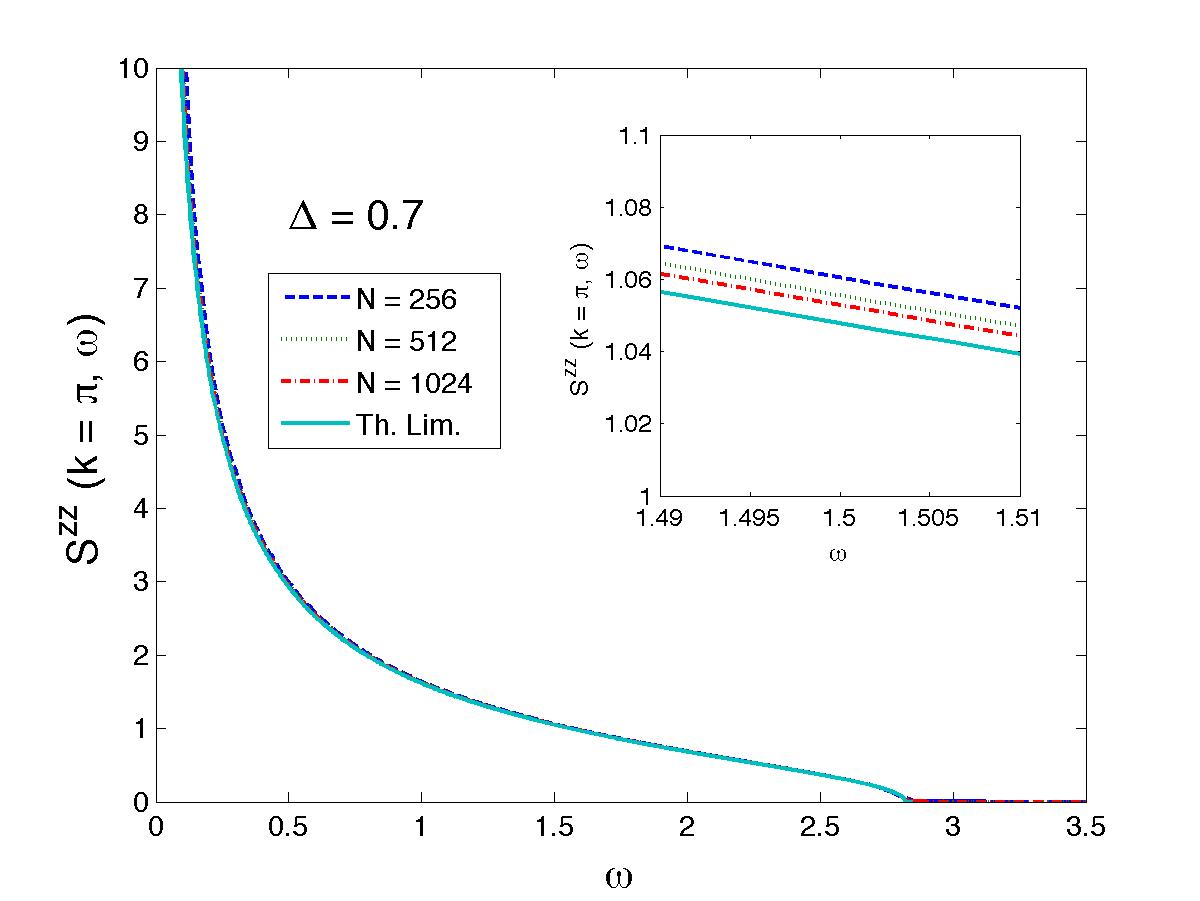}
\end{tabular}
\caption{Comparison between the two-spinon longitudinal structure factor for $\Delta = 0.7$ and momentum $k = \pi/2$ (left)
and $k = \pi$ (right), in the thermodynamic limit (solid curve) and at finite size. 
The finite size results are obtained by summing exactly over the set of $N(N+2)/8$ two-spinon states 
at chain length $N$, and are plotted after gaussian smoothing of the delta function contributions. See main text for more details.}
\label{fig:N}
\end{figure}

Figure \ref{fig:N} presents such a comparison, done at a representative value of anisotropy $\Delta = 0.7$ and two values of momentum, $k = \pi/2$ and $k = \pi$. Other values of anisotropy and momenta give qualitatively similar plots. Besides the thermodynamic limit curve obtained from plotting the two-spinon contribution we have obtained, we also present the equivalent curves for three different system sizes, $N = 256, 512$ and $1024$. The finite-size results must be smoothed with a gaussian, since the correlation function is then a sum of delta-function peaks split in energy by the mean energy level spacing. This smoothing can be sharpened at increasing system size, and this increasing sharpness can be clearly seen, e.g., at the lower boundary of the $k = \pi/2$ plot. It is clear that the finite-size curves tend to the thermodynamic one as system size increases; the inset of each plot offers a magnified view of a selected region away from the singular thresholds, the quantitative differences between the highest system size considered and the infinite size limit being of the order of a percent.

\begin{table}[ht]
\begin{center}
\begin{tabular}{|c|c|c|c|}
\hline
$N$ & $I^{zz}_{2sp}/I^{zz}$ & $I^{zz}_{1,2sp} (\pi/2)/I^{zz}_1$ &  $I^{zz}_{1,2sp} (\pi)/I^{zz}_1$ \\
\hline
64 & 0.9893 & 0.9825 & 0.9852 \\
128 & 0.9843 & 0.9778 & 0.9776 \\
256 & 0.9796 & 0.9733 & 0.9713 \\
512 & 0.9756 & 0.9695 & 0.9668 \\
1024 & 0.9724 & 0.9664 & 0.9636 \\
\mbox{extrap} & 0.963(2) & 0.957(4) & 0.957(4) \\
$\infty$ & 0.9637 & 0.9578 & 0.9578 \\
\hline
\end{tabular}
\end{center}
\caption{Sum rule contributions obtained from the finite-size calculation of the two-spinon contributions to the longitudinal structure factor at the representative value of anisotropy $\Delta = 0.7$, for system sizes up to $N = 1024$ and extrapolated (see text), as compared to the analytical result at infinite system size. See main text for more details.}
\label{tbl:SRN}
\end{table}

In Table \ref{tbl:SRN}, we provide quantitative results for the sum rule contributions for the representative value of anisotropy $\Delta = 0.7$. Similarly to Table \ref{tbl:SR}, we provide both the integrated intensity sum rule contribution (\ref{eq:sr}) as well as the f-sumrule one (\ref{eq:fsumrule}), coming from two-spinon states. The results at finite size are clearly seen to tend to their infinite-size limit. For completeness, we have extrapolated the finite-size results using data at $N = 256, 512, 768$ and $1024$, fitting with a polynomial in $1/\sqrt{N}$. Within the accuracy of the extrapolation, these results coincide with the ones obtained from the analytical form.

\subsection{Threshold behaviour}
\label{subsec:threshold}
The investigation of the precise form of the longitudinal structure factor near the 
upper and lower boundaries of the two-spinon continuum, which our results render possible, 
is of considerable theoretical interest in view of recent developments in the general phenomenology
of one-dimensional quantum liquids \cite{2008_Imambekov_PRL_100,2009_Imambekov_SCIENCE_323,1110.1374}
coming from efforts to calculate dynamical correlations away from the low-energy
limit \cite{2006_Pustilnik_PRL_96,2006_Pereira_PRL_96,2007_Khodas_PRB_76,2007_Pereira_JSTAT_P08022,2008_Cheianov_PRL_100,2008_Pereira_PRL_100,2009_Pereira_PRB_79}. For the special case of the
longitudinal structure factor of the zero-field gapless $XXZ$ antiferromagnet we are considering, 
which is equivalent to the density-density correlator of spinless interacting fermions obtained
via a Jordan-Wigner transformation, the field theory predictions of \cite{2008_Pereira_PRL_100,2009_Pereira_PRB_79} yield a singular LSF in the vicinity of the lower/upper thresholds of the two-spinon continuum. In the vicinity of the upper threshold, for $0 < \Delta < 1$, this singularity is
shown to be of the form $S^{zz} (k, \omega) ~\substack{\vspace{1mm} \\ \xrightarrow{\hspace{14mm}} \\ {\omega \rightarrow \omega_{2,u}(k)}} ~ \sqrt{\omega_{2,u} (k) - \omega}$. For the lower threshold, the power-law becomes anisotropy-dependent, $S^{zz} (k, \omega) ~\substack{\vspace{1mm} \\ \xrightarrow{\hspace{14mm}} \\ {\omega \rightarrow \omega_{2,l}(k)}} ~ (\omega - \omega_{2,l} (k))^{-(1 - K)}$ in which $K$ is the Luttinger parameter, which for the zero field $XXZ$ model takes the value $K = \frac{1}{2} (1 - \frac{\mbox{acos} \Delta}{\pi})^{-1}$. In terms of the parameter $\xi$, the Luttinger parameter becomes $K = \frac{1}{2} (1 + \frac{1}{\xi})$. Our discussion here has two aims: firstly, to reproduce and possibly refine the determination of this threshold behaviour; secondly, to quantify its region of validity, which is very difficult to achieve within nonlinear Luttinger liquid theory.

Let us thus consider evaluating the two-spinon part of the longitudinal structure factor in the vicinity of the excitation thresholds, starting from our exact representation (\ref{eq:Szz2}). 
We consider the explicit evaluation of the fundamental integral (\ref{keyint})
in various limits.  For convenience, we rewrite it as
\begin{eqnarray*}
I_\xi (\rho) = I^{(1)}_\xi (\rho) - I^{(2)}_\xi (\rho),
\end{eqnarray*}
where
\bea
I^{(1)}_\xi (\rho) &:=& \int_0^\infty \frac{dt}{t} \frac{\sinh (\xi + 1)t}{\sinh \xi t} \frac{\sinh t}{\cosh^2 t} \cos 4\rho t, 
\label{eq:Ixi1}
\\ \hb{and}\ws
I^{(2)}_\xi (\rho) &:=& \int_0^\infty \frac{dt}{t} \frac{\sinh (\xi + 1)t}{\sinh \xi t} \frac{\sin^2 2\rho t}{\sinh t \cosh^2 t}
\label{eq:Ixi2}.
\eea
Throughout the discussion below, unless we specifically mention otherwise,
we consider the generic case $\xi = \mbox{O}(1)$.

\subsubsection{The structure factor near the upper threshold.} 
In order to obtain the structure factor near the upper boundary of the two-spinon continuum,
i.e., $\omega \rightarrow \omega_{2,u}(k)$, we must consider the limit $\rho \rightarrow 0$
of the fundamental integral. Let us thus look at the two integrals (\ref{eq:Ixi1}) and
(\ref{eq:Ixi2}) in turn.

%\subsubsection{$I^{(1)}_{\xi} (\rho)$} 
\paragraph{$a)$ $I^{(1)}_{\xi} (\rho)$:} 
For $\rho = 0$, the integrand of $I^{(1)}$ is regular at $t \rightarrow 0$, 
but the integral diverges logarithmically as $t \rightarrow \infty$.  
We thus expect a log divergence as a function of $\rho$,
$I^{(1)}_\xi (\rho) \rightarrow -c \ln \rho + d$ where $c > 0$ and $c, d = \mbox{O}(1)$. 
We can in fact immediately predict the value of the coefficient $c$ by 
looking at the ratio of hyperbolic functions, which tends to $2$ at $t \rightarrow \infty$, so $c = 2$.  
This is easily proved by rewriting $I^{(1)}_\xi$
(defining $f^{(1)}_\xi (t) := \frac{\sinh (\xi + 1)t}{\sinh \xi t} \frac{\sinh t}{\cosh^2 t}$) 
using the cosine integral Ci as
\begin{eqnarray*}
I^{(1)}_\xi (\rho) = I^{(11)}_\xi (\rho | \bar{t}_1) + I^{(12)}_\xi (\rho | \bar{t}_1) - 2 \hb{Ci} (4\rho \bar{t}_1), 
\end{eqnarray*}
in which $\bar{t}_1$ is an arbitrary real number and
\begin{eqnarray*}
I^{(11)}_\xi (\rho | \bar{t}_1) := \int_0^{\bar{t}_1} \frac{dt}{t} f^{(1)}_\xi (t) \cos 4\rho t, 
\quad
I^{(12)}_\xi (\rho | \bar{t}_1) := \int_{\bar{t}_1}^\infty \frac{dt}{t} (f^{(1)}_\xi (t) - 2) \cos 4\rho t,
\end{eqnarray*}
and we have used the identity
\begin{eqnarray*}
\int_{\bar{t}_1}^\infty \frac{dt}{t} \cos 4\rho t = - \hbox{Ci} (4\rho \bar{t}_1).
\end{eqnarray*}
Let us choose $\bar{t}_1 = \mbox{O}(1)$.  We then have 
\begin{eqnarray*}
|I^{(11)}_\xi (\rho)| \leq \int_0^{\bar{t}_1} \frac{dt}{t} f_\xi^{(1)} (t) = \mbox{O}(1).
\end{eqnarray*}
We thus have explicit convergence of the first integral,
$\lim_{\rho \rightarrow 0} I^{(11)}_\xi (\rho | \bar{t}_1) = \mbox{O}(1)$.

$I^{(12)}$ also converges explicitly for generic $\xi = \mbox{O}(1)$,
\begin{eqnarray*}
|I^{(11)}_\xi (\rho)| \leq \int_{\bar{t}_1}^\infty \frac{dt}{t} |(f_\xi^{(1)} (t) - 2)| = \mbox{O}(1).
\end{eqnarray*}
The only problematic terms as $\rho \rightarrow 0$ are therefore relegated to the cosine integral,
which can be rewritten to separate out the singular $\rho$ dependence,
\begin{eqnarray*}
\hbox{Ci} (4 \rho \bar{t}) = \mathbf{C} + \ln \rho + \ln (4\bar{t}) -2 \int_0^{\bar{t}} \frac{dt}{t} \sin^2 2\rho t.
\end{eqnarray*}
We thus obtain the partial result
\begin{eqnarray*}
I^{(1)}_\xi (\rho) 
~\substack{\vspace{1mm} \\ \xrightarrow{\hspace{5mm}} \\ {\rho \rightarrow 0}} ~
%\rightarrow_{\rho \rightarrow 0} 
-2\ln \rho + \mbox{O}(1).
\end{eqnarray*}

%\subsubsection{$I^{(2)}_{\xi} (\rho)$} 
\paragraph{$b)$  $I^{(2)}_{\xi} (\rho)$:} 
The integrand of $I^{(2)}$ vanishes sufficiently rapidly at $t \rightarrow 0$ and $t \rightarrow \infty$, so this 
integral yields a contribution of order $\rho^2 \rightarrow 0$. Thus no nontrivial contribution to the structure factor
comes from this
integral in the limit considered. 

%\subsubsection{The structure factor near the upper threshold} 
%\paragraph{c) The structure factor near the upper threshold.} 
\vspace{5mm}
This means that overall, we have
\begin{equation} 
I_\xi (\rho) 
~\substack{\vspace{1mm} \\ \xrightarrow{\hspace{5mm}} \\ {\rho \rightarrow 0}} ~
-2\ln \rho + \mbox{O}(1).
\end{equation}
We thus have (using $\rho \sim \sqrt{\omega_{2,u}(k) - \omega}$ from (\ref{rhodef})) 
\begin{equation}
%S^{zz}_2 (k, \omega) \rightarrow_{\omega \rightarrow \omega_{2,u}(k)} ~\propto \sqrt{\omega_{2,u}(k) - \omega}.
S^{zz}_2 (k, \omega) 
~\substack{\vspace{1mm} \\ \xrightarrow{\hspace{14mm}} \\ {\omega \rightarrow \omega_{2,u}(k)}} ~
%\rightarrow_{\omega \rightarrow \omega_{2,u}(k)} 
%\sim O(1) ~ \sqrt{\omega_{2,u}(k) - \omega}.
 f_u (\xi) (\sin\frac{k}{2})^{-7/2} \sqrt{\omega_{2,u}(k) - \omega}
\label{eq:uthresh}
\end{equation}
in which $f_u (\xi)$ is a momentum-independent function of anisotropy. The exponent we obtain 
confirms the field theory predictions \cite{2008_Pereira_PRL_100} for the anisotropy-independent square-root cusp at the threshold (for $0 < \Delta \leq 1$). Our results allow us additionally to extract a strongly momentum-dependent prefactor, which greatly enhances the spectral weight around the zone boundaries at $k = 0, 2\pi$, as is also noticeable in the figures. 

For the $\Delta \rightarrow 0$ limit (so $\xi \rightarrow 1$), we have to take the limit more carefully, since
the $\cosh \frac{2\pi \rho}{\xi} + \cos \frac{\pi}{\xi}$ in the denominator of the structure factor now vanishes when $\rho \rightarrow 0$.
%Overall, we get
%\begin{equation}
%S^{zz}_2 (k, \omega) \rightarrow_{\omega \rightarrow \omega_{2,u}(k)} \frac{cst}{\sqrt{\omega_{2,u}(k) - \omega}}, \hspace{5mm} \Delta = 0
%\end{equation}
Overall, in this case one rather obtains a square-root divergence,
\begin{equation}
S^{zz}_2 (k, \omega) 
~\substack{\vspace{1mm} \\ \xrightarrow{\hspace{14mm}} \\ {\omega \rightarrow \omega_{2,u}(k)}} ~
f_u (1) \frac{(\sin\frac{k}{2})^{-1/2}}{\sqrt{\omega_{2,u}(k) - \omega}}
\end{equation}
as expected, since in this non-interacting case the structure factor simply follows the density of
two-spinon states, all two-spinon form factors being energy independent and equal to each other.

\subsubsection{The structure factor near the lower threshold} 
To evaluate the structure factor near the lower threshold of the two-spinon continuum, i.e., for $\omega \rightarrow \omega_{2,l}(k)$, we need to consider the limit $\rho \rightarrow \infty$
of the fundamental integral.
Here, we again consider $\xi = \mbox{O}(1)$, and split the integrals precisely as before, using
(\ref{eq:Ixi1}), (\ref{eq:Ixi2}).
Again choosing $\bar{t}_1 = \mbox{O}(1)$, we can see that $I^{(11)}_\xi (\rho)$ is still bounded by a constant, and so is $I^{(12)}_\xi (\rho)$.
In fact, since the integrands oscillate rapidly, this constant is zero. Moreover, the cosine integral evaluated at infinity also vanishes, so we have
$I^{(1)}_\xi (\rho) 
~\substack{\vspace{1mm} \\ \xrightarrow{\hspace{5mm}} \\ {\rho \rightarrow \infty}} ~
 ~0$.

For $I^{(2)}_\xi (\rho)$, we start by writing
\begin{equation}
I^{(2)}_\xi (\rho) = I^{(21)}_\xi (\rho) + I^{(22)}_\xi (\rho), 
\end{equation}
where we have defined
\begin{equation}
I^{(21)}_\xi (\rho) = \int_0^\infty \frac{dt}{t} \frac{2\sin^2 2\rho t}{\sinh 2t},
\hspace{10mm}
I^{(22)}_\xi (\rho) = \int_0^\infty \frac{dt}{t} \frac{\sin^2 2\rho t}{\tanh \xi t \cosh^2 t}.
\end{equation}
We have
\begin{equation}
I^{(21)}_\xi (\rho) = \ln \cosh \pi \rho \simeq \pi \rho + \mbox{O}(1).
\end{equation}
We can evaluate $I^{(22)}_\xi (\rho)$ for large $\rho$ by splitting it up,
\begin{equation}
I^{(22)}_\xi (\rho) = \int_0^{\bar{t}_2} \frac{dt}{t} \frac{\sin^2 2\rho t}{\tanh \xi t \cosh^2 t}
+ \int_{\bar{t}_2}^\infty \frac{dt}{t} \frac{\sin^2 2\rho t}{\tanh \xi t \cosh^2 t}
\end{equation}
Let us choose $\bar{t}_2 = 1/\sqrt{\rho}$ (any power between 0 and 1 would do).  
For the first integral, we can write
\begin{equation}
\int_0^{1/\sqrt{\rho}} \frac{dt}{t} \frac{\sin^2 2\rho t}{\tanh \xi t \cosh^2 t} = \frac{2\rho}{\xi} (1 + \mbox{O}(\rho^{-2})) \int_0^{2\sqrt{\rho}} dt \frac{\sin^2 t}{t^2}.
\end{equation}
We also have
\begin{equation}
\int_0^{2\sqrt{\rho}} dt \frac{\sin^2 t}{t^2} = \frac{\pi}{2} - \frac{1}{4} \frac{1}{\sqrt{\rho}} + \mbox{O}(1/\rho).
\end{equation}
In the second integral (from $\bar{t}_2 = 1/\sqrt{\rho}$ to $\infty$), $\sin^2 2\rho t$ rapidly oscillates and we can thus replace it by $1/2$
when taking the limit $\rho \rightarrow \infty$.  This yields
\begin{eqnarray*}
\int_{1/\sqrt{\rho}}^\infty \frac{dt}{t} \frac{\sin^2 2\rho t}{\tanh \xi t \cosh^2 t}
= (1 + \mbox{O}(\rho^{-1})) \frac{1}{2} \int_{1/\sqrt{\rho}}^\infty \frac{dt}{t} \frac{1}{\tanh \xi t \cosh^2 t} \nonumber \\
= (1 + \mbox{O}(\rho^{-1})) \left[ \frac{1}{2} \int_{1/\sqrt{\rho}}^\infty \frac{dt}{t} \left( \frac{1}{\tanh \xi t \cosh^2 t} - \frac{1}{\xi t}\right) + \frac{\sqrt{\rho}}{2 \xi} \right] 
= \frac{\sqrt{\rho}}{2 \xi} + \mbox{O}(1).
\end{eqnarray*}
Adding up, we thus get
\begin{equation*}
I^{(2)}_\xi (\rho) 
~\substack{\vspace{1mm} \\ \xrightarrow{\hspace{5mm}} \\ {\rho \rightarrow \infty}} ~
%\rightarrow_{\rho \rightarrow \infty}
 \pi \left(1 + \frac{1}{\xi}\right) \rho + \mbox{O}(1).
\end{equation*}

%\subsubsection{The structure factor near the lower threshold} 
%\paragraph{d) The structure factor near the lower threshold.}
\vspace{5mm}
This means that the overall behaviour of the fundamental integral is
\begin{equation} 
I_\xi (\rho) 
%\rightarrow_{\rho \rightarrow \infty} 
~\substack{\vspace{1mm} \\ \xrightarrow{\hspace{5mm}} \\ {\rho \rightarrow \infty}} ~
-\pi \left(1 + \frac{1}{\xi}\right) \rho + \mbox{O}(1).
\end{equation}
We thus find (using $\rho \sim \frac{1}{2\pi} \ln (\frac{1}{\omega - \omega_{2,l}(k)})$ from (\ref{rhodef})) 
that the structure factor behaves as
%\begin{equation}\label{eq:lthresh}
%S^{zz}_2 (k, \omega) \rightarrow_{\omega \rightarrow \omega_{2,l}(k)} ~\propto \frac{1}{[\omega - \omega_{2,l} (k)]^{\frac{1}{2}(1 - 1/\xi)}}.
%\end{equation}
\begin{equation}
S^{zz}_2 (k, \omega) 
~\substack{\vspace{1mm} \\ \xrightarrow{\hspace{14mm}} \\ {\omega \rightarrow \omega_{2,l}(k)}} 
%~
%\rightarrow_{\omega \rightarrow \omega_{2,l}(k)} ~\propto 
%\frac{O(1)}{[\omega - \omega_{2,l} (k)]^{\frac{1}{2}(1 - 1/\xi)}},
 f_l (\xi) \frac{|\sin k|^{-\frac{1}{2} (1 - \frac{1}{\xi})} (\sin\frac{k}{2})^{-\frac{2}{\xi}}}
{[\omega - \omega_{2,l} (k)]^{\frac{1}{2}(1 - \frac{1}{\xi})}},
\label{eq:lthresh}
\end{equation}
where $f_l (\xi)$ is again a momentum-independent function of anisotropy.
This result assumes that $k \neq \pi$; at the antiferromagnetic point $k = \pi$, 
the behaviour becomes $1/\omega^{1 - \frac{1}{\xi}}$ due to the vanishing of $\omega_{2,l}$. 

For the $\Delta \rightarrow 0$ limit (so $\xi \rightarrow 1$), we thus get
\begin{equation}
S^{zz}_2 (k, \omega) 
~\substack{\vspace{1mm} \\ \xrightarrow{\hspace{15mm}} \\ { w_{\omega \rightarrow \omega_{2,l}(k)}} }~
cst, \hspace{5mm} \Delta = 0
\end{equation}
as expected, since the two-spinon density of states is simply a constant in this region of the continuum, 
and so are the form factors.

Our analytical form for the two-spinon part of the structure factor has thus allowed us to
reobtain the threshold exponents predicted from field theory, and to complement the threshold
behaviour of the longitudinal structure factor with momentum-dependent prefactors hard to access
within that method. 

One final comment here concerns the potential effect of higher-spinon states on the threshold
behaviour. For the generic $0 < \Delta < 1$ case, the obtained threshold exponents would remain
unchanged upon the addition of these contributions, since the power series in the energy
distance $\delta \omega$ to the singularity cannot contain any logarithmic terms which could
re-exponentiate into a different power law. The prefactor might however be corrected.
%; we conjecture
%that this would change the factors $f_{u,l} (\xi)$ in (\ref{eq:uthresh},\ref{eq:lthresh}) 
%but not the momentum dependence of the prefactors.

% %  Mark's edits start here
\subsubsection{Region of validity of threshold behaviour}
We now compare the behaviour at the lower and upper thresholds 
%(\ref{\eq:lthresh} and \ref{eq:uthresh}) 
with the numerical evaluation of expression (\ref{eq:Szz2}) in order to see over what range of frequencies these threshold formulae are valid.  We do this in two different ways. The first way consists in evaluating the ratio $S^{zz}_2/S^{zz}_{thr}$ at fixed momentum as a function of $\omega$, in which $S^{zz}_{thr}$ represents the relevant threshold behaviour in equations (\ref{eq:uthresh},\ref{eq:lthresh}) (the prefactors $f_{u,l}(\xi)$ being obtained numerically directly from the exact representation (\ref{eq:Szz2})), and to find the region of $\omega$ near the singularity for which this ratio remains one within the required accuracy. 
The second way consists in actually fitting a plot of the exact expression (\ref{eq:Szz2}) with the expected threshold power law over a finite but small frequency region near the singularity, and to then check over which interval in frequency this fit remains consistent. The latter method emulates the kind of fitting one might do starting from approximate ab-initio numerical data for the structure factor, and gives an overestimate of the region of validity. The two methods give results consistent with each other when the region of validity is at least of a few percent of the available continuum. Otherwise the stricter first way gives a much smaller region of validity.
%We have that 
%\bea
%S^{zz}_2(k,\omega)\stackrel{\omega \rightarrow \omega_{2,u}(k)}{\longrightarrow} f_u(\xi) \Bigg( \sin \frac{k}{2}\Bigg)^{-7/2}\sqrt{\omega_{2,u(k)}-\omega}
%\eea
%and
%\bea
%S^{zz}_2(k,\omega)\stackrel{\omega \rightarrow \omega_{2,l}(k)}{\longrightarrow} f_l(\xi)\frac{|\sin k|^{(-1/2)(1-1/\xi)}\big(\sin(\frac{k}{2})\big)^{-2/\xi}}{(\omega-\omega_{2,l}(k))^{(1/2)(1-1/\xi)}}
%\eea
%where $f_u(\xi)$ and $f_l(\xi)$ are momentum-independent functions of anisotropy. \\ \\
%

Tables \ref{tab:d1} , \ref{tab:d5} and \ref{tab:d9}  summarise where the difference between the threshold formula and the numerics becomes more than $1\%$ and $10\%$, as a percentage of the numerical result, using the first criterion. 
The entries in the final four columns of these tables show the approximate percentage of the $\omega$ range ($\omega_u-\omega_l$) that the threshold behaviour is valid for (within $1\%$ and $10\%$).  For example, for $\Delta=0.5$ and $k=0.125$ (the first row in Table \ref{tab:d5}), the formula for the upper threshold is within $1\%$ of the result for $\sim 7\%$ of the $\omega$ range, and within $10\%$ of the result for $\sim 47\%$ of the $\omega$ range.

\begin{figure}[h!] 
\centering
\subfigure[$k/2\pi=0.125$]{
%  \resizebox{6cm}{!}{\includegraphics{threshold_d02_k0125}}
%   \label{d2k125}
  \resizebox{7cm}{!}{\includegraphics{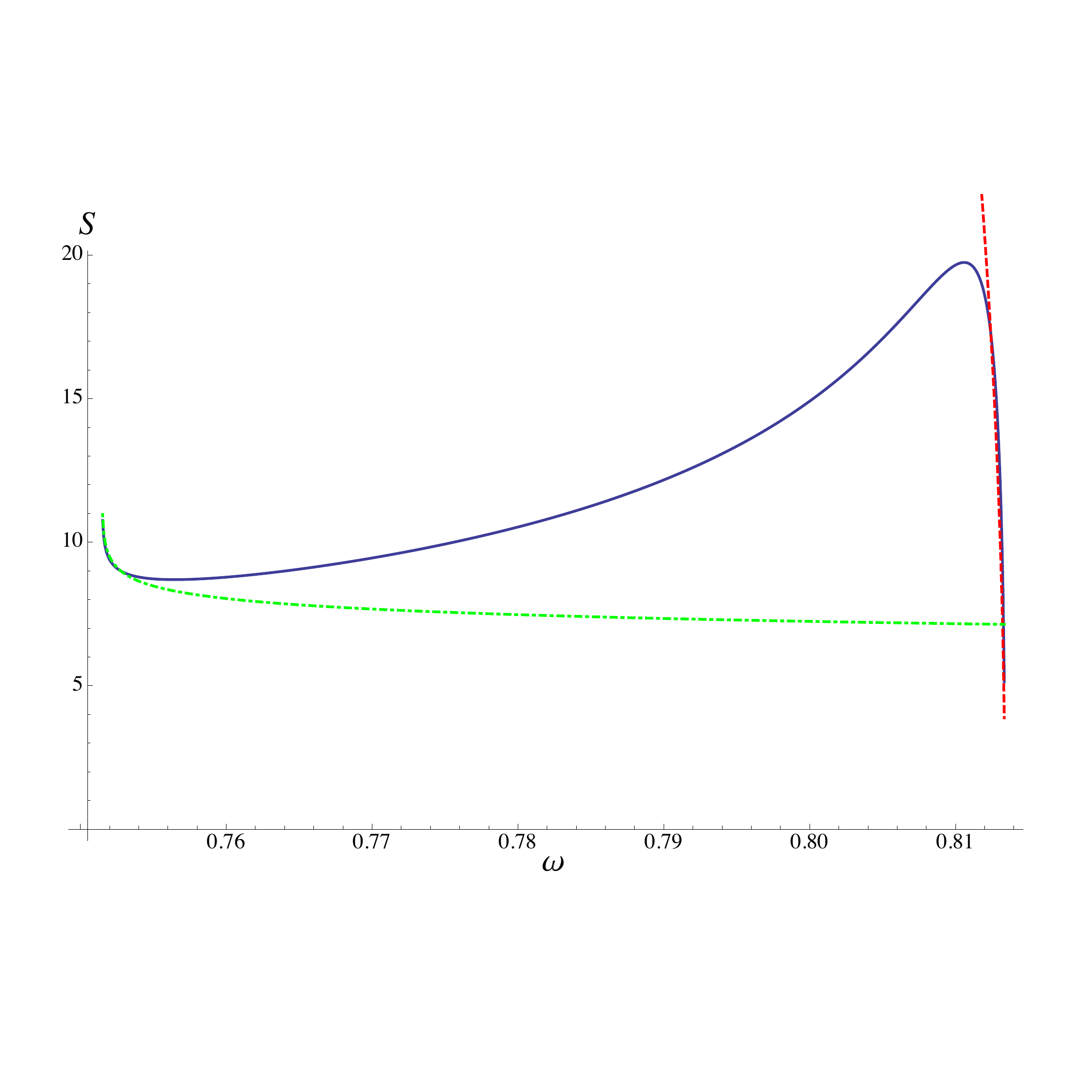}}
   \label{d01k125}
 }
 \subfigure[$k/2\pi=0.25$]{
%  \resizebox{6cm}{!}{\includegraphics{threshold_d02_k0250}}
%   \label{d2k25}
  \resizebox{7cm}{!}{\includegraphics{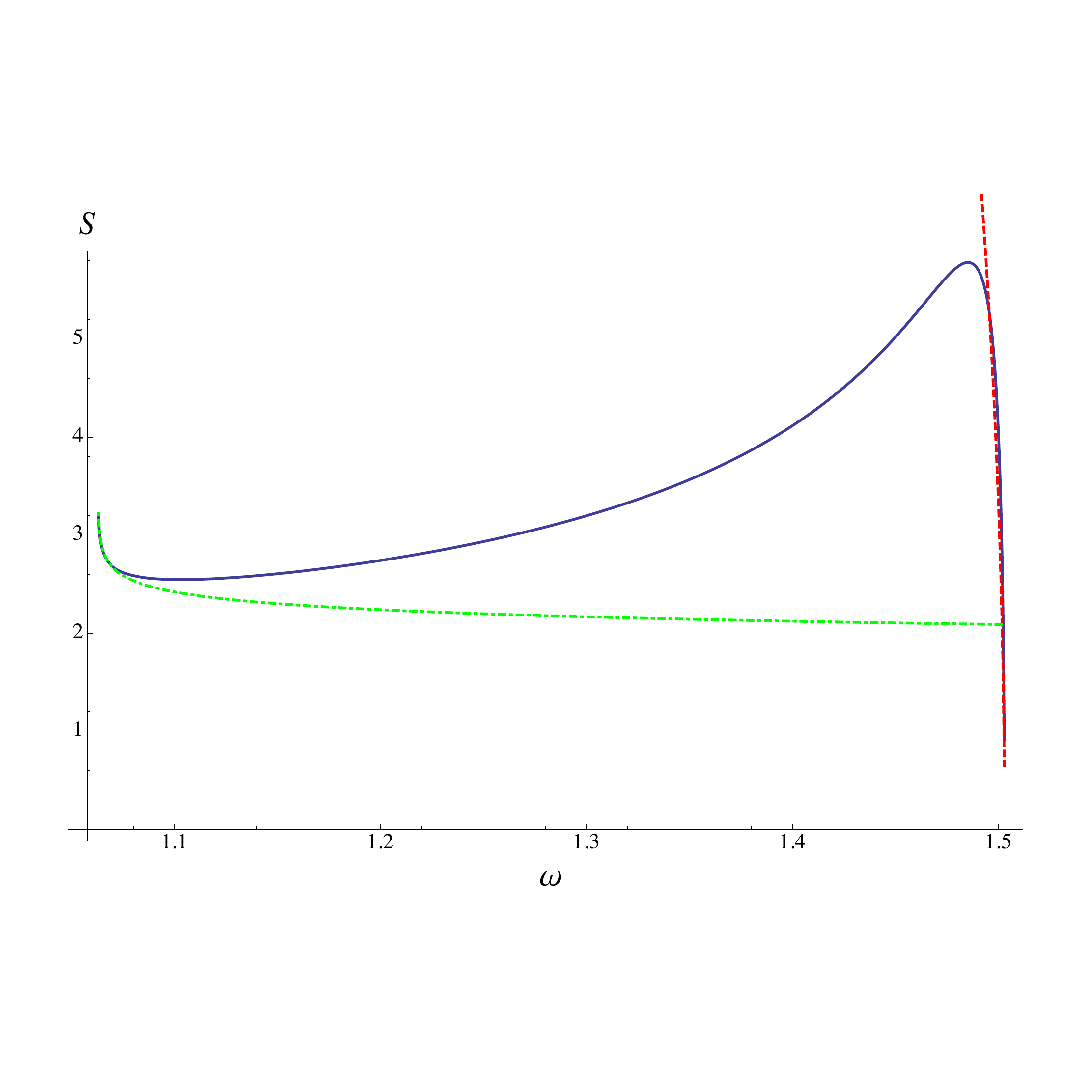}}
   \label{d01k25}
 }
 \subfigure[$k/2\pi=0.375$]{
%  \resizebox{6cm}{!}{\includegraphics{threshold_d02_k0375}}
%   \label{d2k375}
  \resizebox{7cm}{!}{\includegraphics{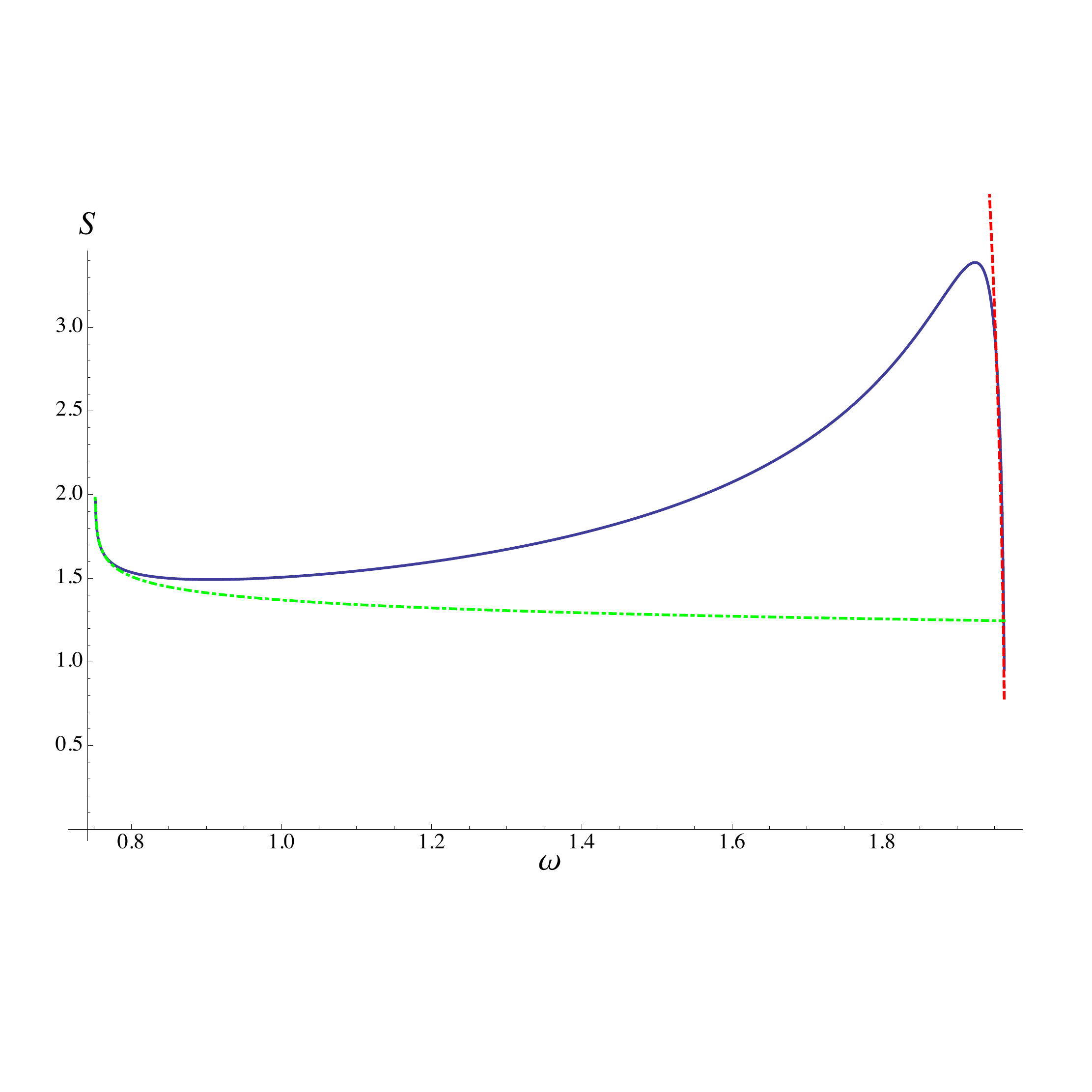}}
   \label{d01k375}
 }
 \subfigure[$k/2\pi=0.5$]{
%  \resizebox{6cm}{!}{\includegraphics{threshold_d02_k0500}}
%   \label{d2k5}
  \resizebox{7cm}{!}{\includegraphics{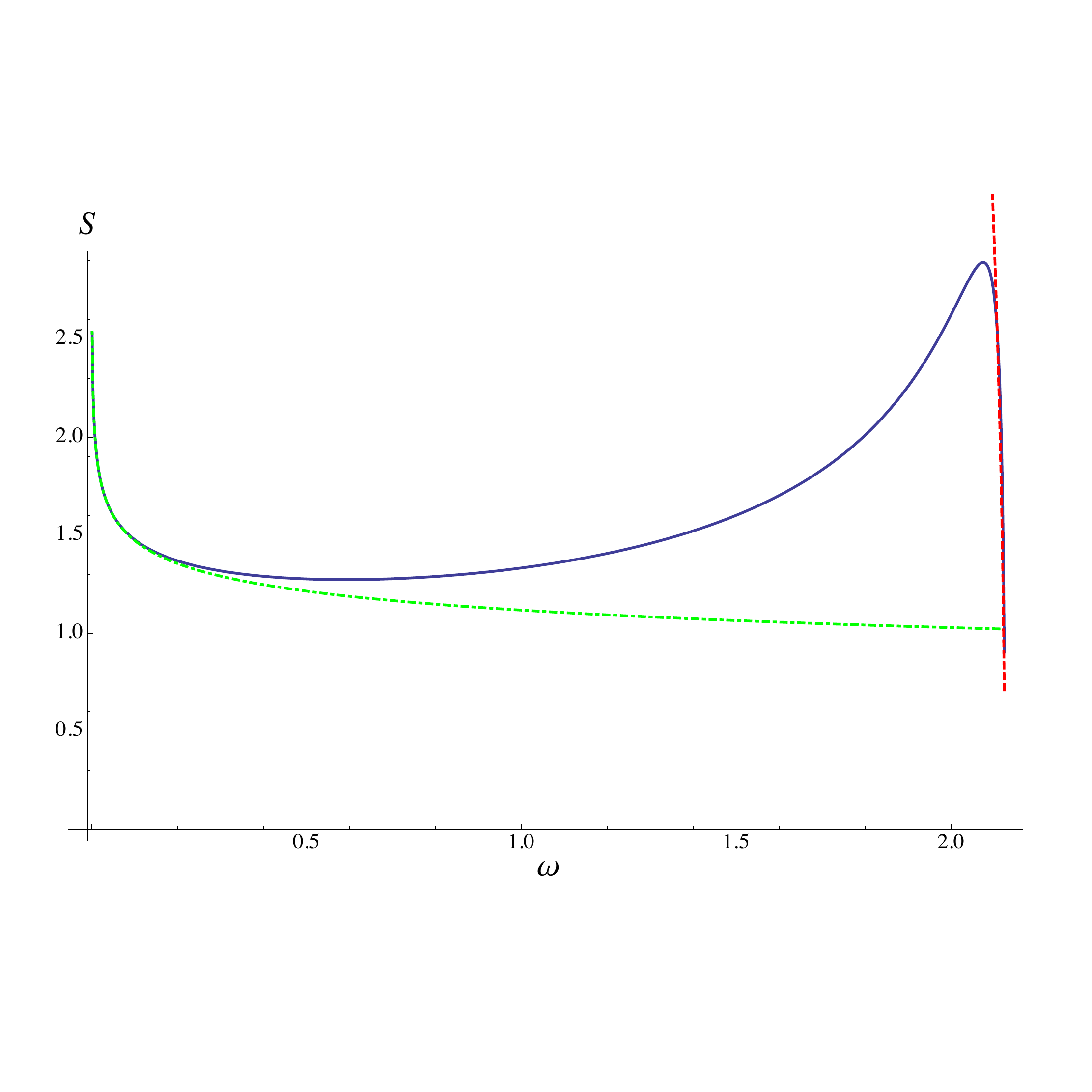}}
   \label{d01k5}
 }
\caption{Threshold behaviour for $\Delta=0.1$, the solid line is the numerical evaluation of $S^{zz}_2$, while the dashed and dotted lines indicate the upper and lower threshold behaviour respectively, fitted using the second method (see main text).}
 \label{fig:d01}
\end{figure}

\begin{figure}[h!]
\centering
\subfigure[$k/2\pi=0.125$]{
%  \resizebox{6cm}{!}{\includegraphics{threshold_d05_k0125}}
%   \label{d5k125}
  \resizebox{7cm}{!}{\includegraphics{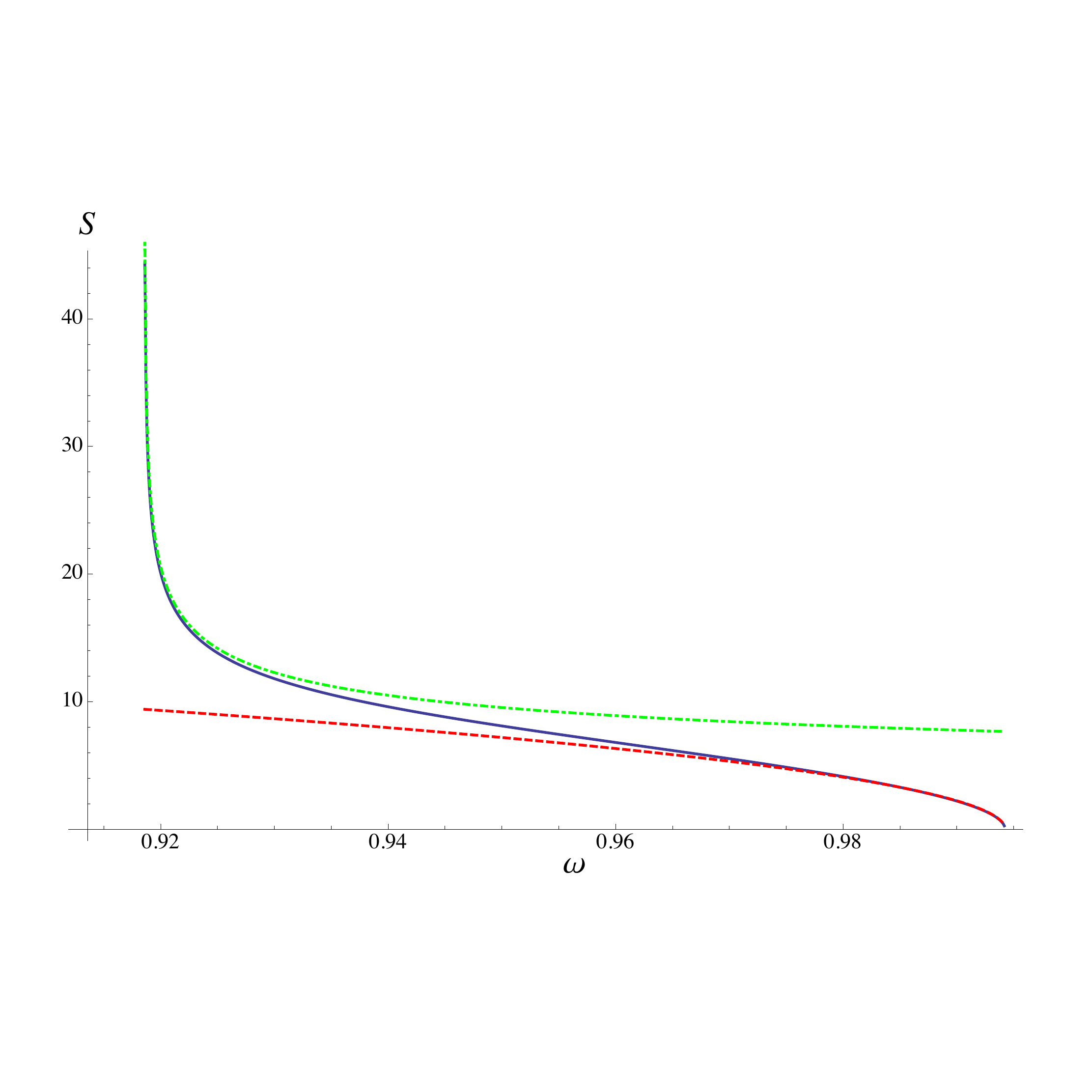}}
   \label{d05k125}
 }
 \subfigure[$k/2\pi=0.25$]{
%  \resizebox{6cm}{!}{\includegraphics{threshold_d05_k0250}}
%   \label{d5k25}
  \resizebox{7cm}{!}{\includegraphics{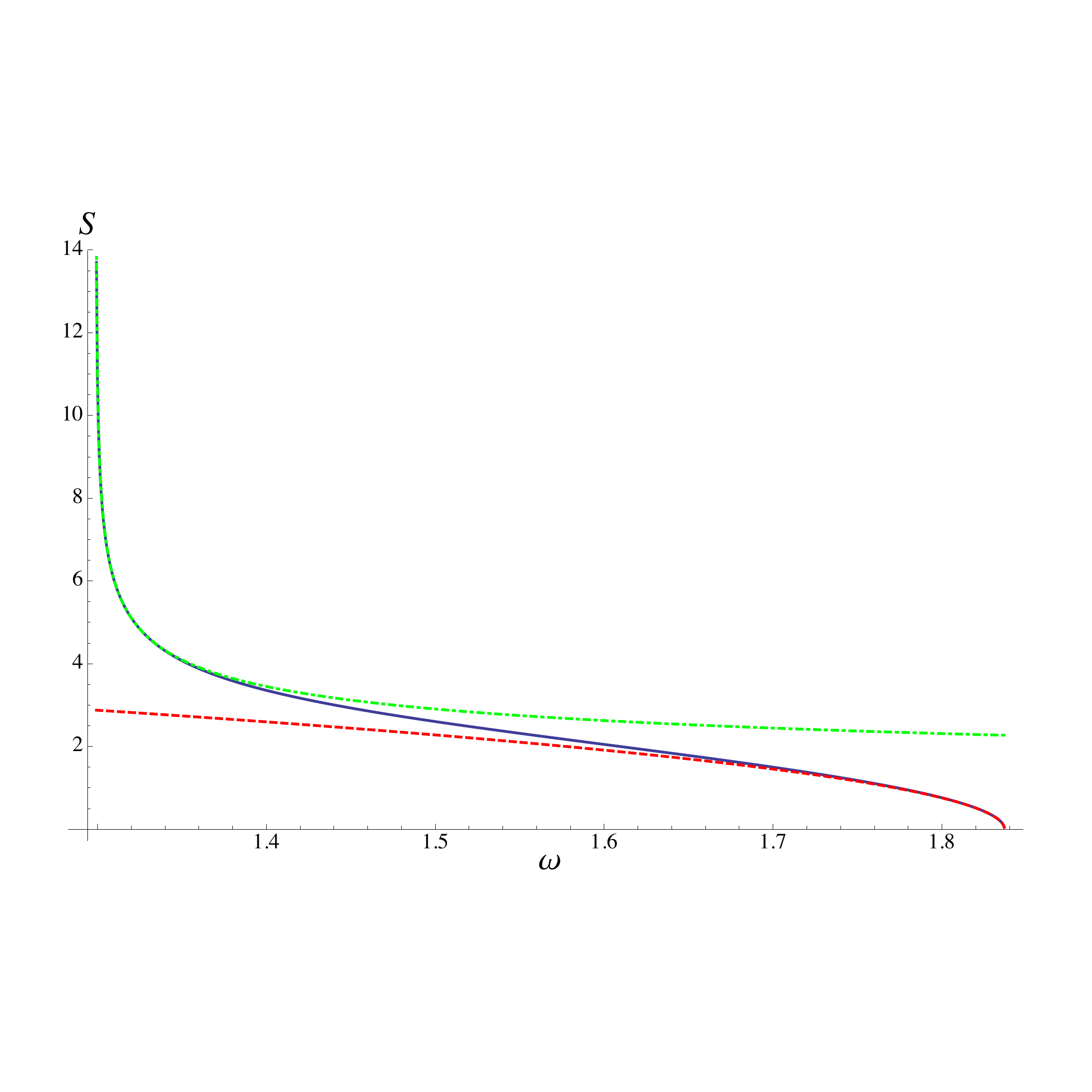}}
   \label{d05k25}
 }
 \subfigure[$k/2\pi=0.375$]{
%  \resizebox{6cm}{!}{\includegraphics{threshold_d05_k0375}}
%   \label{d5k375}
  \resizebox{7cm}{!}{\includegraphics{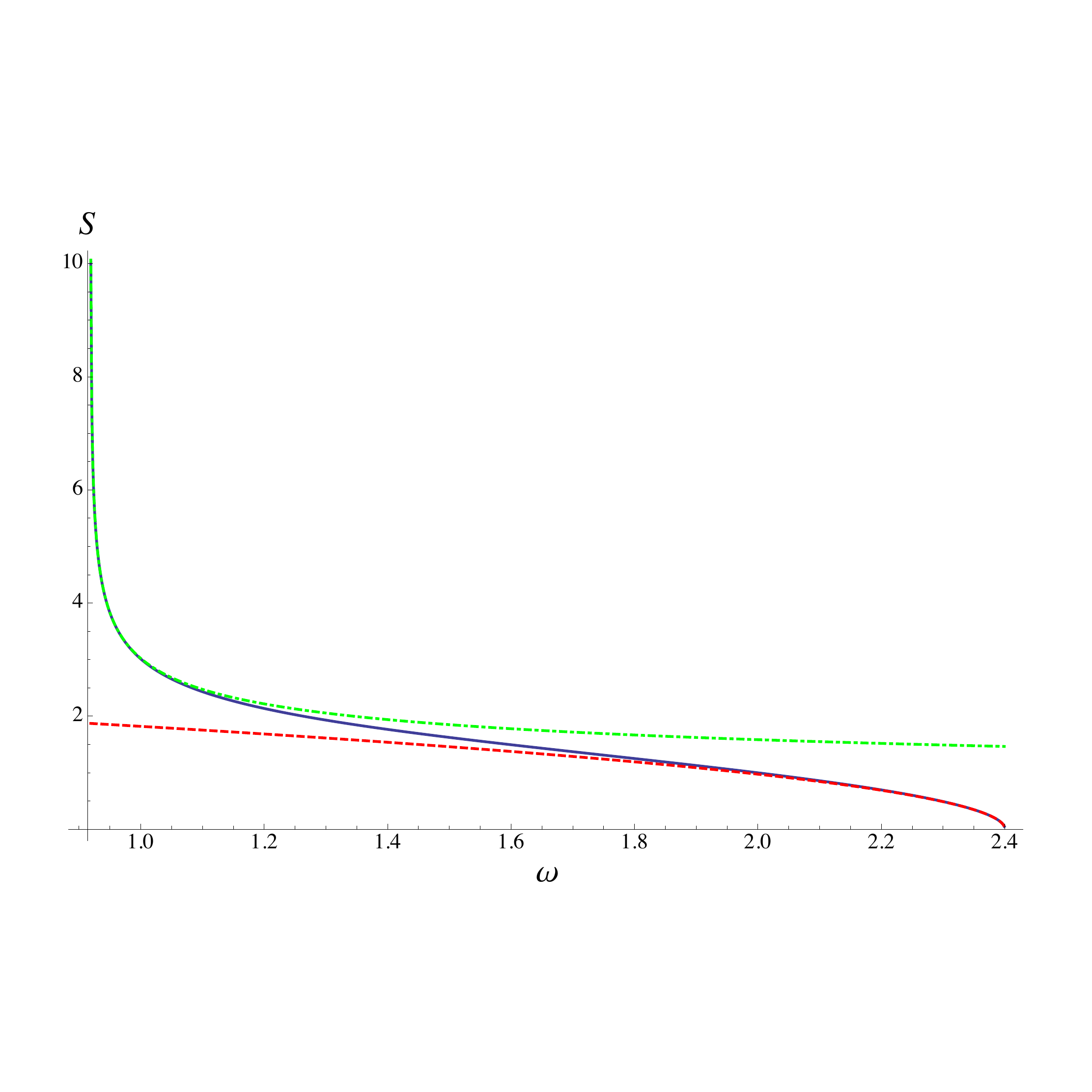}}
   \label{d05k375}
 }
 \subfigure[$k/2\pi=0.5$]{
%  \resizebox{6cm}{!}{\includegraphics{threshold_d05_k0500}}
%   \label{d5k5}
  \resizebox{7cm}{!}{\includegraphics{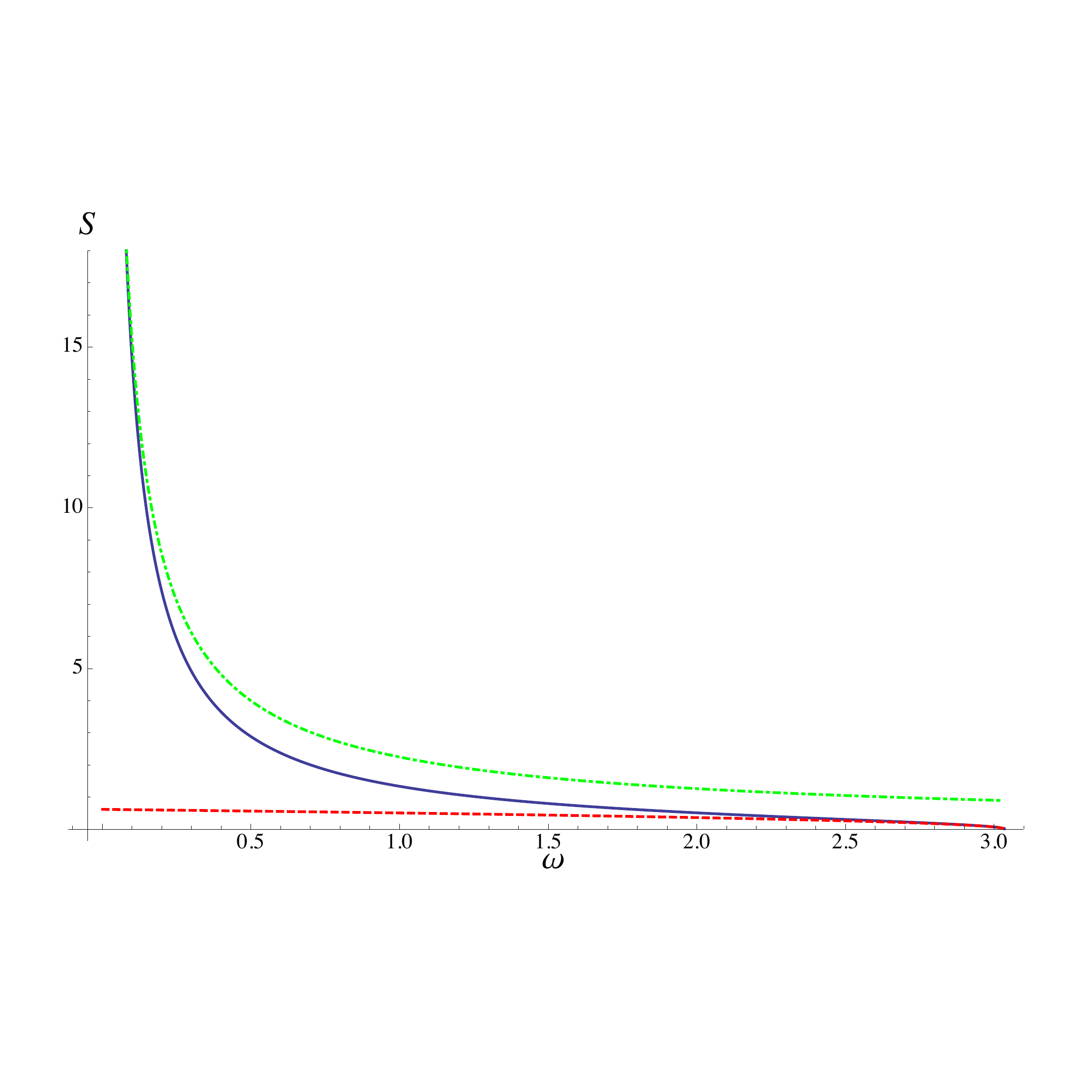}}
   \label{d05k5}
 }
\caption{Threshold behaviour for $\Delta=0.5$, the solid line is the numerical evaluation of $S^{zz}_2$, while the dashed and dotted lines indicate the upper and lower threshold behaviour respectively, fitted using the second method (see main text).}
\label{fig:d05}
\end{figure}

\begin{figure}[h!]
\centering
\subfigure[$k/2\pi=0.125$]{
%  \resizebox{6cm}{!}{\includegraphics{threshold_d08_k0125}}
%   \label{d8k125}
  \resizebox{7cm}{!}{\includegraphics{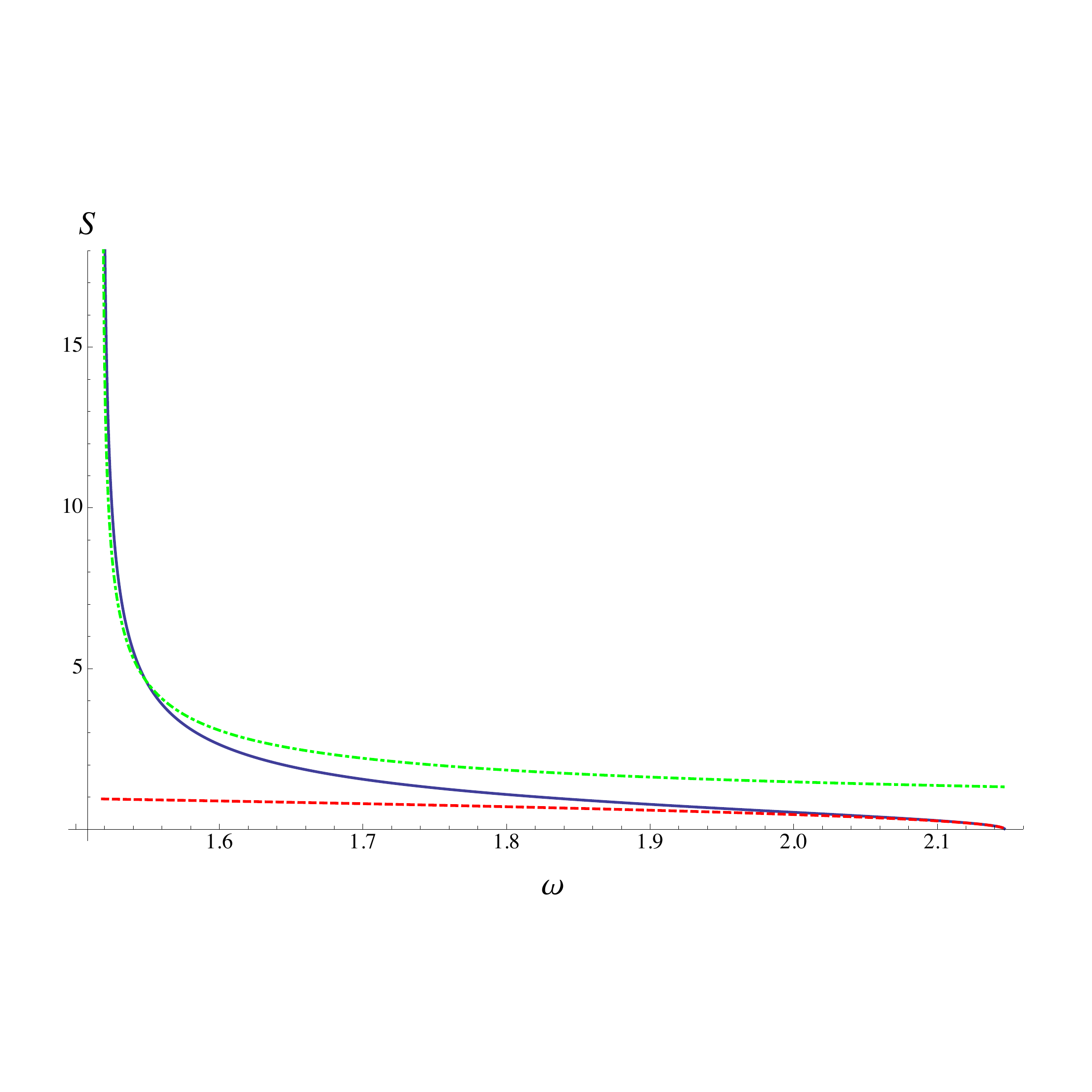}}
   \label{d09k125}
 }
% \subfigure[$k/2\pi=0.125$ { \it zoomed in on lower threshold}]{
%  \resizebox{6cm}{!}{\includegraphics{threshold_d08_k0125_lower}}
%   \label{d8k125low}
% }
 \subfigure[$k/2\pi=0.25$]{
%  \resizebox{6cm}{!}{\includegraphics{threshold_d08_k0250}}
%   \label{d8k25}
  \resizebox{7cm}{!}{\includegraphics{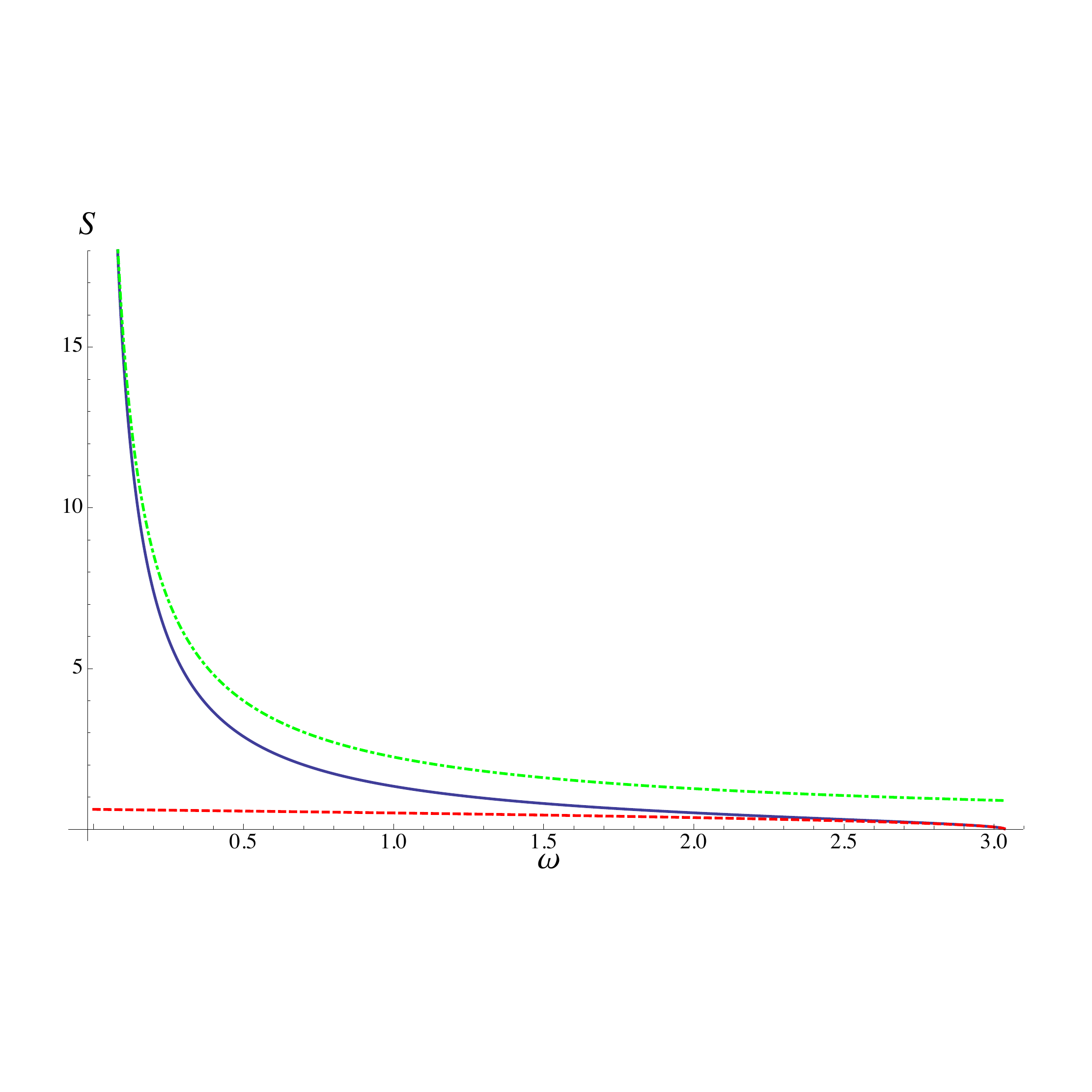}}
   \label{d09k25}
 }
%  \subfigure[$k/2\pi=0.125$ {\it zoomed in on lower threshold}]{
%  \resizebox{6cm}{!}{\includegraphics{threshold_d08_k0250_lower}}
%   \label{d8k25low}
% }
 \subfigure[$k/2\pi=0.375$]{
%  \resizebox{8cm}{!}{\includegraphics{threshold_d08_k0375}}
%   \label{d8k375}
  \resizebox{7cm}{!}{\includegraphics{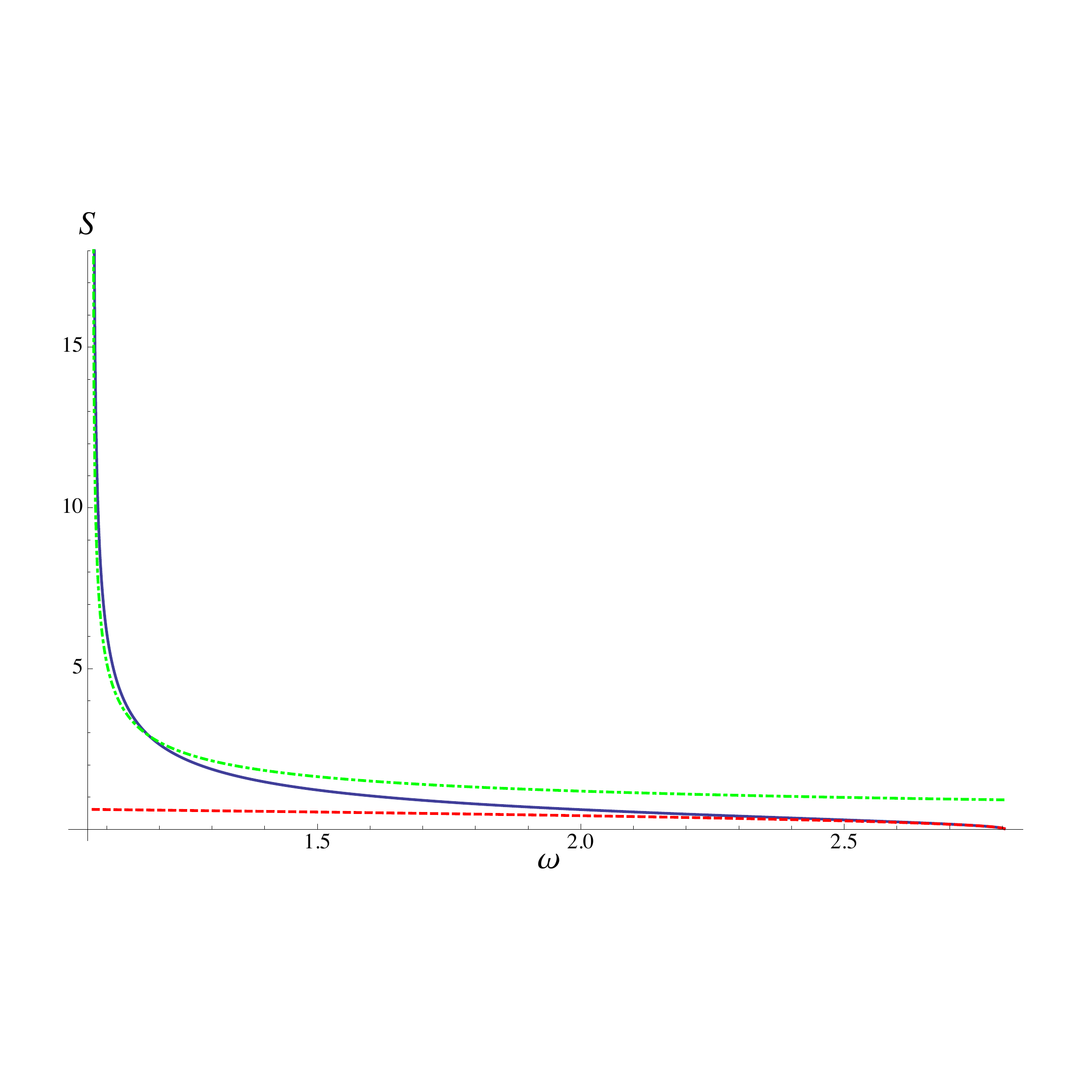}}
   \label{d09k375}
 }
%  \subfigure[$k/2\pi=0.125$ {\it zoomed in on lower threshold}]{
%  \resizebox{6cm}{!}{\includegraphics{threshold_d08_k0375_lower}}
%   \label{d8k375low}
% }
 \subfigure[$k/2\pi=0.5$]{
%  \resizebox{6cm}{!}{\includegraphics{threshold_d08_k0500}}
%   \label{d8k5}
  \resizebox{7cm}{!}{\includegraphics{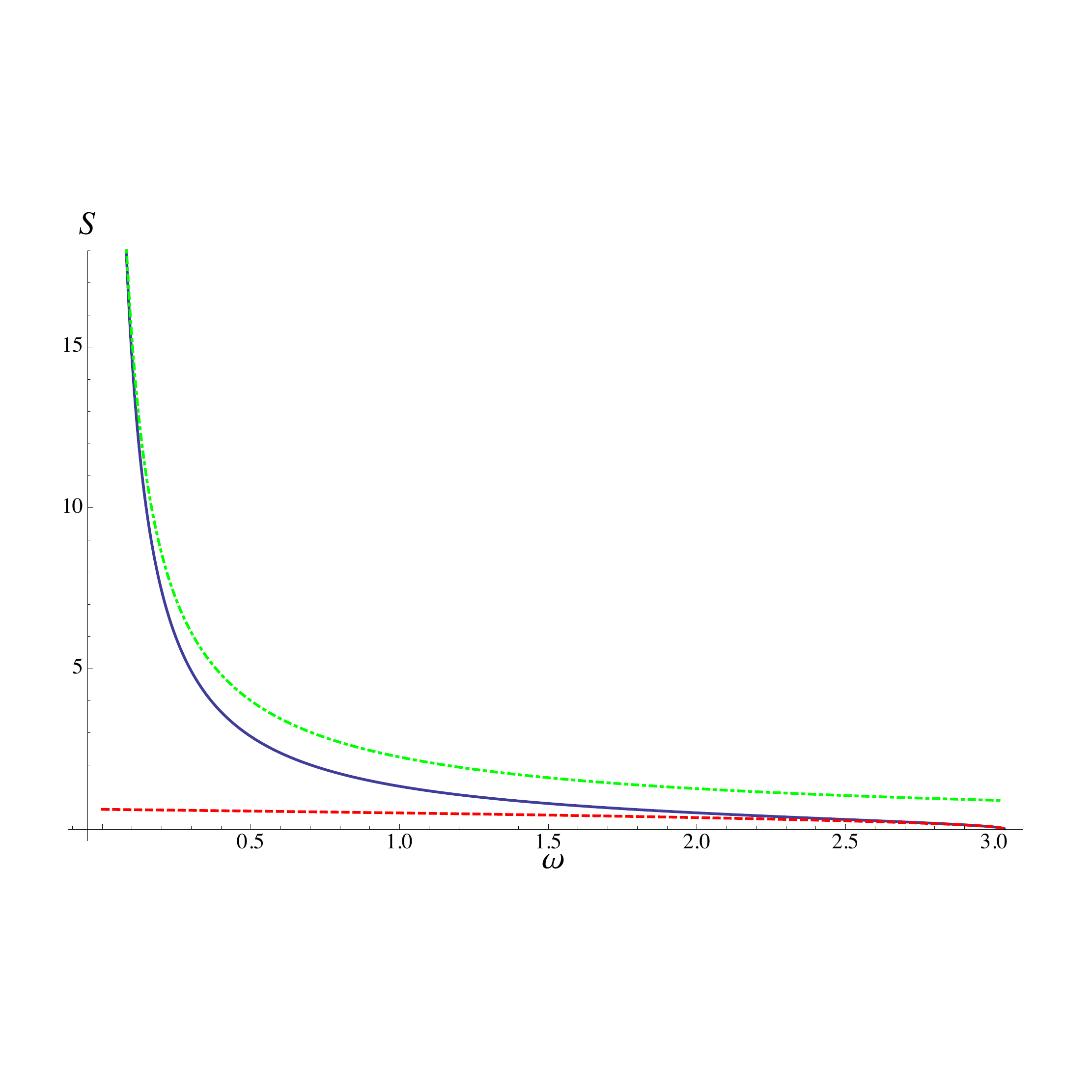}}
   \label{d09k5}
 }
\caption{Threshold behaviour for $\Delta=0.9$, the solid line is the numerical evaluation of $S^{zz}_2$, while the dashed and dotted lines indicate the upper and lower threshold behaviour respectively, fitted using the second method (see main text).}
\label{fig:d09}
\end{figure}

It is immediately obvious that the region of validity of the lower and upper threshold behaviours depends strongly on the anisotropy. For low anisotropy, the upper threshold is not fitted well, the lower one being better described. For intermediate anisotropy the fitting is very reasonable, and covers a substantial range (over half) of the continuum. At high anisotropy, the lower threshold is not well fitted, whereas the upper one is rather well approximated. One point to notice is that the range of correspondence between the lower threshold behaviour and the exact structure factor for momenta at or near $\pi$ is very narrow: in this case, subleading terms correcting the theshold behaviour should not be neglected (the very low percentages presented in some entries in the tables should however be considered as indicative only, in view of numerical difficulties in evaluating the structure factor in the immediate vicinity of thresholds). On the other hand, the less sensitive second way of fitting gives acceptable fits over a wider range of frequencies.
Plots for the structure factor accompanied by the threshold fits using this second way are given for these three values of anisotropy and momentum values in Figures \ref{fig:d01}-\ref{fig:d09}. 
%In general, the region of validity of the lower threshold behaviour is very limited for both high and low anisotropy values. The upper threshold behaviour is somewhat better behaved. At intermediate anisotropies the LSF is in fact relatively well approximated by its threshold behaviour throughout the two-spinon continuum. 
\begin{table}[here!]
\begin{center}
\begin{tabular}{|c|c|c|c|c|c|c|c|}
\hline 
&&& Lower &Lower&Upper&Upper \\
$k/(2 \pi)$ &$w_l$& $w_u$ & $<1\%$ & $<10\%$& $<1\%$ & $<10\%$ \\ \hline \hline
$0.125$ &$0.751$ & $0.813$&$\sim 1\% $&$\sim 12\%$&$< 0.1\%$&$\sim 0.6\%$ \\ \hline
 $0.25$ &$1.06$ &$1.50$ &$\sim 1.2\%$&$\sim 14\%$&$< 0.1\%$&$\sim 0.5\%$ \\ \hline
 $0.375$  &$0.751$ &$1.96$ &$\sim 1.8\%$&$\sim 20\%$&$< 0.1\%$&$\sim 0.4\%$ \\ \hline
  $0.5$  & $0$&$2.13$ & $< 0.1\%$&$\sim 0.9\%$&$< 0.1\%$&$\sim 0.3\%$\\ \hline
\end{tabular}
\caption{Approximate validity of threshold results for anisotropy $\Delta=0.1$. These results are based on the first way of fitting discussed in the text. The low percentages are very approximate.}
\label{tab:d1}
\end{center}
\end{table}
\begin{table}[here!]
\begin{center}
\begin{tabular}{|c|c|c|c|c|c|c|c|}
\hline 
&&& Lower &Lower&Upper&Upper \\
$k/(2 \pi)$ &$w_l$& $w_u$ & $<1\%$ & $<10\%$& $<1\%$ & $<10\%$ \\ \hline \hline
$0.125$ &$0.919$ & $0.994$&$\sim 16\% $&$\sim 38\%$&$\sim 7\%$&$\sim 47\%$ \\ \hline
 $0.25$ &$1.30$ &$1.84$ &$\sim 16\%$&$\sim 38\%$&$\sim 9\%$&$\sim 52\%$ \\ \hline
 $0.375$  &$0.919$ &$2.40$ &$\sim 11\%$&$\sim 35\%$&$\sim 13\%$&$\sim 57\%$ \\ \hline
  $0.5$  & $0$&$2.60$ & $< 0.1\%$&$\sim 0.1\%$&$\sim 14\%$&$\sim 50\%$\\ \hline
\end{tabular}
\caption{Approximate validity of threshold results for anisotropy $\Delta=0.5$. These results are based on the first way of fitting discussed in the text. The low percentages are very approximate.}
\label{tab:d5}
\end{center}
\end{table}
\begin{table}[here!]
\begin{center}
\begin{tabular}{|c|c|c|c|c|c|c|c|}
\hline 
&&& Lower &Lower&Upper&Upper \\
$k/(2 \pi)$ &$w_l$& $w_u$ & $<1\%$ & $<10\%$& $<1\%$ & $<10\%$ \\ \hline \hline
$0.125$ &$1.07$ & $1.16$&$<0.1\% $&$\sim 0.3\%$&$\sim 1.8\%$&$\sim 16\%$ \\ \hline
 $0.25$ &$1.52$ &$2.15$ &$< 0.1\%$&$\sim 0.3\%$&$\sim 1.8\%$&$\sim 16\%$ \\ \hline
 $0.375$  &$1.07$ &$2.81$ &$< 0.1\%$&$\sim 0.25\%$&$\sim 1.7\%$&$\sim 14\%$ \\ \hline
  $0.5$  & $0$&$3.04$ & $< 0.1\%$&$< 0.1\%$&$\sim 1.2\%$&$\sim 11\%$\\ \hline
\end{tabular}
\caption{Approximate validity of threshold results for anisotropy $\Delta=0.9$. These results are based on the first way of fitting discussed in the text. The low percentages are very approximate.}
\label{tab:d9}
\end{center}
\end{table}
%Several other plots for different values of $\Delta$ and $k$ are shown in Figures (\ref{fig:d1k125}) - (\ref{fig:d2k125})
%\begin{figure}[h]
%\begin{center}
%\resizebox{8cm}{!}{\includegraphics{./markfigures/threshold_calculations_D_01_k_0125}}
%\caption{ $\Delta=0.1, k=0.125\times 2 \pi$,{\it The solid line is the numerical evaluation of $S_{zz}^2$, while the dashed and dotted lines indicate the upper and lower threshold behaviour respectively}.
%\label{fig:d1k125}}
%\end{center}
%\end{figure}
%\begin{figure}[p]
%\begin{center}
%\resizebox{8cm}{!}{\includegraphics{./markfigures/threshold_calculations_D_02_k_0125}}
%\caption{ $\Delta=0.2, k=0.125\times 2 \pi$,{\it The solid line is the numerical evaluation of $S_{zz}^2$, while the dashed and dotted lines indicate the upper and lower threshold behaviour respectively}.
%\label{fig:d2k125}}
%\end{center}
%\end{figure}
%\begin{figure}[p]
%\begin{center}
%\resizebox{8cm}{!}{\includegraphics{./markfigures/threshold_calculations_D_02_k_0375}}
%\caption{ $\Delta=0.2, k=0.375\times 2 \pi$,{\it The solid line is the numerical evaluation of $S_{zz}^2$, while the dashed and dotted lines indicate the upper and lower threshold behaviour respectively}.
%\label{fig:d2k375}}
%\end{center}
%\end{figure}
%
%
%
% % Results
\section{Conclusion}
\setcounter{equation}{0} 

In summary, we have presented an analytical expression for the two-spinon contribution
to the longitudinal structure factor of the $XXZ$ chain in the gapless antiferromagnetic
regime $0 \leq \Delta \leq 1$ for zero field in the infinite size limit at zero temperature.
Our results extend to this region previous results for the isotropic or gapped
antiferromagnet. 

The question of the transverse structure factor remains difficult for the methods 
presented here. In the basis we are using, all transverse spin operator form factors
vanish upon taking the gapless limit, and this points to the need for a resummation 
over states including macroscopic numbers of spinons, something which goes beyond 
current capabilities. Extending these results to the case of a finite magnetic field 
faces similar issues; the restriction to zero temperature is even more severe. 
We leave these questions open for the moment.

Another important (but now feasible) extension to our work would be to consider 
higher-spinon contributions
to the longitudinal structure factor. This was actually performed for the isotropic $XXX$ 
antiferromagnet in the recent past \cite{2006_Caux_JSTAT_P12013}; one can expect
that such a calculation would yield the longitudinal structure factor to around 1\% accuracy
for any value of anisotropy in the gapless antiferromagnetic regime in zero field.
It would also allow us to further refine the determination of the threshold
behaviour and of its limit of applicability. 
We will investigate these and other issues in the future.

 % Discussion

\section*{Acknowledgements}
J.-S. C. acknowledges support from the Foundation for Fundamental Research on Matter (FOM), 
which is part of the Netherlands Organisation for Scientific Research (NWO).
H.K is supported by the Grant-in -Aid for Scientific Research (C) 22540022 JSPS, Japan.
M. S. gratefully acknowledges support from the Australian Research Council (ARC).
R. W. would like to thank Nikolai Kitanine for useful discussions.

\newpage
\begin{appendix}
\section{Elliptic Functions}\label{app:ellipticfn}
In this appendix, we collect together the notational conventions and identities for the various elliptic functions that we use in the paper. Many further properties of elliptic functions can be found for example in \cite{MR1773820}. 
We firstly make use of the Jacobi elliptic functions defined by
\bea \sn(u)=\sin(\am(u)),\quad \cn(u)=\cos(\am(u)),\quad \dn(u)=\sqrt{1-k^2\sin^2(\am(u))},\label{eqn:sncn}\eea
where $\am(u)$ is the amplitude function defined in terms of the elliptic modulus $k$ by
\bea u=\int_0^{\am(u)} \frac{dx}{\sqrt{1-k^2 \sin^2(x)}}.\label{eqn:am}\eea
Defining the conjugate modulus by $k'=\sqrt{1-k^2}$, the complete elliptic integrals $K$, and $K'$ are defined in terms of the same modulus $k$ by the integrals:
\bea K=\int_0^{\frac{\pi}{2}} \frac{dx}{\sqrt{1-k^2 \sin^2(x)}},\quad K'=\int_0^{\frac{\pi}{2}} \frac{dx}{\sqrt{1-k'^2 \sin^2(x)}}.\eea
We sometimes make the $k$ dependence of these various functions explicit by writing them as $\am(u,k)$, $\sn(u,k)$, $\cn(u,k)$ and $\dn(u,k)$.
We also define the following functions 
\bea \hbox{snh}(u)=-i\,\hbox{sn}(iu),\quad\hbox{cnh}(u)=\hbox{cn}(iu),\quad\hbox{dnh}(u)=\hbox{dn}(iu).\label{snhcnh}\eea

The other type of elliptic function we use are theta functions. These are defined  in terms of a parameter $q$ called the elliptic nome by
\begin{eqnarray*}
\vartheta_{1}(u,q )&=&2q^{1/4} (q^2;q^2)_{\infty} \sin(\pi u)
\prod_{n=1}^\infty (1-2q^{2n}\cos(2\pi u) +q^{4n}) \\[-1mm]
\vartheta_{2}(u,q )&=&2q^{1/4} (q^2;q^2)_{\infty} \cos(\pi u)
\prod_{n=1}^\infty (1+2q^{2n}\cos(2\pi u) +q^{4n}) \\[-1mm]
\vartheta_{3}(u,q )&=& (q^2;q^2)_{\infty} 
\prod_{n=1}^\infty (1+2q^{2n-1}\cos(2\pi u) +q^{4n-2}) \\[-1mm]
\vartheta_{4}(u,q )&=&(q^2;q^2)_{\infty} 
\prod_{n=1}^\infty (1-2q^{2n-1}\cos(2\pi u) +q^{4n-2}) 
\end{eqnarray*}
If we identify the elliptic nome as $q=e^{-\pi \frac{K'}{K}}$, then the Jacobi elliptic functions and theta functions are related by 
\ben 
\hbox{sn}(u)= \frac{1}{\sqrt{k}} \frac{\vartheta_{1}\left(\frac{u}{2K},q \right)}{\vartheta_{4}\left(\frac{u}{2K},q \right)},\quad
\hbox{cn}(u)= \frac{\sqrt{k'}}{\sqrt{k}} \frac{\vartheta_{2}\left(\frac{u}{2K},q \right)}{\vartheta_{4}\left(\frac{u}{2K},q \right)},\quad
\hbox{dn}(u)= \sqrt{k'} \frac{\vartheta_{3}\left(\frac{u}{2K},q \right)}{\vartheta_{4}\left(\frac{u}{2K},q \right)}.
\een

\nin In this paper, we use various identities that we now list.\vspace*{5mm}

\nin{\bf Identity 1} -  the conjugate modulus transformation for Jacobi elliptic functions \cite{MR1773820}:
\bea \sn(iu,k)=i \frac{\sn(u,k')}{\cn(u,k')},\quad 
\cn(iu,k)=i \frac{1}{\cn(u,k')},\quad 
\dn(iu,k)=i \frac{\dn(u,k')}{\cn(u,k')}.\lb{cmtrans}\eea

\nin{\bf Identity 2} - the half-period property of theta functions \cite{MR1773820}:
\bea &&\vartheta_a(u\pm \frac{\tau}{2},q)=(\pm i)^{g_a} e^{-i\pi \tau/4} e^{\mp i \pi u} \vartheta_{\bar{a}}(u,q),\ws\nn\\
&&\hb{where}\ws q=e^{i\pi \tau}\ws \hbox{and} \ws
\bar{1}:=4,\, \bar{2}:=3, \,\bar{3}:=2, \,\bar{4}:=1,\quad g_1=g_4=1, \,g_2=g_3=0.\lb{halfperiod}\eea

\nin{\bf Identity 3} - \cite{MR1424469}:  %{\bf[ Does this follow from a Landen transformation Hitoshi? Should we state it in the general form depending on $q$ and include a proof?]}
\bea
&&i\frac{\vartheta_1\left(\frac{1}{4}-\frac{i\gt}{\pi}, 
p^{\frac{r}{4}}\right)}{\vartheta_1\left(\frac{1}{4}
+\frac{i\gt}{\pi},p^{\frac{r}{4}}\right)}=\sn\left(\frac{2I'\gt}{\pi},k_I\right)+i\cn\left(\frac{2I'\gt}{\pi},k_I\right).\label{hitid}\eea

\nin{\bf Identity 4} - the limiting behaviour of elliptic functions (which follows straight from the above definitions):
\bea &&\am(u,k=1)=2\arctan(e^u)-\frac{\pi}{2}.\label{eqn:am0limit}\\
&&\am(u,k=0)=u,\ws \sn(u,k=0)=\sin(u),\ws \cn(u,k=0)=\cos(u),\ws \dn(u,k=0)=1.\label{eqn:jac1limit}
\eea

\section{Derivation of \eqref{eqn:fftrans} in the Vertex Operator Picture}\lb{VOA}

A key observation is an identification of the type I vertex operators 
with the half-transfer matrices on the lattice. 
Then applying the gauge transformations \eqref{URU} 
to each $R$ matrix constituting the half-transfer matrix, one can reach the following definition of 
 the type I vertex operators in the disordered regime. 
\begin{eqnarray}
&&\tPhi^{(j)}_{\vep;\ell}(u):=%(-i)^{\delta_{j+\ell,0}}
\sum_{\vep'=\pm} (U_{j+\ell})_{\vep\vep'}\G_{j+\ell+1} \Phi^{(\ell,1-\ell)}_{\vep'}(u)\G_{j+\ell}^{-1}\ :\ 
%&&\hspace{-2cm}\tPhi^{'(j)}_\vep(u):=\sum_{\vep'} (U_1)_{\vep\vep'}
%\G_0 \Phi^{(j,1-j)}_{\vep'}(u)\G_1^{-1}\ : \G_1\H^{(1-j)}\to \G_0\H^{(j)}. 
\G_{j+\ell}\H^{(1-\ell)}\to 
\G_{j+\ell+1}\H^{(\ell)}.
%\nn\\&&
\lb{tPhi}
\ena
Note that from \eqref{Hdis} these are linear operators on $\H^{(j)}_{dis}$. 
Accordingly, one can realise a local operator as an operator on 
$\H^{(j)}_{dis}=\G_{j+\ell}\H^{(1-\ell)}$ by 
\be
&&\cO'(E_{\vep\vep'})^{(j)}_\ell:=\tPhi^{(j)}_{-\vep;1-\ell}(u-1)\tPhi^{(j)}_{\vep';\ell}(u)\Bigl|_{u=0}.
\en
Then it follows from \eqref{Eee} and \eqref{tPhi} that we have the gauge 
transformation of the spin operators as 
\bea
&& \cO'(\sigma^x)^{(j)}_\ell=(-)^{j+\ell}
\G_{j+\ell}\cO(\sigma^z)^{(1-\ell)}\G_{j+\ell}^{-1},\qquad 
\cO'(\sigma^y)^{(j)}_\ell=
\G_{j+\ell}\cO(\sigma^x)^{(1-\ell)}\G^{-1}_{j+\ell},\nn\\ 
&&\cO'(\sigma^z)^{(j)}=
(-)^{j+\ell}\G_{j+\ell}\cO(\sigma^y)^{(1-\ell)}\G_{j+\ell}^{-1}.\lb{OpO}
\ena
This is consistent to the transformation of $\sigma^\alpha$ 
as a $2\times 2$ matrix: $\Ad U_j^{-1}: \sigma^x, \sigma^y, \sigma^z \mapsto 
(-1)^j\sigma^z, \sigma^x,  (-1)^j\sigma^y$.

In the vertex operator picture, the gauge transformation of the vacuum vectors $\ket{{\rm vac};pr}^{(j)}=x^{2H^{(j)}}/(Z^{(j)})^{1/2}\in \F^{(j)}\cong \End(\H^{(j)})$ follows from  the fact that  $x^{2H^{(j)}}$ originates as the product of two corner transfer matrices. 
Hence the vacuum vectors as well as other physical $2n$-spinon excited states should have the same transformation property 
as the corner transfer matrix. 
Therefore we identify new vacuum vectors as 
\bea
\vac^{(j)}=\G_{j+\ell} \frac{1}{(Z^{(1-\ell)})^{1/2}} x^{2H^{(1-\ell)}} \G_{j+\ell}^{-1}\quad (\ell=0,1)
\lb{disVac}
\ena
in  $\F^{(j)}_{dis}\cong \End(\H^{(j)}_{dis})$.  
Similarly to the type I case, the type II vertex operators $\Psi^{*(1-j,j)}_\vep(i\gt/\pi)$ in the principal regime are mapped to linear operators on $\H^{(j)}_{dis}$ 
by
\be
\widetilde{\Psi}^{*(j)}_{\vep;\ell}(\gt)=\G_{j+\ell+1}\Psi^{*(\ell,1-\ell)}_\vep(i\gt/\pi)\G_{j+\ell}^{-1} 
\ :\ \G_{j+\ell}\H^{(1-\ell)}\to \G_{j+\ell+1}\H^{(\ell)}.
\en
One should note that the new type II vertex operators $\widetilde{\Psi}^{*(j)}_{\vep;\ell}(\gt)$ commute with the new type I vertex operators $\tPhi^{(j)}_{\vep;\ell}(u)$ 
in pairs. Namely, 
\be
\tPhi^{(j)}_{\vep;\ell}(u){\tPsi}^{*(j)}_{\mu_2;1-\ell}(\gt_2){\tPsi}^{*(j)}_{\mu_1;\ell}(\gt_1)
=\tau(u-i\gt_1/\pi)\tau(u-i\gt_2/\pi){\tPsi}^{*(j)}_{\mu_2;\ell}(\gt_2){\tPsi}^{*(j)}_{\mu_1;1-\ell}(\gt_1)\tPhi^{(j)}_{\vep;\ell}(u).
\en
Hence we obtain the following identification of the disordered $2n$-spinon states 
\bea 
&&\ket{\gt_1,\gt_2\cdots \gt_{2n}}^{(j)}_{\ep_1,\ep_2,\cdots,\ep_{2n}}\nn\\
&&= \widetilde{\Psi}^{*(j)}_{\ep_{2n};1-\ell}(\gt_{2n}) \cdots  \widetilde{\Psi}^{*(j)}_{\ep_2;1-\ell}(\gt_2) \widetilde{\Psi}^{*(j)}_{\ep_1;\ell}(\gt_1)\vac^{(j)}\nn\\
&&=\G_{j+\ell}\Psi^{*(1-\ell,\ell)}_{\vep_{2n}}(i\gt_{2n}/\pi)\cdots \Psi^{*(1-\ell,\ell)}_{\vep_{2}}(i\gt_{2}/\pi)\Psi^{*(\ell,1-\ell)}_{\vep_{1}}(i\gt_{1}/\pi)\frac{1}{(Z^{(1-\ell)})^{1/2}} x^{2H^{(1-\ell)}}\G_{j+\ell}^{-1}.
\lb{disPsi}
\ena
Combining \eqref{OpO}, \eqref{disVac} and \eqref{disPsi}, we then obtain the form factor of the spin operator $\gs^\alpha$ in the disordered regime as 
\bea
&&{}^{(j)}\bra{{\rm vac}}\sigma^{\alpha}
\ket{\gt_1,\cdots,\gt_{2n}}^{(j)}_{\vep_1,\cdots,\vep_{2n}}\nn\\
&&=\frac{1}{Z^{(1-\ell)}}{\tr_{{\cal H}^{(j)}_{dis}}(
\G_{j+\ell}x^{2 H^{(1-\ell)}}\G_{j+\ell}^{-1} \cO'(\sigma^{\alpha})^{(j)}
\widetilde{\Psi}^{*(j)}_{\vep_{2n};1-\ell}(\gt_{2n})\widetilde{\Psi}^{*(j)}_{\vep_{2n-1};\ell}(\gt_{2n-1})\cdots
\widetilde{\Psi}^{*(j)}_{\vep_{1};\ell}(\gt_{1})\G_{j+\ell}x^{2 H^{(1-\ell)}}\G_{j+\ell}^{-1})}\nn\\
&&=\frac{1}{Z^{(1-\ell)}}{\tr_{{\cal H}^{(1-\ell)}}(x^{4 H^{(1-\ell)}} \cO(\Ad U_{j+\ell}^{-1}(\sigma^{\alpha}))^{(1-\ell)}
\Psi^{*(1-\ell,\ell)}_{\vep_{2n}}(i\gt_{2n}/\pi)\Psi^{*(\ell,1-\ell)}_{\vep_{2n-1}}(i\gt_{2n-1}/\pi)\cdots
\Psi^{*(\ell,1-\ell)}_{\vep_{1}}(i\gt_{1}/\pi))}\nn\\
&&={}^{(1-\ell)}\bra{{\rm vac};pr}\Ad U_{j+\ell}^{-1}(\sigma^{\alpha})
\ket{\gt_1,\cdots,\gt_{2n};pr}^{(1-\ell)}_{\vep_1,\cdots,\vep_{2n}}
\label{disFF}
\ena
with $\ell=0,1$.

\end{appendix}

%%%%%%%%%%%%%%%%%%%%%%%%%
\renewcommand{\baselinestretch}{0.7}
%\bibliography{ff}   % Bibliography
%\bibliography{xxz2}
\bibliography{xxz2}

\end{document}